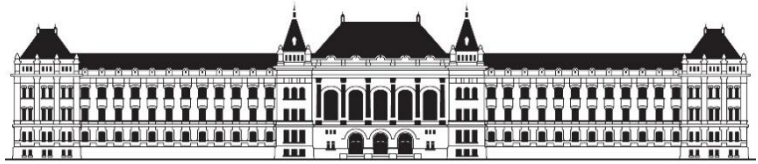

Budapest University of Technology and Economics
Faculty of Electrical Engineering and Informatics
Department of Telecommunications and Media Informatics

# High-Quality Vocoding Design with Signal Processing for Speech Synthesis and Voice Conversion

*Ph.D. Dissertation*
*Doctoral School of Computer Engineering*

**Mohammed Salah Hamza Al-Radhi**
M.Sc. in Communication Systems Engineering

Supervisors:
Tamás Gábor Csapó, Ph.D.
Géza Németh, Habil, Ph.D.

Budapest, Hungary
2020

# Declaration

I, *Mohammed Salah Hamza Al-Radhi*, hereby declare, that this dissertation, and all results claimed therein are my own work, and rely solely on the references given. All segments taken word-by-word, or in the same meaning from others have been clearly marked as citations and included in the references.

Mohammed Salah Hamza Al-Radhi
---
Name

20 April 2020
---
Date

# Abstract

This Ph.D. thesis focuses on developing a system for high-quality speech synthesis and voice conversion. Vocoder-based speech analysis, manipulation, and synthesis plays a crucial role in various kinds of statistical parametric speech research. Although there are vocoding methods which yield close to natural synthesized speech, they are typically computationally expensive, and are thus not suitable for real-time implementation, especially in embedded environments. Therefore, there is a need for simple and computationally feasible digital signal processing algorithms for generating high-quality and natural-sounding synthesized speech. In this dissertation, I propose a solution to extract optimal acoustic features and a new waveform generator to achieve higher sound quality and conversion accuracy by applying advances in deep learning. The approach remains computationally efficient. This challenge resulted in five thesis groups, which are briefly summarized below.

I introduce firstly a new method to shape the high-frequency component of the unvoiced excitation by estimating the temporal envelope of the residual signal. I showed experimentally that this approach is helpful in achieving accurate approximations compared to natural speech. Secondly, I propose a new type of noise masking to reduce the perceptual effect of the residual noise and allowing a proper reconstruction of noise characteristics. The results suggest that the continuous masking approach gives better quality speech than traditional binary techniques of the literature.

Next, I concern with estimating the fundamental frequency (also known as pitch tracking or F0) on clean and noisy speech signals, which acts as a key in speech processing applications. I describe novel approaches which can be used to enhance and optimize some other existing F0 estimator algorithms. Three adaptive techniques based on Kalman-filter, time-warping, and instantaneous-frequency have been developed in order to achieve a robust and accurate continuous F0. As a result, these approaches achieve higher accuracy and smoother continuous F0 trajectory on noisy and clean speech. In addition, I propose and perform an experiment showing that adding a new excitation harmonic-to-noise ratio (HNR) parameter to the voiced and unvoiced components can indicate the degree of voicing in the excitation and reduced the influence of buzziness caused by the vocoder.

Later on, I build and implement deep learning based acoustic modeling using deep feed-forward and sequence-to-sequence recurrent neural networks. A perception and acoustic experiments have shown that the developed vocoder can be applied by the proposed learning framework and showed its superiority against hidden Markov-model based text-to-speech (HMM-TTS).

Afterwards, I propose a new continuous sinusoidal model (CSM) that is applicable in statistical frameworks, which can provide a vocoder with a fixed and low number of parameters and generate high quality synthetic speech compared to state-of-the-art models of speech. I also apply CSM with deep learning based on bidirectional long short-term memory (LSTM) to provide more natural and intelligible TTS capabilities.

Finally, I apply the two vocoders using continuous parameters (source-filter and sinusoidal models) within a voice conversion framework. I experimentally proved that the suggested models give state-of-the-art similarity results.

Overall, this Ph.D. dissertation has established competitive alternative vocoders for speech analysis and synthesis systems. The utilization of proposed models and methods clearly demonstrates that it is compelling to apply them for the statistical parametric speech synthesis and voice conversion.

# Table of Contents

# Chapter 1

# Introduction

> *"Success is a science; if you have the conditions, you get the result."*
> Oscar Wilde (1854 – 1900)

## 1.1 Background and Problem Definition

The quote "*All we need to do is make sure we keep talking*", said by Stephen Hawking, is to draw the encouraging line of all the speech technology. With the fast growth of computer technology to become more functional and prevalent, a wide range of the speech processing area is becoming a core function for establishing a human-computer communication interface. An excellent example of this, among other multimedia applications, is known both as *speech synthesis*, i.e. artificial generation of speech waveforms, and as *text–to–speech*, i.e. building natural-sounding synthetic voices from text. Both are of great current interest and are still receiving much attention from researchers and industry.

State-of-the-art text-to-speech (TTS) synthesis is either based on unit selection [1] or statistical parametric methods [2]. In the last two decades, particular attention has been paid to hidden Markov model (HMM) [3], which has gained much popularity due to its advantages in flexibility, smoothness, and small footprint. Deep neural networks (DNNs) have also become the most common types of acoustic models used in TTS for obtaining high-level data representations and with the availability of multi-task learning a significant improvement in speech quality can be achieved [4]. In view of these systems the speech signal is decomposed to parameters representing excitation (e.g. fundamental frequency, F0) and spectrum of speech, these are fed to a machine learning system. After the statistical model is generated using training data, during synthesis, the parameter sequences are converted back to speech signal with reconstructing methods (e.g. speech vocoders).

Although nowadays TTS systems are intelligible, a limitation of current parametric techniques does not allow full naturalness yet and there is room for improvement in being close to human speech. Mainly, there are three factors in statistical parametric speech synthesis (SPSS) that are needed in order to achieve as high quality synthesized speech as unit selection: improved vocoder techniques, acoustic modeling accuracy, and over-smoothing during parameter generation. Since the design of a vocoder-based SPSS depends on speech characteristics, the preservation of voice quality in the analysis/synthesis phase is the main problem of the vocoder. Although there are vocoding methods which yield close to natural synthesized speech (e.g. the current de facto method, STRAIGHT [5]), their high computational complexity and variable parameters are still considered challenging issues, which present some speech quality degradation in the TTS and other speech applications. Moreover, recent work has demonstrated that a generative WaveNet model [6] yields state-of-the-art performance and gives a good sounding speech in a variety of voices. However, it requires a large quantity of data and computation power which makes it difficult to train for real-time implementation, especially in embedded environments. Therefore, vocoder-based SPSS still provides a quick and flexible solution that can capture high quality synthesized speech. Besides, it gives controllability, which usually is not fully supported by neural vocoders. Hence, there is a need for simple and computationally feasible digital signal processing algorithms for generating natural-sounding synthetic speech.

The goal of another related field of speech technology, *Voice Conversion* (VC), is to modify the speech of a source speaker with digital speech signal processing methods in a way that it will be similar to the speech of a target speaker, while the linguistic content must remain the same. Although there has been a long research in voice conversion [7], current methods lack the flexibility to convert the speech of any source speaker to any other target speaker. Moreover, the naturalness of the converted voice still deteriorates compared to the target speaker due to over-smooth phenomenon or discontinuity problems which make the converted speech sound muffled. Besides, improving the performance of converted voice is still a challenging research question. Therefore, there is also a need to develop advanced adaptable vocoder based VC for achieving high-quality converted speech.

In general, this dissertation proposes a solution to achieve higher sound quality and conversion accuracy with machine learning advances, while its approach remains simple, flexible, and efficient.

## 1.2 Research Objective

The main goal of my Thesis work is to construct a vocoder that is very flexible (whose parameters can be controlled) with respect to achieving high quality synthesized speech. This challenge required five major contributions of the work presented in this dissertation, as depicted in Figure 1:

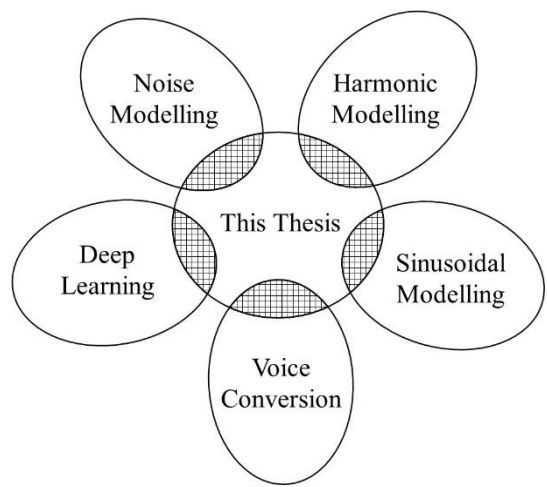

**Figure 1:** Thesis contributions

The *first research* objective is **modulating the noise component of the excitation signal**. It was argued that the noise component is not accurately modelled in modern vocoders (even in the widely used STRAIGHT vocoder). Therefore, two new techniques for modelling voiced and unvoiced sounds are proposed in this part of the research by: 1) estimating the temporal envelope of the residual signal that is helpful in achieving accurate approximations compared to natural speech, and 2) noise masking to reduce the perceptual effect of the residual noise and allowing a proper reconstruction of noise characteristics.

The *second research* objective is **harmonic modelling**. This study focuses on improving the accuracy of the continuous fundamental frequency estimation and the naturalness of the speech signal by proposing three different adaptive techniques based on Kalman filter, time warping, and instantaneous frequency. A clear advantage of the proposed approaches is its robustness to additive noise. Moreover, Harmonic-to-Noise ratio technique is added as a new vocoded-parameter to the voiced and unvoiced excitation signal in order to reduce the buzziness caused by the vocoder.

The *third research* objective is **acoustic modelling design based on deep learning**. In this part of the research, the novel continuous vocoder was applied in the acoustic model of deep learning based speech synthesis using feedforward and recurrent neural networks as an alternative to HMMs. Here, the objective is two-fold: (a) to overcome the limitation of HMM which often generate over-smoothing and muffled synthesized speech, (b) to ensure that all parameters used by the proposed vocoder were taken through training that could synthesize very high quality TTS.

The *fourth research* objective is **designing a sinusoidal modelling system**. Here, a new continuous vocoder was built using a sinusoidal model that is applicable in statistical frameworks which decomposes the speech frames into a harmonic component lower band and a stochastic component upper band based on maximum voiced frequency. The objective is to consider whether a different synthesis technique can produce more accurate synthesis of speech than the source-filter model.

The *fifth research* objective is **proposing a novel model applied for voice conversion with parallel training data**. This part of research includes investigating the novel continuous vocoders in a VC framework. The vocoders are tested both in same-gender and cross-gender scenario.

As a final point, this dissertation provides a detailed and complete study on several speech analysis and synthesis techniques and their applications in text-to-speech and voice conversion.

## 1.3 Methodology

I validated the proposed research by experiments and analytical examinations, in which I developed and improved a novel continuous vocoder for SPSS. The applied methodology of this dissertation follows the international standards. In the following, speech databases, conditions, and evaluation methods are detailed.

### 1.3.1 Continuous Vocoder: Baseline

The first version of our residual-based vocoder was proposed in [8]. Using a continuous F0 (contF0) [9], maximum voiced frequency (MVF) [10], and 24-order Mel-generalized cepstral analysis (MGC) [11] is performed on the speech signal with $alpha = 0.42$ and $gamma = -1/3$. In all steps, $5\ ms$ frame shift is used. The results are the contF0, MVF and MGC parameter streams.

During the synthesis phase, voiced excitation is composed of principle component analysis (PCA) residuals overlap-added pitch synchronously, depending on the contF0. After that, this voiced excitation is low-pass filtered frame by frame at the frequency given by the MVF parameter. In the frequencies higher than the actual value of MVF, white noise is used. Voiced and unvoiced excitation are added together. Finally, a Mel generalized-log spectrum approximation (MGLSA) filter is used to synthesize speech from the excitation and the MGC parameter stream [12].

### 1.3.2 Speech Corpora

In order to evaluate the performance of the suggested models, a database containing a few hours of speech from several speakers was required for giving indicative results. Five English speakers were firstly chosen from the CMU-ARCTIC[1] database [13], denoted BDL (American English, male), JMK (Canadian English, male), AWB (Scottish English, male), CLB (US English, female), and SLT (American English, female). Each one produced one hour of speech data segmented into 1132 sentences, restricting their length from 5 to 15 words per sentence (a total of 10045 words with 39153 phones). Moreover, CMU-ARCTIC are phonetically-balanced utterances with 100% phoneme, 79.6% diphone, and 13.7% triphone coverage. The waveform sampling rate of this database is 16 kHz.

My second database is the corpus created by my co-authors in [J1]. It was motivated by the fact that it builds the first modern standard Arabic audio-visual expressive corpus which is annotated both visually and phonetically. It contains 500 sentences with 6 emotions (Happiness – Sadness – Fear – Anger – Inquiry – Neutral), and recorded by a native Arabic male speaker (denoted ARB). The waveform sampling rate of this database is 48 kHz.

The third corpus is the one based on Hungarian language. Two Hungarian male and two female subjects (4 speakers) with normal speaking abilities were recorded while reading sentences aloud (altogether 209 sentences each). The ultrasound data and the audio signals were synchronized using the tools provided by Articulate assistant advanced software [C6]. The waveform sampling rate of this database is 44 kHz.

### 1.3.3 Reference Systems

The proposed vocoders based TTS and VC systems were evaluated by comparing them with strong reference systems. STRAIGHT [5] and WORLD [14] are high-quality vocoders and widely regarded as the state-of-the-art models in SPSS. MagPhase [15] and log domain pulse model (PML) [16] are considered as modern sinusoidal models. Sprocket [17] is a vocoder-free VC system that was used recently for the voice conversion challenge in 2018. YANGsaf algorithm [18] is an F0 estimator method that can be compared along with adaptive contF0 in Chapter 4. The choice of YANGsaf is confirmed by the fact that it was recently shown in [19] to outperform other well-known F0 estimation approaches found in the current literature (like YIN, RAPT, or DIO).

---

[1] http://www.festvox.org/cmu_arctic/

### 1.3.4 Experimental Conditions

I used the open source Merlin[2] [20] toolkit to implement machine learning methods introduced in Part III and V. Constructive changes are also introduced in this toolkit to be able to adapt the proposed vocoder. The training sets contain 90% of the speech corpora, while the rest were used for testing. The training procedures were conducted on a high-performance NVidia Titan X GPU. In the vocoding experiments, 100 sentences from each speaker were analyzed and synthesized with the baseline and proposed vocoder(s).

With the purpose of assessing true performance of the continuous pitch tracking presented in Chapter 4, a reference pitch contour (ground truth) is estimated from the electro-glottograph (EGG) as it is directly derived from glottal vibration and is largely unaffected by the nonharmonic components of speech. In my evaluation, the ground truth is extracted from EGG signals using Praat's autocorrelation-based pitch estimation algorithm [21]. The analysis of the measurements and the statistical examinations were made by Python 3.5 and MATLAB 2018b.

### 1.3.5 Perceptual Listening Tests

I conducted several web-based MUSHRA (MUlti-Stimulus test with Hidden Reference and Anchor) listening tests [22] in order to evaluate which system is closer to the natural speech. I compared natural sentences with the synthesized sentences from the baseline, proposed, and a hidden anchor system (different for each test). Listeners were asked before the test to listen to an example from a speaker to adjust the volume. In the test, the listeners had to rate the naturalness of each stimulus relative to the reference (which was the natural sentence), from 0 (highly unnatural) to 100 (highly natural). The utterances were presented in a randomized order (different for each participant).

Besides, Mean Opinion Score (MOS) test was also carried out in Chapter 10 and 11. In the MOS test, I compared three variants of the sentences: 1) Target, 2) Converted speech using the baseline systems, and 3) Converted speech using the proposed vocoder. Similarly to the MUSHRA test, the listeners had to rate the naturalness of each stimulus, from 0 (highly unnatural) to 100 (highly natural).

About 20 participants (for each test) between the age of 23-40 (mean age: 30 years), mostly with engineering background, were asked to conduct the online listening tests. Altogether, 10 MUSHRA with 2 MOS tests were performed during my research work to evaluate my dissertation. On average, the MUSHRA test took 14 minutes, while the MOS test was 12 minutes long.

---

[2] https://github.com/CSTR-Edinburgh/merlin

# Part I
# Excitation Noise Modelling

# Chapter 2

# Temporal Envelopes

> *"If your experiment needs statistics, you ought to have done a better experiment."*
> Ernest Rutherford (1871 – 1937)

## 2.1 Introduction

A statistical speech synthesis framework is guided by the vocoder (which is also called speech analysis/synthesis system) to reproduce human speech. Although there are several different types of vocoders that use analysis/synthesis, they follow the same main strategy. The analysis stage is used to convert the speech waveform into a set of parameters which represent separately the vocal-folds excitation signal (sound is voiced or unvoiced) and vocal-tract filter transfer function to filter the excitation signal (vocal-folds movements), whereas in the synthesis stage, the entire parameter set is used to reconstruct the original speech signal.

Since the design of a vocoder-based SPSS depends on speech characteristics, the preservation of voice quality in the analysis/synthesis phase and the irregular "buzzy" synthetic speech sounds are the main problems of several vocoders. Although some other vocoder-based methods have been recently developed and applied to produce high quality speech synthesis (for a comparison, see [23]), they are not successful in synthesizing high-quality and natural-sounding speech. The reason for this is the inaccurate composition and estimation of the vocoder parameters which leads to a degradation in the speech signal. Therefore, this chapter considers the above issues by suggesting a robust method for advanced modelling of the noise excitation which can yield an accurate noise component.

To reconstruct the time-domain characteristics of the noise part, temporal envelope based-technique gives a reliable estimation that follows closely sudden variations in amplitude and avoids ripples in more stable regions. While the voiced parts describe efficiently the periodicities in continuous vocoder, modeling of the noise part introduces artifacts because of the specific time-domain characteristics of noise might appear in voiced speech. Thus, I propose four time-envelope to model the temporal characteristics of noise in the context of continuous vocoder. Figure 2 shows the whole parts of the proposed continuous vocoder in this dissertation, in which the dashed box (envelope estimation) in the lower half of the figure is related to this chapter.

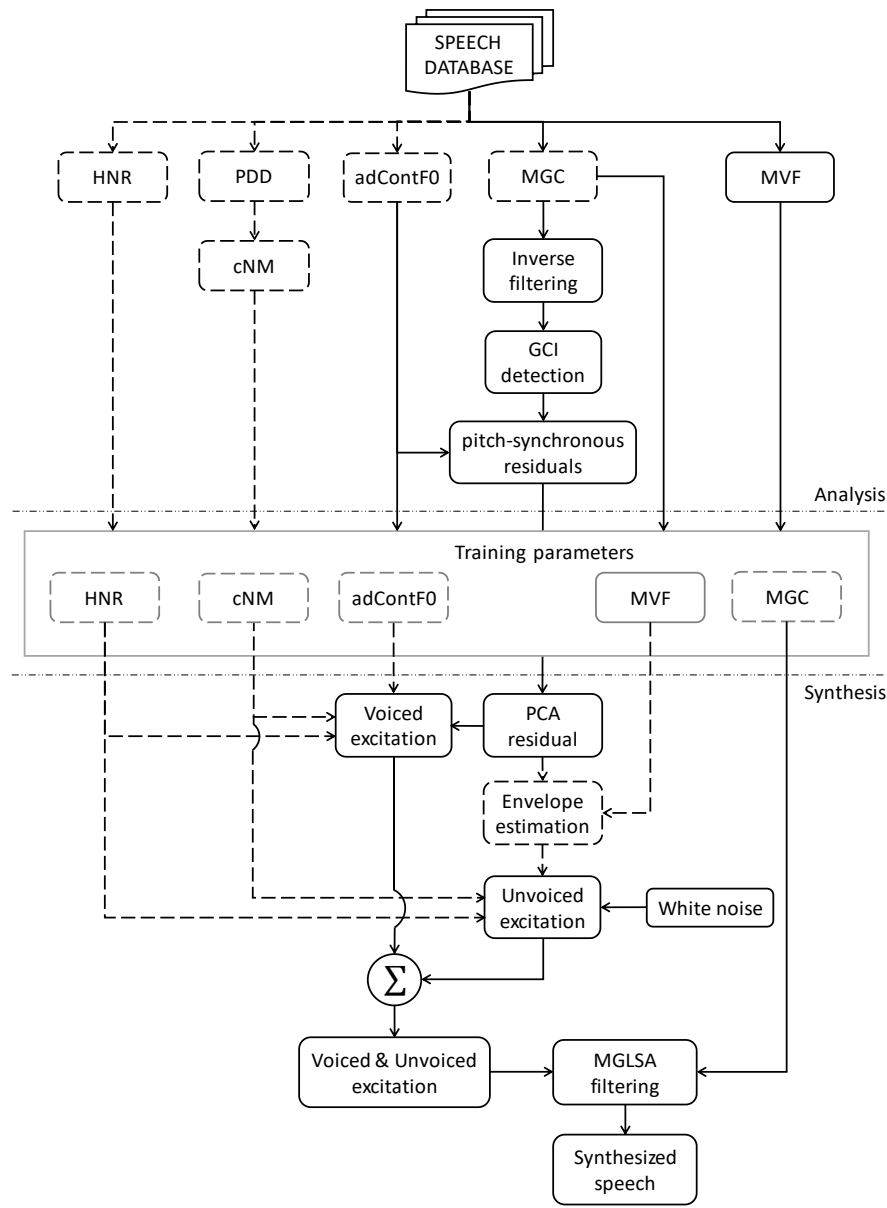

**Figure 2:** Schematic diagram of the developed continuous vocoder. Additions and refines are marked with dashed lines.

## 2.2 Modeling of the Residual Signal

Accurate modeling of the residual signal (excitation) has been shown to significantly enhance the synthesis quality (in terms of buzziness) than the traditional pulse-noise excitation [23] [24]. Here, the residual signal is obtained by MGLSA inverse filtering [12]. Then, the Glottal Closure Instant (GCI) detection algorithm [24] is used to find the glottal period boundaries in the voiced parts of the residual signal. GCIs indicate the instants at which significant excitation of vocal-tract take place during voicing. Finally, the pitch-synchronous residuals are isolated by a GCI-centered two-period long excitation frames.

In order to model their low-frequency contents, pitch-synchronous residual frames can be decomposed by principle component analysis (PCA). PCA is an orthogonal linear transformation which applies a rotation of the axis system in order to find the best representation of lower dimensionality input data in a way that maximizes the data dispersion along the new axes. It makes the large data set simpler, easy to explore, and visualize. PCA can be run numerous times with variables being removed or added at every run, only if those manipulations are validated in the analyses. As a result, PCA reduces the computational complexity of the model which makes machine learning algorithms run faster and reduce memory usage.

PCA can be achieved firstly by calculating the eigenvalues and eigenvectors of the data covariance matrix. The top eigenvector that correspond to the largest eigenvalue is then chosen. Next, the projection matrix $W$ can be constructed from the selected eigenvector. Lastly, transform the original $d$-dimensional dataset via $W$ to obtain the new $m$-dimensional feature subspace (where $m \ll d$).

If we have a dataset consists of $N$ residual frames, then the PCA computation will lead to $k$ eigenvalues $\lambda_i$ with their corresponding eigenvectors $\mu_i$. $\lambda_i$ represents the data dispersion along axis $\mu_i$ and is consequently a measure of the information this eigenvectors conveys on the dataset. To give an idea of what a PCA residual looks like in a continuous vocoder, Figure 3 exhibits the first eigenvector, interpreted as the principal pattern arising from the data, for the female SLT speaker ($k = 168$) and for the male AWB speaker ($k = 258$). Instead of impulses, this PCA residual will be used for the synthesis of the voiced frames.

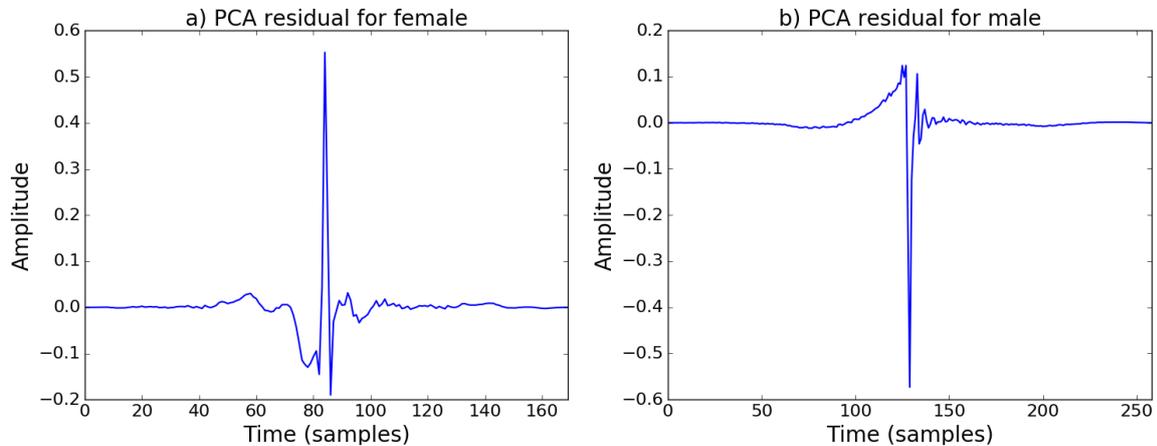

**Figure 3:** Illustration of the first eigenvector $\boldsymbol{\mu_1(n)}$ for a given speaker from the CMU-ARCTIC database.

## 2.3 Parameterizing the Noise Components

Degottex and Erro [25] argue that the noise component is not accurately modelled in modern vocoders (even in the widely used STRAIGHT vocoder). It was also shown that in natural speech, the high-frequency noise component is time-aligned with the F0 periods [26]. Therefore, I designed a temporal envelope to shape the high-frequency component (above MVF) of the excitation by estimating the envelope of the PCA residual; that is helpful in achieving accurate approximations compared to natural speech. I added a time-domain envelope to the unvoiced excitation to make it more similar to the residual of natural speech. The advantage of this approach is a parametric description of speech that tries to refine the time accuracy to obtain valuable moments during transient events like noise bursts (fricatives, affricates, or stop bursts), and fits a curve that approximately matches the peaks of the residual frame[3].

### 2.3.1 Amplitude Envelope

The amplitude envelope refers here to the shape of sound energy over time. It is usually calculated as filtering the absolute value of the voiced frame $v(n)$ by moving the average filter to the order of $2N + 1$ [27], where N is chosen to be 10. The amplitude envelope is given by

$$A(n) = \frac{1}{2N+1} \sum_{k=-N}^{N} |v(n-k)| \tag{1}$$

Previous work showed that by down-sampling the amplitude envelope to a different number of samples will reduce the relative time square error [28] during parameterizing the noise components. Figure 6b shows the effects of applying the amplitude envelope on the PCA residual signal.

### 2.3.2 Hilbert Envelope

Another method of calculating an envelope is based on the Hilbert transform technique [29], which has been used first to obtain an analytical signal in the complex-valued time-domain. Here, the analytic signal $v_a(n)$ can be defined as a complex function of time derived from a real voiced frame $v(n)$, and can be written as

$$v_a(n) \stackrel{\text{def}}{=} v(n) + j\mathcal{H}\{v(n)\} \tag{2}$$

where $j$ is the imaginary unit $\sqrt{-1}$, and $j\mathcal{H}\{\cdot\}$ denotes the operation of the Hilbert transform which is equivalent to the integration form [30]

$$\mathcal{H}\{v(n)\} = \frac{1}{\pi} \int_{-\infty}^{+\infty} v(\tau) \frac{1}{n-\tau} d\tau = \frac{1}{\pi n} * v(n) \tag{3}$$

where $*$ stands for convolution symbol. Thus, the Hilbert envelope $H(n)$ can be estimated by taking the magnitude of the analytical signal to capture the slowly varying features of the sound signal (see Figure 6c) as

$$H(n) = |v_a(n)| = \sqrt{v(n)^2 + \mathcal{H}\{v(n)\}^2} \tag{4}$$

### 2.3.3 Triangular Envelope

A further time domain parametric envelope that can be easily applied to each frame signal is the triangular envelope. It was proposed in [31] by using four parameters as it assumes the triangle to be symmetric. In [32], a polynomial curve was used to detect these parameters. In this work, our approach for estimating the triangular envelope $T(n)$, which is slightly different from [31], is only using three parameters (a, b, and c) obtained by detecting them directly on the envelope. Here, the design parameters have been given as: $a = 0.35L_f$, $b = 0.65L_f$, $c =$

---

[3] This section describe four different temporal envelopes based on [J1] [C1], while others based on discrete all-pole and frequency domain linear predication approaches are investigated in [J5].

$\frac{a+b}{2}$, and we set $A = 1$; where $L_f$ is the frame length. These parameters are illustrated in Figure 4, and the performance of the Triangular envelope can be observed in Figure 6d.

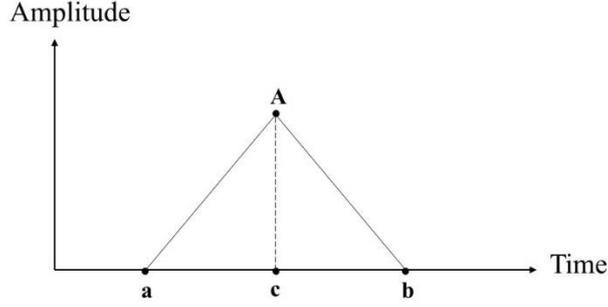

**Figure 4:** Triangular time-domain envelope estimation.

### 2.3.4 True Envelope

Another new approach, which can be used for estimating the time domain envelope, is called the true envelope (TE). It is based on cepstral smoothing of the amplitude spectrum [33] [34]. In an iterative procedure, the TE algorithm starts with estimating the cepstrum and updating it with the maximum of the original spectrum signal and the current cepstral representation. To have an efficient real-time implementation, [35] proposed a concept of a discrete cepstrum which consists of a least mean square approximation, and [36] added a regularization technique that aims to improve the smoothness of the envelope. In this study, the procedures for estimating the TE is shown in Figure 5 in which the cepstrum $c(n)$ can be calculated as the inverse Fourier transform of the log magnitude spectrum $S(k)$ of a signal frame $v(n)$

$$c(n) = \sum_{k=0}^{N-1} S(k) * e^{j\left(\frac{2\pi}{N}\right)kn} \tag{5}$$

$$S(k) = log|V(k)| \tag{6}$$

where $V(k)$ is $N$-point discrete Fourier transform of a $v(n)$, and can be found mathematically as

$$V(k) = \sum_{n=0}^{N-1} v(n) * e^{-j\left(\frac{2\pi}{N}\right)nk} \tag{7}$$

Next, the algorithm iteratively update $M(k)$ with the maximum of $S(k)$ and the Fourier transform of the smoothing cepstrum $C_i(k)$, that is the cepstral representation of the spectral envelope at iteration $i$.

$$C(k) = \sum_{n=0}^{N-1} c(n) * e^{-j\left(\frac{2\pi}{N}\right)nk} \tag{8}$$

$$M_i(k) = \max\bigl(S_{i-1}(k), C_{i-1}(k)\bigr) \tag{9}$$

It can be noted that the TE with weighting factor $w_f$ will bring a unique time envelope which makes the convergence more close to natural speech. In practice, the $w_f$ which was found to be the most successful is 10. Thus, TE envelope $T(n)$ is proposed here as

$$T(n) = \sum_{k=0}^{N-1} w_f * M(k) * e^{j\left(\frac{2\pi}{N}\right)kn} \tag{10}$$

Despite the good performance, TE makes oscillations whenever the change in $S(k)$ is fast. This can be seen in Figure 6e.

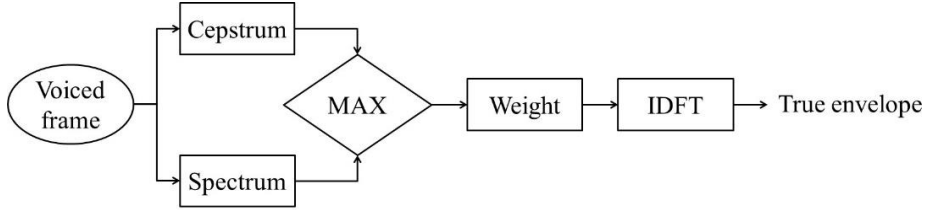

**Figure 5:** Procedures for estimating the true envelope.

## 2.4 Evaluation

In order to achieve our goals and to verify the effectiveness of the proposed methods, objective and subjective evaluations were carried out.

### 2.4.1 Phase Distortion Deviation

Recent progress in the speech synthesis field showed that the phase distortion of the signal carries all of the crucial information relevant to the shape of glottal pulses [25]. As the noise component in the developed continuous vocoder is parameterized in terms of time envelopes and computed for every pitch-synchronous residual frame, I compared the natural and vocoded sentences by measuring the phase distortion deviation (PDD). PDD can be estimated in this experiment at 5 $ms$ frame shift by

$$PDD = \sigma_i(f) = \sqrt{-2\log\left|\frac{1}{N}\sum_{n \in C} e^{j(PD_n(f) - \mu_n(f))}\right|} \tag{11}$$

$$\mu_n(f) = \angle\left(\frac{1}{N}\sum_{n \in C} e^{jPD_n(f)}\right) \tag{12}$$

where $C = \left\{i - \frac{N-1}{2}, \ldots, i + \frac{N-1}{2}\right\}$, $N$ is the number of frames, PD is the phase difference between two consecutive frequency components, and we denote the phase by $\angle$. As I wanted to quantify the noisiness in the higher frequency bands only, I zeroed out the PDD values below the MVF contour.

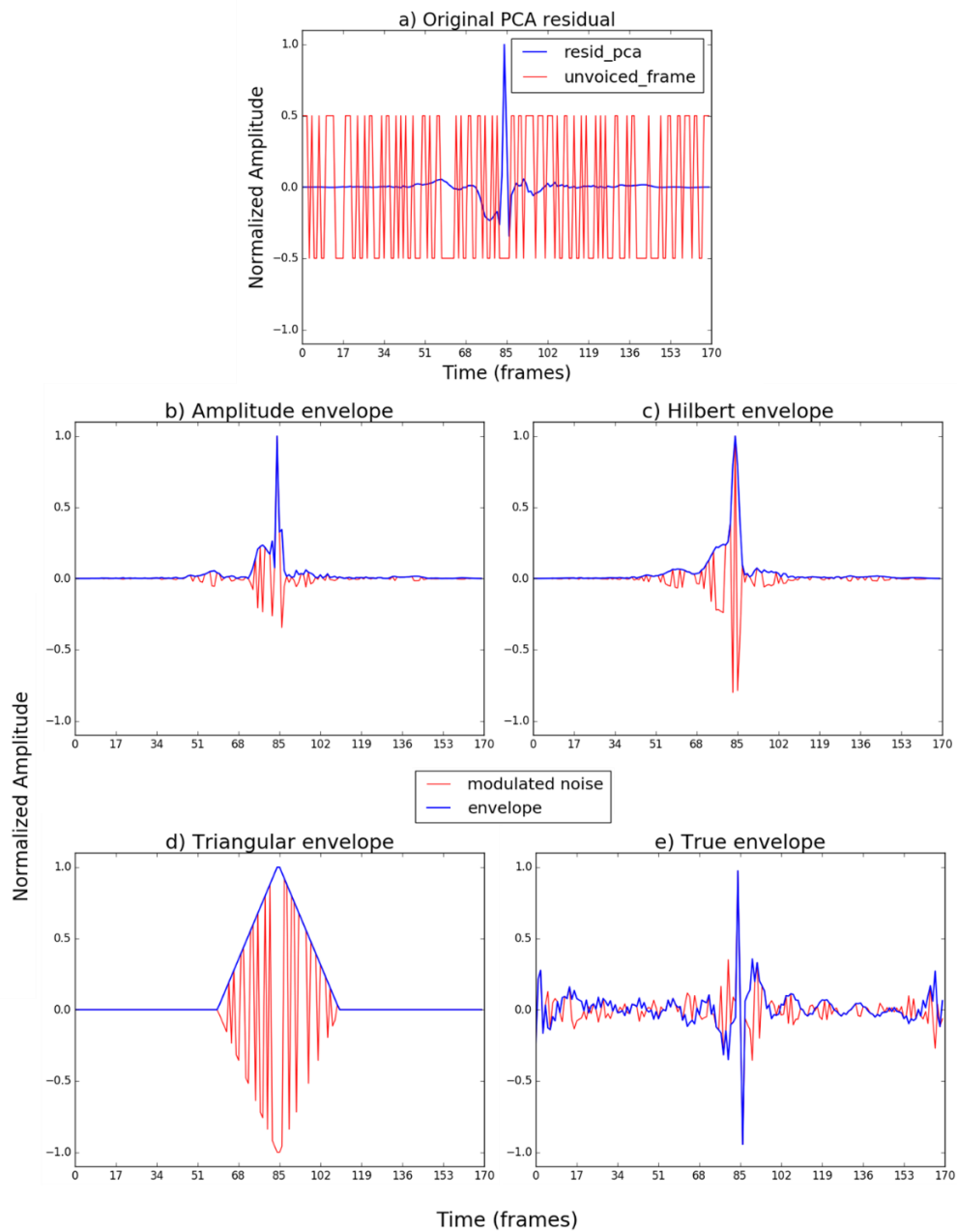

**Figure 6:** Illustration of the performance of the time envelopes. "unvoiced_frame" is the excitation signal consisting of white noise, whereas "resid_pca" is the first eigenvector resulting from the PCA compression on the voiced excitation frame, so here I have 1$^{st}$ PCA component.

Samples for the PDD of one natural and five vocoded utterances in comparison to the high-quality STRAIGHT vocoder are shown in Figure 7. For the four methods, significant differences between the vocoded samples of the different envelope types can be noted. As can be seen, the baseline vocoding sample has too much noise component compared to the natural sample (e.g. see the colors between 1 – 1.7s in English and 1.4 – 2s in Arabic sentences). On the other hand, the proposed systems with envelopes and STRAIGHT have PDD values (i.e., colors in the figure) closer to the natural speech. In particular, the 'Amplitude' envelope system results in too low PDD values, meaning that the noisiness is too low compared to natural speech. Otherwise, in general the proposed framework using 'Hilbert' and 'True' envelopes provide high-quality vocoding for SPSS.

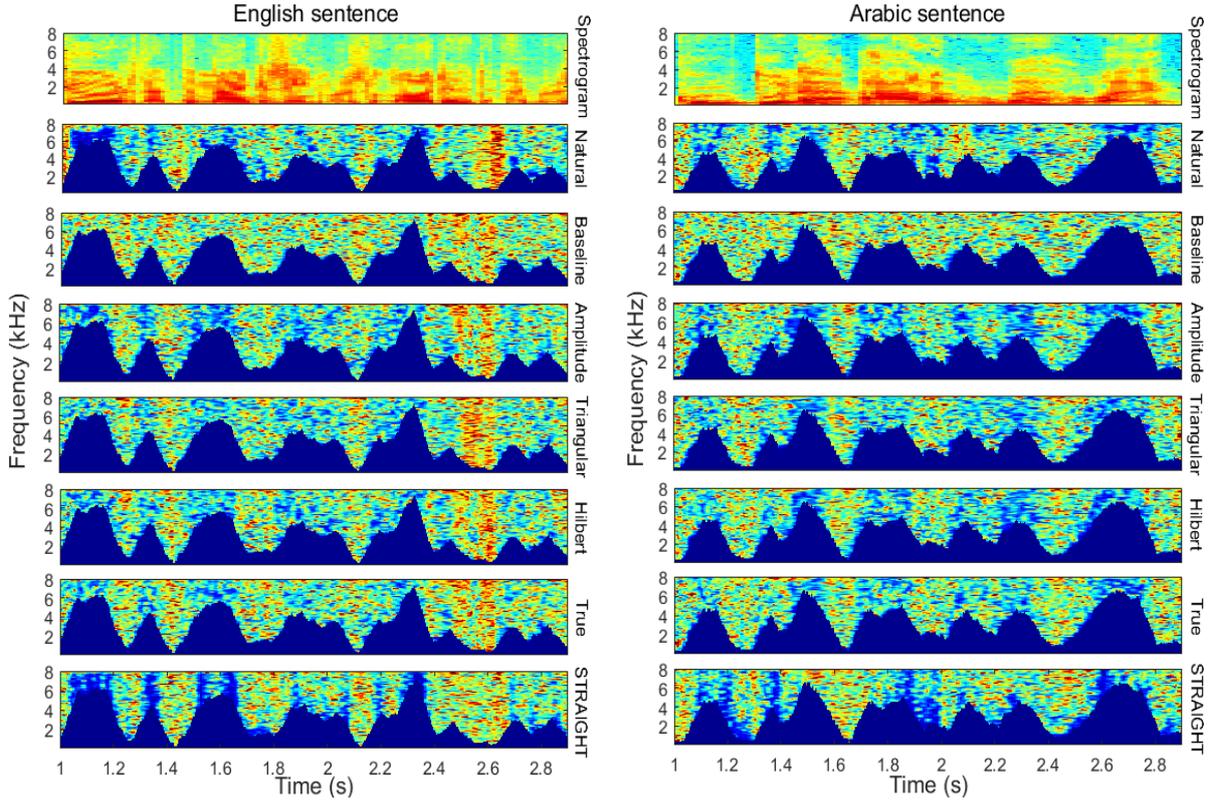

**Figure 7:** Phase Distortion Deviation of natural and vocoded speech samples above the Maximum Voiced Frequency region. The top row shows the spectrogram of the natural utterances. English sentence: "He made sure that the magazine was loaded, and resumed his paddling.", from speaker AWB; and an Arabic sentence "كنت لا أعلم ماهو كما كنت لا أهتم به" translated as "I did not know what it was as I did not care about it" and the Latin transcription is "knt la ahlm mahowo kma knt la ahtm bh". The warmer the color, the bigger the PDD value and the noisier the corresponding time-frequency region.

### 2.4.2 Objective Measurement

A range of speech quality and intelligibility metrics are considered to evaluate the quality of the proposed model. A calculation is done frame-by-frame, and the results were averaged over the test utterances for each speaker. The following four evaluation metrics were used:

a) **Weighted Spectral Slope (WSS)** [37]: The algorithm first decomposes the frame signal into a set of frequency bands. The intensities within each critical band are measured. Then, a weighted distance between the measured slopes of the log-critical band spectra are computed

$$WSS = \frac{1}{N} \sum_{j=1}^{N} \left( \frac{\sum_{i=1}^{K} W_{i,j}(Y_{i,j} - X_{i,j})^2}{\sum_{i=1}^{K} W_{i,j}} \right) \qquad (13)$$

where $N$ is the number of frames in the utterance, and $K$ is the number of sub-bands. $W_{i,j}$, $X_{i,j}$, and $Y_{i,j}$ denote the weight, the spectral slope of natural and synthesized speech; respectively, at the $i^{th}$ frequency band and $j^{th}$ frame.

b) **Normalized Covariance Metric (NCM)** [38]: It is based on a Speech Transmission Index (STI) [39], which uses covariance coefficient $r$ of the Hilbert envelope between the natural and vocoded frame signal

$$NCM = \frac{1}{N} \sum_{j=1}^{N} \left( \frac{\sum_{i=1}^{K} W_{i,j} \cdot \log \frac{r_{i,j}^2}{1 - r_{i.j}^2}}{\sum_{i=1}^{K} W_{i,j}} \right) \quad (14)$$

where $W$ is the weight vector applied to the STI of $K$ bands and can be found by the articulation index [40].

c) **frequency-weighted segmental SNR (fwSNRseg)** [38]**:** Similarly to Equation (14), $fwSNR_{seg}$ can be estimated by

$$fwSNR_{seg} = \frac{1}{N} \sum_{j=1}^{N} \left( \frac{\sum_{i=1}^{K} W_{i,j} \cdot \log \frac{X_{i,j}^2}{X_{i,j}^2 - Y_{i.j}^2}}{\sum_{i=1}^{K} W_{i,j}} \right) \quad (15)$$

where $X_{i,j}^2$, $Y_{i.j}^2$ are critical-band magnitude spectra in the $j^{th}$ frequency band of the natural and synthesis frame signals respectively, $K$ is the number of bands, $W$ is the weight vector defined in [40].

d) Jensen and Taal [41] introduced an **Extended Short-Time Objective Intelligibility (ESTOI)** measure that can be used here to calculate the correlation between the temporal envelopes of natural and synthesized speech.

Table 1 displays the results of the evaluation of four methods in comparison to the STRAIGHT vocoder. As Table 1 shows, the proposed methods tend to significantly outperform the baseline approach among all metrics, suggesting the superiority of these techniques. In particular, we see that NCM and ESTOI measures display that the proposed vocoder based on Hilbert and True envelopes are closer to STRAIGHT in all speakers. Hence, we can conclude that the temporal envelope based approaches were beneficial to model the noise component. But it should be pointed out that there is no guarantee that better objective measures yield a better model as synthetic speech quality is an inherently perceptual study.

**Table 1:** Average scores performance of resynthesized speech signal per each speaker. The bold font shows the best performance among the proposed vocoder variants.

| Metric | Speaker | Models | | | | | |
|---|---|---|---|---|---|---|---|
| | | Baseline | Amplitude | Hilbert | Triangular | True | STRAIGHT |
| **fwSNRseg** | AWB | 6.971 | 9.638 | 9.665 | 9.566 | **9.693** | 12.209 |
| | SLT | 8.020 | 10.820 | **10.949** | 10.775 | 10.919 | 15.427 |
| | ARB | 9.288 | 12.732 | 12.748 | 12.664 | **12.770** | 14.248 |
| **NCM** | AWB | 0.642 | 0.865 | **0.872** | 0.863 | 0.871 | 0.990 |
| | SLT | 0.665 | 0.867 | **0.883** | 0.863 | 0.880 | 0.981 |
| | ARB | 0.682 | 0.864 | 0.875 | 0.857 | **0.876** | 0.978 |
| **ESTOI** | AWB | 0.532 | 0.785 | 0.785 | 0.782 | **0.786** | 0.796 |
| | SLT | 0.665 | 0.868 | **0.872** | 0.865 | 0.871 | 0.943 |
| | ARB | 0.599 | 0.847 | **0.853** | 0.845 | 0.851 | 0.880 |
| **WSS** | AWB | 54.162 | **37.158** | 37.362 | 37.524 | 37.404 | 35.586 |
| | SLT | 58.449 | 40.168 | 40.401 | **40.150** | 40.488 | 21.719 |
| | ARB | 43.359 | 26.977 | **26.932** | 27.051 | 27.066 | 24.589 |

### 2.4.3 Subjective Listening Test

In order to evaluate which proposed vocoder variant is closer to the natural speech, I conducted a web-based MUSHRA listening test. I compared natural sentences with the synthesized sentences from the baseline, proposed, STRAIGHT, and a hidden anchor system (the latter being a vocoder with simple pulse-noise excitation). In this section, I show the results of two perceptual listening tests.

**Listening test #1: English corpus**

A total number of 19 participants (8 males and 11 females) were asked to conduct the online listening test. The listening test samples can be found online[4].

The MUSHRA scores of the listening test are presented in Figure 8. It can be observed that all of the proposed systems significantly outperformed the baseline. For the male speaker in this experiment (see Figure 8a), out of the proposed versions, the Amplitude, Hilbert, and True reached the highest naturalness scores in the listening test. It is worth mentioning that the proposed systems have significantly higher ratings than those of STRAIGHT for the female voice (see Figure 8b). Overall, our model contributes notably to the synthetic quality of the proposed vocoder than other systems (see Figure 8c).

---

[4] http://smartlab.tmit.bme.hu/vocoder_Arabic_2018

I therefore draw the conclusion that the average scores achieved by the proposed vocoder based Hilbert and True envelope significantly outperformed STRAIGHT in case of the female speaker, while they reached almost the highest naturalness for the male speaker. This means that the approach presented in this work is an interesting alternative to the earlier version of the residual-based vocoder [8], and at least for the female voice in the STRAIGHT vocoder.

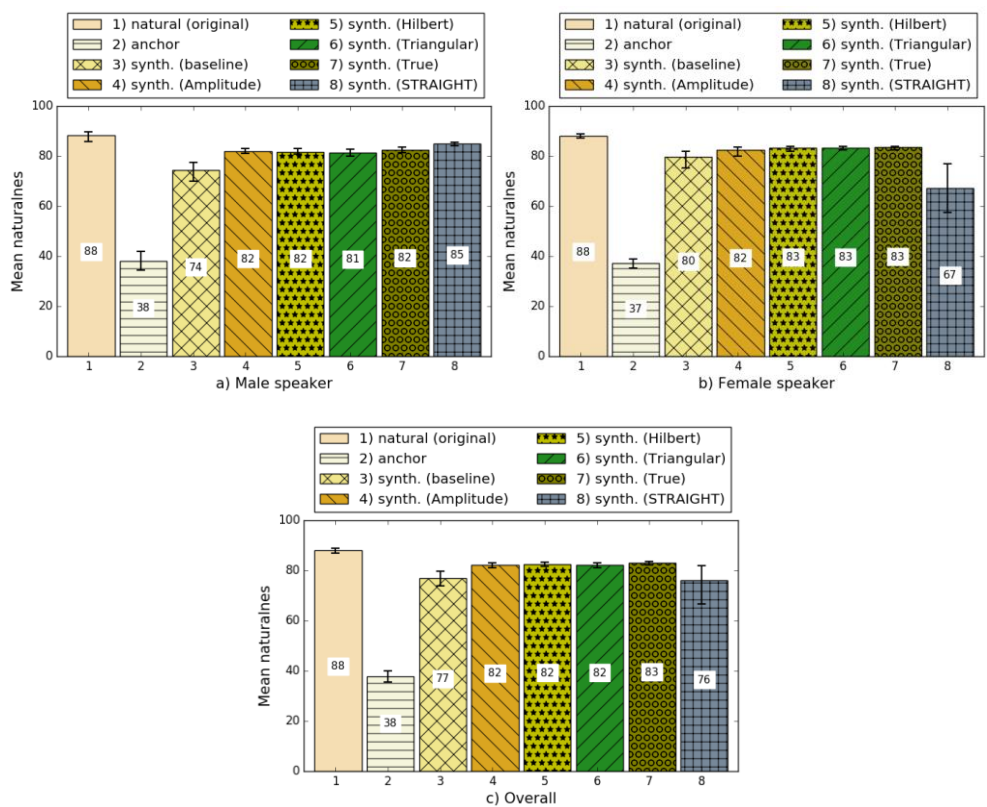

**Figure 8:** Results of the subjective evaluation #1 (English samples) for the naturalness question. Higher value means better naturalness. Error bars show the bootstrapped 95% confidence intervals.

**Listening test #2: Arabic corpus**

For a second MUSHRA test, twelve sentences were selected from the Arabic corpus (two from each emotion). Altogether, 84 utterances were included in the test (1 speaker x 7 types x 6 emotions x 2 sentences). Another set of 21 participants (8 males and 13 females) were asked to conduct the online listening test. All of them were native Arabic speakers and none of them reported any hearing loss. On average, the test took 20 minutes to fill. The listening test samples can be found online[4].

The MUSHRA scores of the listening test are presented in Figure 9. Here, a number of observations can also be made. The results show that all of the proposed systems are significantly better than the baseline. It is also important to note that the difference between the proposed systems using the envelopes and the STRAIGHT vocoder is significant, meaning that our system could reach the quality of state-of-the-art vocoders. Focusing on the Neutral and Sad types, it is obvious that STRAIGHT works better (with mean naturalness of 86%) than other methods. For the other emotions, the proposed vocoder based on the envelopes was superior with mean naturalness of 86% in Anger, 85% in Fear, 87% in Happy, and 85% in Question. Overall, our proposed vocoder is preferred in synthesized Arabic speech and reached the highest rate (85%) in the listening test, being evaluated higher than the STRAIGHT (78%) and baseline (77%) vocoders.

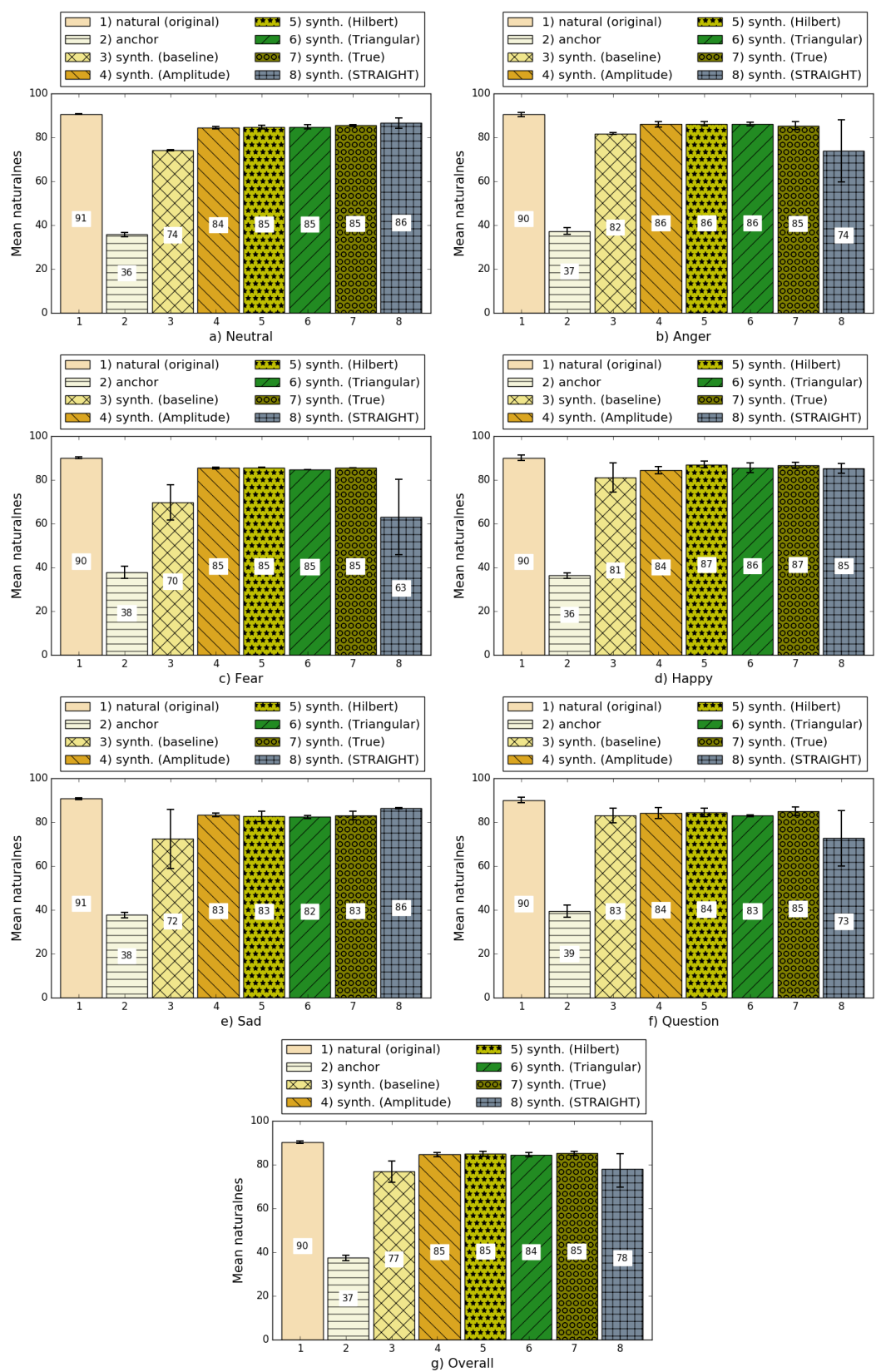

**Figure 9:** Results of the subjective evaluation #2 (Arabic samples) for the naturalness question. Higher value means better naturalness. Error bars show the bootstrapped 95% confidence intervals.

## 2.5 Discussion

To discuss the trend of why STRAIGHT synthesizer scores fell below 70 in English female speaker and below 80 in some Arabic emotions, PDD samples of natural and vocoded utterances by STRAIGHT are shown in Figure 10. The main cause seems to be the error that the voiced segment was wrongly affected by higher frequency harmonics (e.g. above 5 kHz between 0.2-0.5s on Figure 10 left), which degrades the quality of the synthesized speech; thus explaining the lower value for the English female speaker and Anger, Fear, Question for the Arabic male speaker. Conversely, the synthetic speech of the proposed technique exceeds this limitation by controlling the harmonic frequencies and improves speech quality as previously described and shown in Section 2.3.

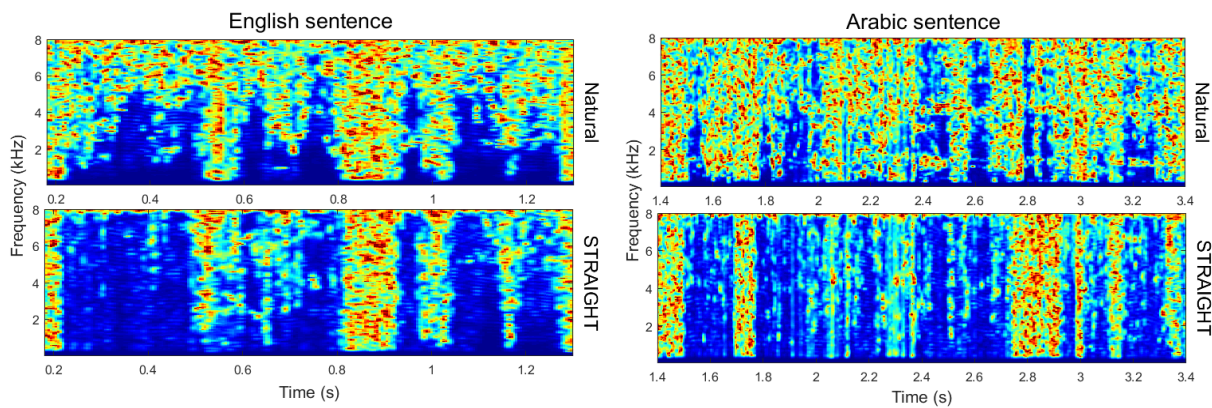

**Figure 10:** Phase Distortion Deviation of a natural and vocoded speech samples based STRAIGHT vocoder. English sentence: "Will we ever forget it.", from speaker SLT; and an Arabic sentence: "يحسن ان ابدأ مذاكرتي الان حتى لايختلط علي الامر" translated as "it would be better to start studying now as I don't want to lose time" and the Latin transcription is "yuhsen en abda mothakeraty alaan heta la yakhtalet aly alamer". The warmer the color, the bigger the PDD value and the noisier the corresponding time-frequency region.

Some confusion of Arabic emotion types of speech synthesized by our methods and STRAIGHT was observed during the results of the listening test #2. Therefore, the empirical cumulative distribution function of phase distortion mean values are calculated and displayed in Figure 11 to see whether these systems can be normally distributed and how far they are from the natural signal. The empirical cumulative distribution function $F_n(PDM)$ is defined as

$$F_n(x) = \frac{\#\{X_i : X_i \leq x\}}{n} = \frac{1}{n}\sum_{i=1}^{n} I_{X_i \leq x}(X_i) \tag{16}$$

where $X_i$ is the PDM variables with density function $f(x)$ and distribution function $F(x)$, $\#A$ symbolizes the number of elements in the set $A$ ($X_i \leq x$), $n$ is the number of experimental observations, $I$ is the indicator of event $A$ given as

$$I_A(x) = \begin{cases} 1, & x \in A \\ 0, & x \notin A \end{cases} \tag{17}$$

It can be noticed that the higher mode of the distribution (positive x-axis in Figure 11b-f) corresponding to STRAIGHT's PDMs are clearly higher than that of the original signal. This also demonstrates why the synthesized speech for the STRAIGHT ranked lower in the perception test. On the contrary, the higher mode of the distribution corresponding to the proposed configurations are better reconstructed especially in the emotions of Anger, Fear, and Question. These results can be explained by the fact that modulating high frequencies based on time envelope is still beneficial and can substantially reduce any residual buzziness. Focusing on the lower mode of the distribution (negative x-axis in Figure 11a, e), STRAIGHT's PDMs gives better synthesized performance than other systems for the Neutral and Sad emotions, whereas the proposed vocoder almost reaches the natural distribution for fear and question emotions (Figure 11c, f).

Consequently, the experimental results verify the effectiveness of the proposed vocoder in terms of speech naturalness; and it is comparable, or even better in some cases, than STRAIGHT. In particular, our emotional Arabic utterances are also more suitable to model with the continuous vocoder applying envelopes and provide a better performance in Arabic speech re-synthesis.

## 2.6 Summary

This chapter has presented a new approach for modelling unvoiced sounds in a novel continuous vocoder, and evaluating it using English and Arabic speech samples. The main idea was to further control the time structure of the high-frequency noise component by estimating a suitable temporal envelope.

Using a variety of measurements, the performance strengths and weaknesses of each of the proposed methods for different speakers were highlighted. From the objective experiments, it was shown that the proposed vocoders have a better capability for modelling the time structure of the noise component than the baseline. The Hilbert and True envelopes are the best when combined with the continuous vocoder (i.e. they are close to the natural sentences in terms of PDD). Furthermore, the results of the MUSHRA test demonstrated the effectiveness of the proposed approaches for improving the quality of synthetic speech. It was shown that the proposed vocoder outperformed the state-of-the-art (STRAIGHT) models in Arabic and female English speakers.

The results obtained in this chapter will allow us to enhance the performance of other types of vocoders in order to yield a more natural synthetic signal.

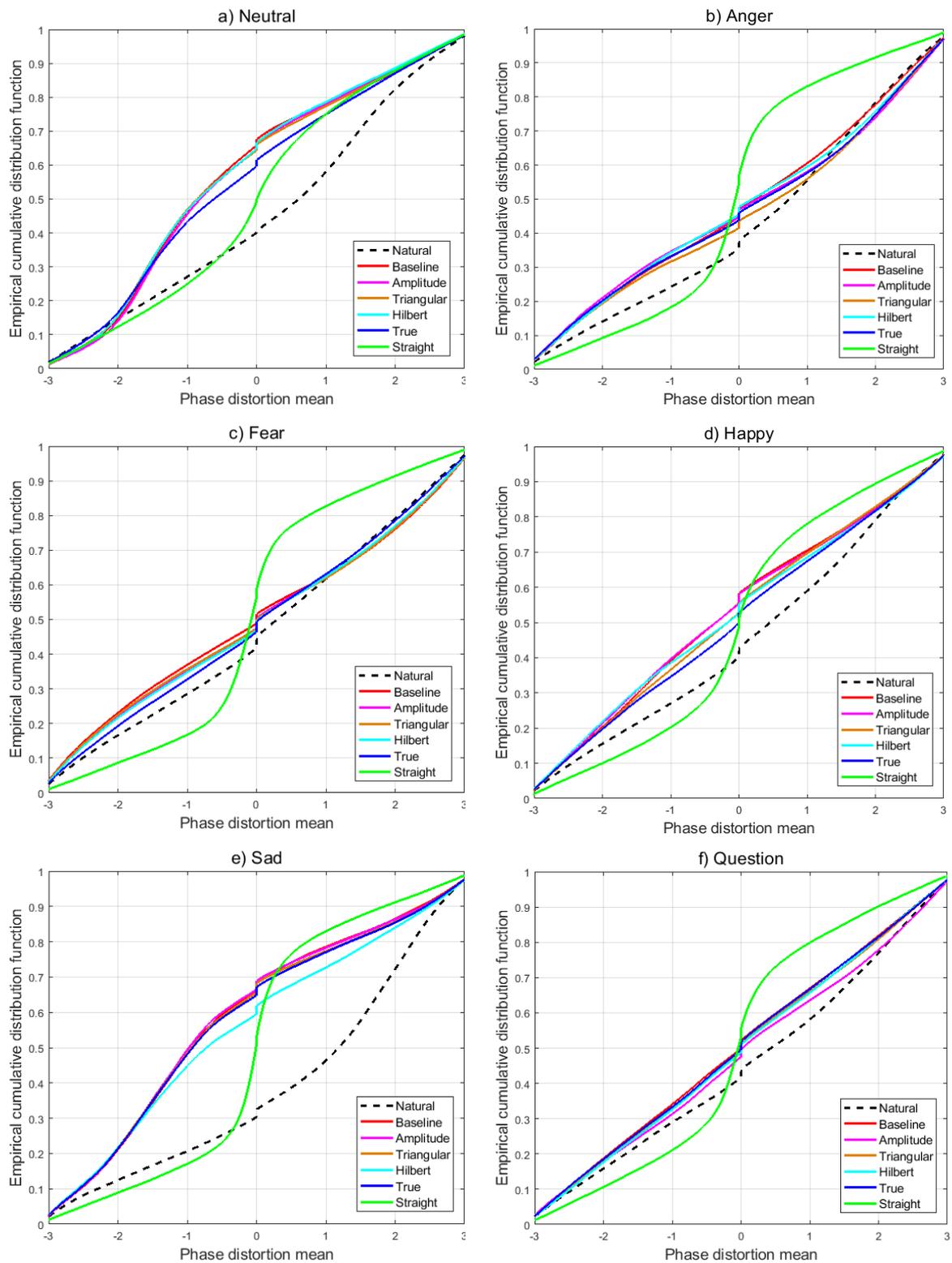

**Figure 11:** Empirical cumulative distribution function of PDMs using 6 vocoders based emotions compared with the PDM measure on the natural speech signal.

# Chapter 3

# Continuous Noise Masking

*"If I have seen further, it is by standing upon the shoulders of giants."*
Isaac Newton (1643 – 1727)

## 3.1 Motivation

Traditional parametric vocoders generally show a perceptible deterioration in the quality of the synthesized speech due to different processing algorithms. Similarly to other vocoders (e.g. a lack of proper noise modelling in STRAIGHT [5]), the noise component in the baseline continuous vocoder is still not accurately modelled that limits the overall perceived quality. Furthermore, an inaccurate noise resynthesis (e.g. in breathiness or hoarseness) is also considered to be one of the main underlying causes of performance degradation, leading to noisy transients and temporal discontinuity in the synthesized speech [42] [43].

To mitigate the problem above, I propose in this chapter a continuous noise masking (cNM) approach with the aim of improving the naturalness of synthetic speech. This is shown by the dashed box (PDD and cNM) in the left upper half of Figure 2. This method has a twofold advantage: a) it allows to mask out most of the noise residuals; and b) it attempts to reproduce the voiced and unvoiced (V/UV) regions more precisely, that is, resembles the natural sound signal. Thus, proper reconstruction of noise in voiced segments (like in breathiness parts) is necessary for the synthetic speech to achieve a quality closer to that of the natural sound.

## 3.2 Proposed Method

Noise masking is a fundamental technique to improve the performance of the speech synthesizer by reducing the number of noise artifacts in the time-frequency domain. It has been widely used in earlier studies. One simple method is presented in [44] as a small amount of artificial noise is added to the clean speech to improve the noise immunity of the model and reach the desired signal-to-noise ratio (SNR). Another method with similar goals is capable of lowering the statistical mismatch of acoustic features in the training and testing conditions [45]. Moreover, a good degree of noise robustness in both filter bank and Mel-frequency cepstral domains can be found in [46]. Recently a binary noise mask (bNM) was proposed for improving both speech intelligibility based on noise distortion constraints [47], and parametric speech synthesis based on thresholding the PDD [25]. However, forcing the PDD values below thresholding to zero might lack a minimum of randomness in the voiced segments [16] [48]. Therefore, by considering that both PDD and bNM help in decreasing the influence of variability in the speech signals, I introduce a new masking approach to avoid any residual buzziness, improve creakiness, and ensure the proper randomization of the noise segments in the parametric vocoders.

In principle, cNM changes from 0 to 1 (or 1 to 0) rather than a binary 0 or 1 as in the bNM, and hence preserves the quality of the voiced segments. In order to compute the cNM, we should first compute the PDD. Originally, PDD can be calculated based on early Fisher's standard-deviation [49] and as defined in Section 2.3. Unlike in bNM which was just a thresholded version of PDD, cNM can be estimated here as

$$cNM = 1 - P\acute{D}D(f) \quad (18)$$

where $P\acute{D}D$ is a normalized PDD value using nearest-neighbor resampling method. Then, to model the speech signal in the continuous vocoder, the following formulas are applied in the synthesis phase $s(t)$ as shown in Figure 2:

$$s(t) = \sum_{n=1}^{N} v_n(t) + u_n(t) \quad (19)$$

where $v(t)$ and $u(t)$ are the voiced and unvoiced speech components at frame $n$, respectively. Thus, for $\forall t$

$$v_n(t) = \begin{cases} v_n(t), & cNM \leq threshold \\ 0, & cNM > threshold \end{cases} \quad (20)$$

$$u_n(t) = u_n(t) * cNM(t) \quad (21)$$

The masking algorithm developed here is to carry out the masking in the voiced and unvoiced segments of the continuous vocoder. To better understand how to approach the above conditions, the suggested model shall satisfy the properties: If the value of the cNM estimate for the voiced frame is greater than the threshold, then this value is replaced (masked) in order to reduce the perceptual effect of the residual noise as may appear in the voiced parts of the cNM (lower values), whereas Equation (21) controls the unvoiced frame based on the unvoiced part of the cNM (higher values). This means that cNM can save parts of speech component in the weak-voiced and unvoiced segments by using a smaller value instead of 0 or 1 caused by the bNM estimation.

Accordingly, cNM improves the synthesis robustness to noise generated in creaky voice segments and closely resembles natural background noise (such as breathy voice). In informal listening tests, we experimented with several continuous values (from 0 to 1), and selected 0.77 as the one producing the best results for indication of presence/absence of voicing in respective voiced/unvoiced frames. This threshold is supported by the experiment in Subsection 3.4 (Figure 14) showing that the probability kernel density function of the proposed model (blue line) starts to match the natural one (black dash line) at PDD 0.77, which then is confirmed as a confidence threshold in this study to avoid any other erroneous estimates. Nevertheless, the results are not to be very sensitive to this threshold as it is more like a clipping needed to account for a low and high level estimation issue in the voiced and unvoiced frames.

An example of cNM estimation on a female speech sample and masking threshold compared with the MVF contour is shown in Figure 12. It can be seen that the cNM also follows the actual voiced/unvoiced regions of the MVF. In other words, if the segment is voiced, then the cNM must be lower to give indication to the synthesis process that this region is voiced and should discard any other noise artifacts depending on the threshold. On the contrary, if the segment is unvoiced, then the cNM must be higher to give indication to the synthesis process that this region is unvoiced and should mask any other higher harmonics frequencies depending also on the threshold. Consequently, it possible for this method to reduce the effect of residual noise, and thus yielding to save parts of speech components.

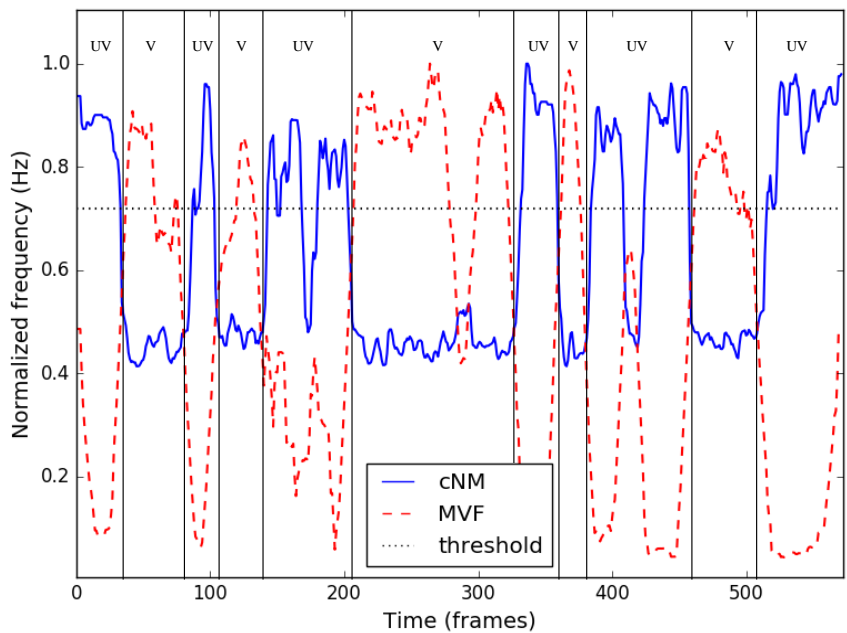

**Figure 12:** Illustration of the performance of the continuous noise mask (blue line) plotted across the maximum voiced frequency (red dashed line), where threshold = 0.77 (black dotted line) is obtained after informal listening tests; UV and V are the unvoiced and voiced segments, respectively. English sentence: "I was not to cry out in the face of fear." from a female speaker.

## 3.3 Maximum Voiced Frequency Algorithm

During the production of voiced sounds, MVF is used as the spectral boundary separating low-frequency periodic and high-frequency aperiodic components. MVF has been used in various speech models [24] [26] [50], that yield sufficiently better quality in synthesized speech. Our vocoder follows the algorithm proposed by [10] which has the potential to discriminate harmonicity, exploits both amplitude and phase spectra, and use the maximum likelihood criterion as a strategy to derive the MVF estimate. The performance of this algorithm has been previously assessed by comparing it with two state-of-the-art methods, namely the Peak-to-Valley (P2V) used in [26] and the Sinusoidal Likeness Measure (SLM) [50]. Based on Receiver Operating Characteristic (ROC) curve and Area Under the Curve (AUC), the algorithm proposed by [10] objectively outperforms both P2V and SLM methods. Moreover, a substantial improvement was also observed over the state-of-the-art techniques in a subjective listening test using male, female, and child speech.

The method consists of the following steps. First, 4 period-long Hanning window is applied to exhibit a good peak structure. Then, the frequencies of the spectral peaks are detected using a standard peak picking function. Amplitude spectrum, phase coherence, and harmonic-to-noise ratio are extracted in the third step for each harmonic candidate which convey some relevant statistics to predict the strategy decision by using the maximum likelihood criterion. Time smoothing step is finally applied to the obtained MVF trajectory in order to remove unwanted spurious values. An example of spectrogram of the natural waveform with the MVF contour is shown in Figure 13. Here, the duration of this sentence is about 3s, and was sampled at 16 kHz with a 16-bit quantization level. It is windowed by Hanning window function in duration of 25 ms, shifted by 5 ms steps. The thresholds for the pitch tracking are set from 80 to 300 Hz. Thus, the MVF parameter models the voicing information: for unvoiced sounds, the MVF is low (around 1 kHz), for voiced sounds, the MVF is high (above 4 kHz).

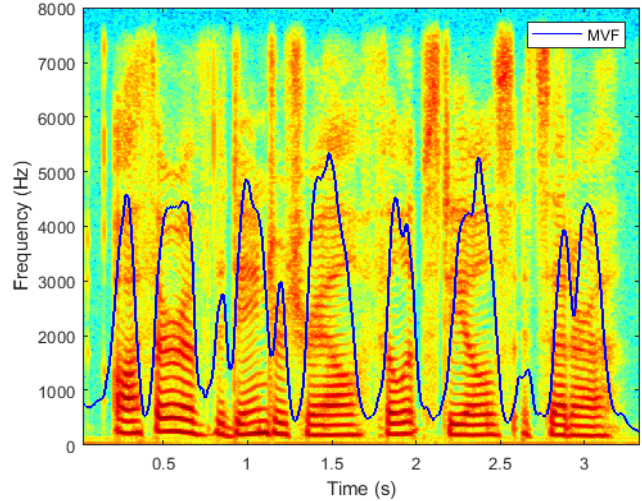

**Figure 13:** Example of spectrogram of the natural waveform and MVF contour (blue). Sentence: "Author of the danger trail, Philip Steels, etc.", from a female speaker.

## 3.4 Evaluation

### 3.4.1 Objective Measurement

Finding a meaningful objective metric is always a challenge in evaluating the performance of speech quality, similarity, and intelligibility. In fact, one metric might be possibly suitable for a few systems but not convenient for all. The reason for that may be returned to some factors which are influenced by the speed, complexity, or accuracy of the speech models. Speaker types and environmental conditions should also be taken into account when choosing these metrics. Therefore, four objective speech quality measures are considered to evaluate the quality of the proposed model. Frequency-weighted segmental signal-to-noise ratio (fwSNRseg) was firstly calculated as described in Subsection 2.3.2.

Secondly, coherence and speech intelligibility index (SII) [51] was employed to evaluate the noise and distortion of the synthetic speech. The coherence SII (CSII) measure was chosen here because it has been shown in [38] to be one of the best predictors for speech intelligibility in fluctuating noise conditions. In this Thesis, the CSII is obtained for each frame $m$ as:

$$CSII_j(m) = 10 \log_{10} \frac{\sum_{k=0}^{l-1} \hat{X}(m,k) \cdot W_j(k)}{\sum_{k=0}^{l-1} \hat{S}(m,k) \cdot W_j(k)} \qquad (22)$$

where $W_j(k)$ is the filter window function, $k$ is the FFT bin index, $\hat{X}(m,k)$ and $\hat{S}(m,k)$ are estimations of the natural and synthesized speech power spectra, respectively. These are obtained as

$$\hat{X}(m,k) = |\gamma(k)|^2 \cdot |S(mT,k)|^2 \tag{23}$$

$$\hat{S}(m,k) = (1-|\gamma(k)|^2) \cdot |S(mT,k)|^2 \tag{24}$$

where $S(mT,k)$ is the short-time Fourier transform of the synthesized speech, $T$ is the frameshift, and the magnitude squared coherence $\gamma$ of the cross-spectral density $S_{xs}$ between natural speech $x(n)$ and synthesized speech $s(n)$, both having spectral densities $S_{xx}$ and $S_{ss}(k)$ respectively, is given by

$$|\gamma(k)|^2 = \frac{|S_{xs}|^2}{S_{xx}(k)S_{ss}(k)}, \qquad 0 \le |\gamma(k)|^2 \le 1 \tag{25}$$

Additionally, the density estimate using a kernel smoothing method [52] [53] was calculated to show how the reconstruction of the noise component in the state-of-the-art vocoders behaved in comparison to the proposed model. The probability kernel density function is given by

$$\hat{f}_h(s) = \frac{1}{nh}\sum_{i=1}^{n} K\left(\frac{s-y_i}{h}\right) \tag{26}$$

where $s$ is the synthesized speech signal, $\{y_i\}_{i=1}^{n}$ are finite random samples drawn from some distribution with an unknown density, $K(\cdot)$ is the kernel function, and $h > 0$ is a smoothing parameter to adjust the width of the kernel. A more detailed case-by-case analysis by fwSNRseg and CSII are shown in Table 2. The results were averaged over 25 synthesized test utterances for each speaker, and a calculation is done frame-by-frame.

**Table 2:** Average scores based on re-synthesized speech for male and female speakers. The bold font shows the best performance of each column.

| Metric | Speaker | Models | | | |
|---|---|---|---|---|---|
| | | Baseline | PML | STRAIGHT | Proposed |
| **fwSNRseg** | JMK | 6.083 | 9.959 | **14.436** | 11.661 |
| | BDL | 6.449 | 13.578 | **16.371** | 12.298 |
| | CLB | 7.559 | 13.752 | **16.583** | 9.789 |
| | SLT | 6.771 | 13.538 | **15.742** | 10.938 |
| **coherence SII** | JMK | 0.048 | 0.208 | 0.252 | **0.271** |
| | BDL | 0.044 | 0.191 | 0.244 | **0.248** |
| | CLB | 0.043 | 0.199 | **0.226** | 0.204 |
| | SLT | 0.065 | 0.236 | 0.252 | **0.263** |

First, it could be observed that the proposed method significantly outperforms the baseline vocoder in both metrics. In particular, it can be seen from the fwSNRseg measure that the proposed vocoder is also better than PML in the JMK speaker. On the contrary, STRAIGHT vocoder still gives better metric results than other systems. Second, for both male and SLT female speakers, the coherence SII values indicate that the proposed system obviously outperforms all systems. In a sense, there is a tendency to increased CSII when considering continuous noise masking in the proposed method. It is interesting to emphasize that the baseline does not at all meet the performance of the other vocoders in all speakers. In other words, the results reported in Table 2, strongly support the use of proposed vocoder than others in terms of coherence SII measure. I can conclude that the approach reported in this work is beneficial and can substantially reduce any residual buzziness.

Probability kernel density function of PDD values for all systems are also estimated and shown in Figure 14 compared to the PDD measure on the natural speech signals. It can be shown that the proposed vocoder based cNM start to match the natural PDD values at a threshold of 0.77, whereas other systems (like STRAIGHT) presents more deviation from the natural one. This indicated that the proposed cNM method gives a better synthesis of the noise in voiced and unvoiced segments than, for example, the bNM in PML.

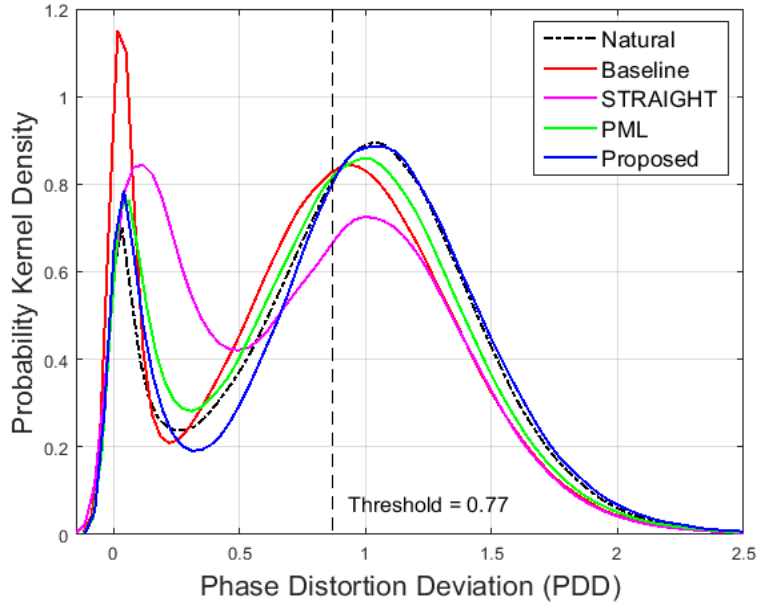

**Figure 14:** Estimation of the probability kernel density functions of PDDs using 4 vocoders compared with the PDD measure on the natural speech signal. The threshold = 0.77 is shown in the vertical dashed line.

Finally, the empirical cumulative distribution function [54] of phase distortion mean values are calculated and displayed in Figure 15 to see whether these systems can be normally distributed and how far they are from the natural signal. The empirical cumulative distribution function $F_n(PDM)$ defined as

$$F_n(x) = \frac{\#\{X_i : X_i \leq x\}}{n} = \frac{1}{n}\sum_{i=1}^{n} I_{X_i \leq x}(X_i) \qquad (27)$$

where $X_i$ is the PDM variables with density function $f(x)$ and distribution function $F(x)$, $\#A$ symbolizes the number of elements in the set $A$ ($X_i \leq x$), $n$ is the number of experimental observations, $I$ is the indicator of event $A$ given as

$$I_A(x) = \begin{cases} 1, & x \in A \\ 0, & x \notin A \end{cases} \qquad (28)$$

It can be noticed that the higher mode of the distribution (positive x-axis in Figure 15) corresponding to STRAIGHT's PDMs is clearly higher than that of the original signal, while the PML's PDMs is lower. This also demonstrates why the synthesized speech for them ranked lower in the perception test (see Subsection 3.4.2). On the contrary, the higher mode of the distribution corresponding to the proposed configuration is better synthesized performance with almost matching the natural speech signal. The performance of STRAIGHT and the baseline vocoders appear considerably worse than PML. Focusing on the lower mode of the distribution (negative x-axis in Figure 15), PML's PDMs gives the second better synthesized performance behind the proposed model. This result is probably explained by the fact that cNM can substantially reduce any residual buzziness.

Hence, the experimental results confirm the effectiveness of the proposed vocoder in terms of speech naturalness to be comparable, or even better, to the STRAIGHT and PML vocoders.

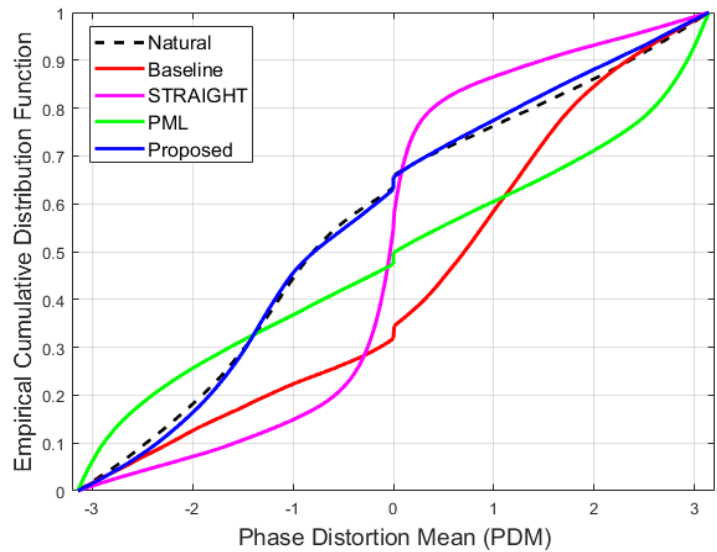

**Figure 15:** Empirical cumulative distribution function of PDMs using 4 vocoders compared with the PDM measure on the natural speech signal.

### 3.4.2 Subjective Listening Test

In order to evaluate the perceptual quality of the proposed systems, we conducted a web-based MUSHRA listening test. We compared natural sentences with the synthesized sentences from the baseline, proposed, STRAIGHT, PML, and an anchor system. The anchor type was the re-synthesis of the sentences with a standard MGLSA vocoder using pulse-noise excitation [12] implemented in the speech signal processing toolkit (SPTK)[5]. The listening test samples can be found online[6], and 18 participants (9 males, 9 females) with a mean age of 29 years were asked to conduct the online listening test. We evaluated 16 sentences (4 from each speaker). Altogether, 96 utterances were included in the test (4 speaker x 6 types x 4 sentences). The MUSHRA scores for all the systems are shown in Figure 16, showing both speaker by speaker and overall results.

According to the results, the proposed vocoder clearly outperformed the baseline system (Mann-Whitney-Wilcoxon ranksum test, $p<0.05$). Particularly, one can see that in the case of both male speakers (BDL and JMK) the proposed method is significantly better than the PML and STRAIGHT vocoders. In terms of the female speakers (Figure 16c, d), we can see that the proposed vocoder is ranked as the second best choice. In other words, the vocoder based cNM is superior to the method based on bNM in PML and the method based on voice decision in STRAIGHT vocoder in case of CLB and SLT speakers, respectively. This unexpected difference (specially in Figure 16d) probably might be due to one of two concerns. First, SLT under-articulates, speaks with a low vocal effort, and exhibit a pressed voice quality [55]. Alternatively, the female SLT speaker has a rather modal phonation with a bit of nasality, which is affecting the evaluation scores. Second, the voiced/unvoiced decision was also left up to the maximum voiced frequency parameter in our study, whereas other systems have separate complex parameters to model this (e.g. aperiodicity parameter in STRAIGHT). Therefore, some possibly inaccurate decisions might have also occurred (especially in unvoiced regions). Listeners seem to prefer the female voices of PML and the male voices of the proposed model. But our system is simpler, i.e. uses less parameters compared to STRAIGHT and PML vocoders.

---

[5] http://sp-tk.sourceforge.net/
[6] http://smartlab.tmit.bme.hu/cNM2019

Based on the overall results, we can conclude that among the techniques investigated in the study of noise reconstruction, cNM performs well in the continuous vocoder when compared with other approaches (Figure 16e). When taking these overall results, the difference between STRAIGHT, PML and the proposed system is not statistically significant (Mann-Whitney-Wilcoxon ranksum test, $p<0.05$), meaning that our methods reached the quality of other state-of-the-art vocoders. This positive result was confirmed by a coherence SII measure in the statistical aspects of the objective's experimental test.

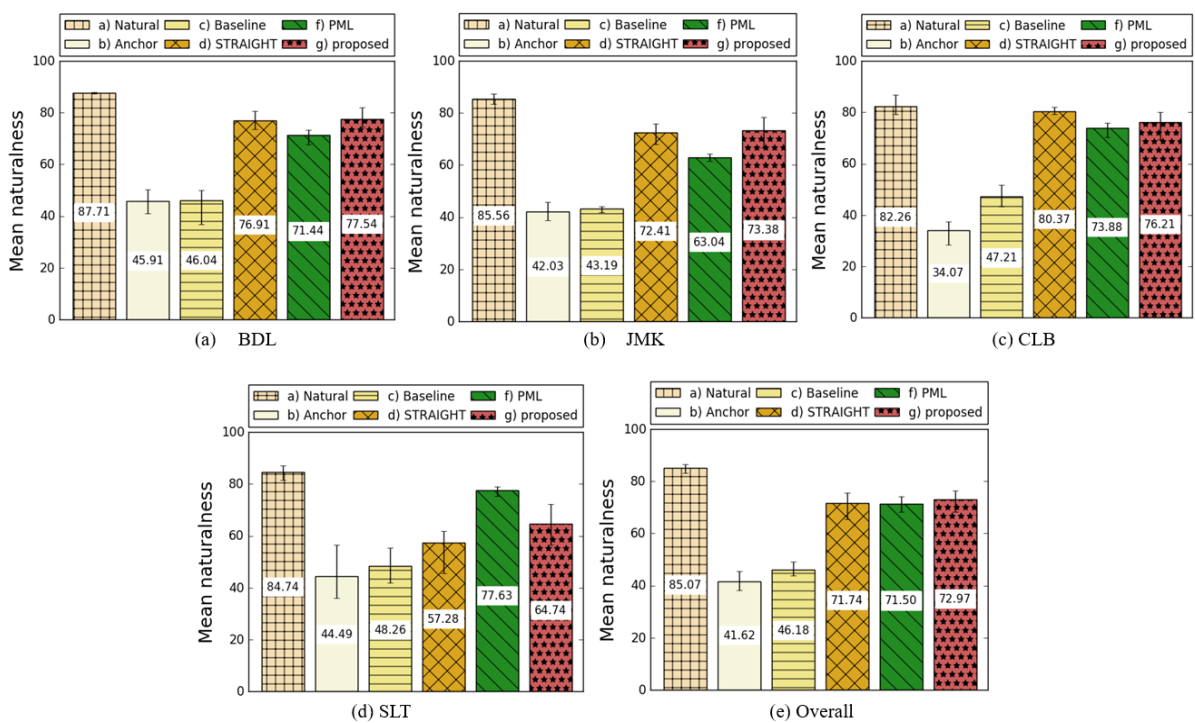

**Figure 16:** Results of the subjective evaluation for the naturalness question. A higher value means larger naturalness. Error bars show the bootstrapped 95% confidence intervals.

## 3.5 Summary

This chapter has developed an encouraging alternative method to reconstruct the noisiness of the speech signal in a continuous vocoder. I have described an implementation of how to generate such a continuous noise masking to avoid any residual buzziness. It was also shown in a subjective listening test that the continuous vocoder allows better ability to synthesize the speech compared to the PML and STRAIGHT vocoders, in case of male voices. Moreover, the continuous synthesizer was also found to have similar or slightly worse quality than state-of-the-art vocoder in female speaker. Therefore, cNM offers a good alternative method to reconstruct noise than other approaches (for instance, bNM).

As the cNM parameter is not limited only to our novel vocoder, it is recommended to apply it to other types of modern parametric vocoders (such as Ahocoder [50] as well as PML [16]) to deal with the case of noisy conditions.

# Part II
# Excitation Harmonic Modelling

# Chapter 4

# Adaptive Continuous Pitch Tracking Algorithm

*"The whole of science is nothing more than a refinement of everyday thinking."*
Albert Einstein (1879 – 1955)

## 4.1 Introduction

Parametric representation of speech often implies fundamental frequency (also referred to as F0 or pitch) contour as a parameter of TTS synthesis. During voiced speech such as vowels, pitch values can be successfully estimated over a short-time period (e.g., a speech frame of 25ms). Pitch observations are continuous and usually range from 60Hz to 300Hz for voiced human speech [56]. But in unvoiced speech such as unvoiced consonants, the long term spectrum of turbulent airflow tends to be a weak function of frequency [57], which suggests that the identification of a single reliable F0 value in unvoiced regions is not possible. Thus, a commonly accepted assumption is that F0 values in unvoiced speech frames are undefined and must instead be represented by a sequence of discrete unvoiced symbols [58]. In other words, F0 is a discontinuous function of time and voicing classification is made through pitch estimation.

In standard TTS with the binary decision excitation system, frames classified as voiced will be excited with a combination of glottal pulses and noise while frames classified as unvoiced will just be excited with noise. Consequently, any hard voiced/unvoiced (V/UV) classification gives two categories of errors: false voiced, i.e. setting frames to voiced that should be unvoiced, and false unvoiced, i.e. setting frames to unvoiced that are voiced. Perceptually, the synthesized speech with false voiced produces buzziness mostly in the higher frequencies, while false unvoiced introduce a hoarse quality in the speech signal. Generally, both of them sound unnatural [59].

One solution is to directly model the discontinuous F0 observation with multi-space probability distribution using hidden Markov models (MSD-HMM) [60]. However, MSD-HMM has some restrictions with dynamic features that cannot be easily calculated due to the discontinuity at the boundary between V/UV regions. Hence, separate streams are normally used to model static and dynamic features [61]. But this also limits the model ability to correctly capture F0 trajectories. An alternative solution, random values generated from a probability density with a large variance have been used for unvoiced F0 observations [62], while setting all unvoiced F0 to be zero has been investigated in [63]. Once again, both of these techniques are inappropriate for the TTS system since it would lead to a synthesis of random or meaningless F0 [9].

In recent years, there has been a rising trend of assuming that continuous F0 observations are present similarly in unvoiced regions and there have been various modelling schemes along these lines. It was found in [58] that a continuous F0 creates more expressive F0 contours with HMM-based TTS than one based on the MSD-HMM system. Zhang et al. [64] introduce a new approach to improve modeling piece-wise continuous F0 trajectory with voicing strength and V/UV decision for HMM-based TTS. Garner et al. [9], the baseline method in this thesis, proposed a simple continuous F0 tracker, where the measurement distribution is determined from the autocorrelation coefficients. This algorithm is better suited to the Bayesian pitch estimation of Nielsen et al. [65]. Tóth and Csapó [66] have shown that continuous F0 contour can be approximated better with HMM and deep neural network than traditional discontinuous F0. In [8], an excitation model was proposed which combines continuous F0 modeling with MVF. This model produced more natural synthesized speech for voiced sounds than traditional vocoders based on standard pitch tracking. However, continuous F0 is still sensitive to additive noise in speech signals and suffers from short-term errors (when it changes rather quickly over time). To alleviate these issues, three adaptive techniques have been developed in this chapter for achieving a robust and accurate contF0. This is shown by the dashed box (adContF0) in the middle upper half of Figure 2.

## 4.2 F0 Detection and Refinement

This section is comprised of a brief background of the continuous F0 (contF0) estimation algorithm, and a description of three powerful adaptive frameworks for refining it. The effectiveness of these proposed methods is evaluated in Section 4.3.

### 4.2.1 contF0: Baseline

The contF0 estimator introduced in this chapter as a baseline is an approach proposed by Garner et al. [9]. The algorithm starts simply with splitting the speech signal into overlapping frames. The result of windowing each frame is then used to calculate the autocorrelation. Identifying a peak between two frequencies and calculating the variance are the essential steps of the Kalman smoother to give a final sequence of continuous pitch estimates with no voiced/unvoiced decision.

In view of this, contF0 is still sensitive to additive noise in speech signals and suffers from short-term errors (when it changes rather quickly over time). Moreover, it can cause some tracking errors when the speech signal amplitude is low, voice is creaky, or low HNR. Therefore, further refinements were developed in this chapter.

### 4.2.2 Adaptive Kalman Filtering

To begin with, the Kalman filter in its common form can be mathematically described as a simple linear model

$$x_t = A_t x_{t-1} + w_{t-1} \quad , \quad w_t \sim N(0, Q_t) \tag{29}$$

$$y_t = B_t x_t + v_t \quad , \quad v_t \sim N(0, R_t) \tag{30}$$

Here $t$ is a time index, $x_t$ is an unobserved (hidden) state variable, $A_t$ is the state transition model to update the previous state, $w_t$ (state noise with zero mean) and $v_t$ (measurement noise with zero mean) are independent Gaussian random variables with covariance matrices $Q_t$ and $R_t$ respectively; $y_t$ is the measurement derived from the observation state $x_t$, $B_t$ is the measurement model which maps the underlying state to the observation. Alternatively, the Kalman filter operates by propagating the mean and covariance of the state through time. In recent times, this method has been used for obtaining smoothed vocal-tract parameters [67], and in speech synthesis systems [68] [69].

It is known from the literature that the Kalman filter is one of the best state estimation methods in several different senses when the noise of both $w_t$, $v_t$ are Gaussian, and both covariance $Q_t$, $R_t$ is expected to be known. However, this can be very difficult in practice. If the noise statistics (estimates of the state and measurement noise covariance) are not as expected, the Kalman filter will be unstable or gives state estimates that are not close to the true state [70]. One promising approach to overcome this problem is the use of adaptive mechanisms into a Kalman filter. In particular, signal quality indices (SQIs) have been proposed by [71], and recently used in [72], which give the confidence in the measurements of each source. When the SQI is low, the measurement should not be trusted; this can be achieved by increasing the noise covariance. Tsanas et al. [73] proposed an approach to consider both the state noise and the measurement noise covariance which are adaptively determined based on the SQI (but in [71] and [72], the state noise was a priori fixed). Therefore, to improve the contF0 estimation method, we used the SQI algorithm reported in [73] in order to compute the confidence in both state noise and measurement noise covariance. So that, their covariance matrices $Q_t$ and $R_t$ are updated appropriately at each time step until convergence. Detailed steps of this algorithm are summarized simply in Figure 17. In this formulation, the aim of the adaptive Kalman filter is to use the measurements $y_t$ to update the current state $\tilde{x}_t = x_{t-1}$ to the new estimated state $x_t$ when $Q_t$ and $R_t$ are given at each time step.

Figure 18a shows the performance of this adaptive methodology. However, in some cases, this approach may over-fit to the speech dataset due to the number of manually-specified parameters which is required for tuning. Thus, this technique should be used carefully.

### 4.2.3 Adaptive Time-Warping

In speech signal processing, it may be necessary that harmonic components are separated from each other with the purpose of being easily found and extracted. Once F0 rapidly changes, harmonic components are subject to overlap each other and make it difficult to separate these components; or the close neighboring components make the separation through filtering very hard especially with a low-pitched voice (such as male pitch) [74]. To overcome this problem, previous work in the literature has provided methods by introducing a time-warping based approach [75] [76].

Abe et al. [77] incorporate time-warping into the instantaneous frequency spectrogram frame by frame according to the change of the harmonic frequencies. In view of that, the observed F0 is seen to be constant within each analysis frame. More recently, a time-warping pitch tracking algorithm was also proposed by [78] which apparently had a significant positive impact on the voicing decision error and led to good results even in very noisy conditions. There was another approach introduced by Stoter et al. [79] based on iteratively time-warping the speech signal and updating F0 estimate on time-warped speech, which has a nearly constant F0 over a segment of short duration that sometimes leads to inaccurate pitch estimates. To achieve a further reduction in the amount of contF0 trajectory deviation (deviate from their harmonic locations) and to avoid additional sideband components generation when a fast movement of higher frequencies occurs, adaptive time warping approach combined with the instantaneous frequency can be used to refine the contF0 algorithm.

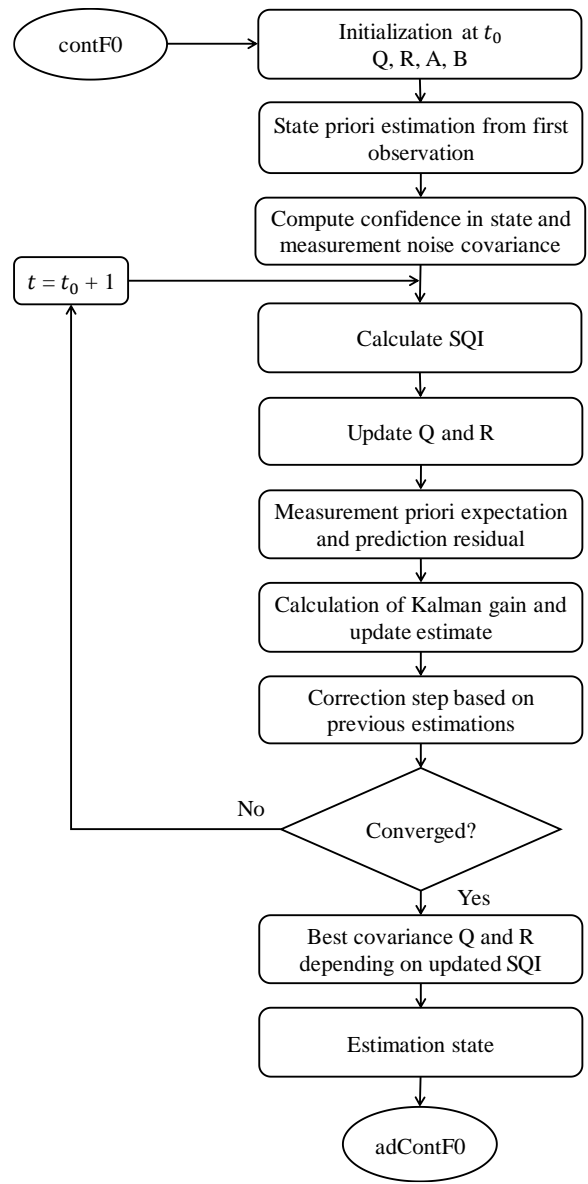

**Figure 17:** Structure chart of adaptive Kalman filter based contF0 (adContF0).

We refer to the warping function as $p$ which defines the relationship between two axes

$$\tau = p(t) \quad , \quad t = p^{-1}(\tau) \tag{31}$$

where $\tau$ represents a time stretching factor. The first step is to stretch the time axis in order to make the observed contF0 value in the new temporal axis stay unchanged and preserves the harmonic structure intact [75] [76]. As the initial estimate of the contF0 is available, the second step of the refinement procedure is that the input waveform is filtered by bandpass filter bank $h(\tau)$ with different center frequencies $f_c$ multiplied by Nuttall window $w(\tau)$ [80] to separate only the fundamental component in the range near $f_c$

$$h(\tau) = w(\tau)\cos(2j\pi f_c \tau) \tag{32}$$

$$w(\tau) = 0.338946 + 0.481973\cos\left(\frac{j\pi}{2}f_c\tau\right) + 0.161054\cos(j\pi f_c\tau) \\ + 0.018027\cos\left(\frac{3j\pi}{2}f_c\tau\right) \tag{33}$$

Next, instantaneous frequencies $IF(\tau)$ of $h(\tau)$ have to be calculated. Flanagan's equation [81] is used to extract them from both the complex-valued signal and its derivative

$$IF_k(\tau) = \frac{a\frac{db}{d\tau} - b\frac{da}{d\tau}}{a^2 + b^2} \tag{34}$$

where $a$ and $b$ are the real and imaginary parts of the spectrum of $h(\tau)$, respectively. $k$ represents the harmonic number. As the $IF(\tau)$ indicates the value close to F0, the $\acute{cont}F0$ is thus refined to a more accurate F0 by using a linear interpolation between $IF(\tau)$ values and contF0 coordinates. Then, using a weighted average

$$\sum_{k=1}^{N} w_k \frac{\acute{cont}F0_k}{k} \tag{35}$$

where $\sum_{k=1}^{N} w_k = 1$, provides a new $contF0_\tau$ estimate on the warped time axis. The last step is unwarped in time to return the estimated value to the original time axis. Recursively applying these steps gives a final adaptive contF0 estimate (adContF0). An example of the proposed refinement based on the time-warping method is depicted in Figure 18b. It can be seen that the adContF0 trajectory given by the time-warping method is robust to the tracking error (dip at frame 30 and frame 138) to make it a more accurate estimation than the baseline. Despite the good performance, this technique requires a little tweaking the time-warp to achieve the desired results.

### 4.2.4 Adaptive Instantaneous Frequency

Another method used to improve the noise robustness of the result estimated by contF0 is based on the instantaneous frequency which is defined as the derivative of the phase of the waveform. This approach (named as StoneMask) is also used in WORLD [14], that is a high-quality speech analysis/synthesis system, to adjust its fundamental frequency (DIO) algorithm [82]. Flanagan's equation defined as in Equation (34) is used to calculate the instantaneous frequency $IF(t)$. Here, $a$ and $b$ are respectively the real and imaginary parts of the spectrum of a waveform $S(w)$ windowed by a Blackman window function $w(t)$ defined in $[-T_0, T_0]$ with the following form

$$w(t) = 0.42 + 0.5\cos\frac{\pi t}{NT_0} + 0.08\cos\frac{2\pi t}{NT_0} \tag{36}$$

where $N$ is a positive integer, and $T_0$ is the inverse of the contF0 candidate. Hence, contF0 can be further refined using recursively a formula given by

$$adContF0 = \frac{\sum_{k=1}^{k}|S(kw_0)|\,IF(kw_0)}{\sum_{k=1}^{k}k|S(kw_0)|} \tag{37}$$

where $w_0$ represents the angular frequency of the contF0 candidate at a temporal position, and $k$ represents the harmonic number (we set $k = 6$ for further refinement of the methodology).

In this work, The $IF(t)$ of the periodic signal shows the value close to contF0 when the frequency is around contF0. Because the spectrum around contF0 has a larger power, this approach is more robust than others. Therefore, the contF0 is refined to a more accurate one even if the contF0 candidate has some error.

The impact of the proposed method on contF0 performance is illustrated in Figure 18c. It is quite obvious that the adContF0 obtained by StoneMask almost matches the reference pitch contour, and it is much better than the others. It can be also seen here that the proposed adContF0 in the unvoiced region (frames from 170 to 202 in Figure 18c) is significantly much smaller than for the baseline, which is not the case with previous refined methods.

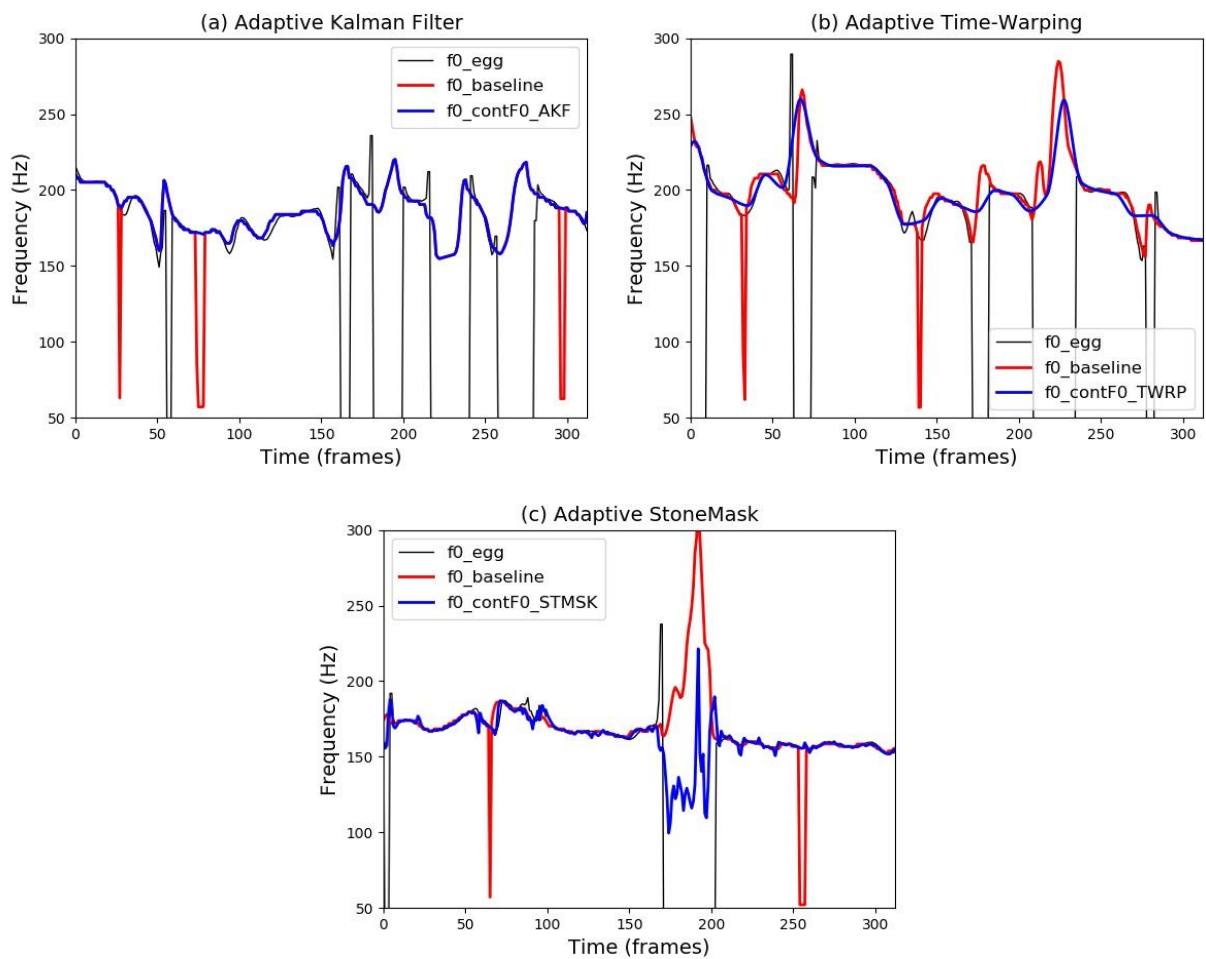

**Figure 18:** Examples from a female speaker of F0 trajectories estimated by the baseline (red) and ground truth (black) plotted along with proposed refined contF0 (adaptive Kalman filter (AKF), time-warping (TWRP), and StoneMask (STMSK)) methods.

## 4.3 Evaluation

The experimental evaluation aims to evaluate the accuracy of the adaptive contF0 using several measurement metrics. The objective evaluation is discussed in the subsections below whereas the subjective listening test is discussed in Chapter 5.

### 4.3.1 Error Measurement Metrics

I try to adopt a series of distinct measurements in accordance with [83] [84] to assess the accuracy of the adContF0 estimation. The results were averaged over the utterances for each speaker. The following three evaluation metrics were used:

1) **Gross Pitch Errors:** GPE is the proportion of frames considered voiced $N_v$ by both estimated and referenced F0 for which the relative pitch error $e(n)$ is higher than a certain threshold (usually set to 20% for speech). The $e(n)$ can be calculated as:

$$e(n) = \frac{F0_{n,refined}}{F0_{n,referenced}} - 1, \quad n = 1, \dots, N_v \tag{38}$$

where $n$ is the frame index. If $|e(n)| > 0.2$, we classified the frame as a gross error $N_{GE}$. Thus, GPE can be defined as

$$GPE = \frac{N_{GE}}{N_v} * 100\% \tag{39}$$

2) **Mean Fine Pitch Errors:** Fine pitch error is referred to all pitch errors that are not classified as GPE. In other words, MFPE can be derived from Equation (38) when $|e(n)| < 0.2$

$$MFPE = \frac{1}{N_{FE}} \sum_{n=1}^{N_{FE}} (F0_{n,refined} - F0_{n,referenced}) \tag{40}$$

where $N_{FE}$ is the number of remaining voiced frames that do not have gross error $(N_v - N_{GE})$.

3) **Standard Deviation of the Fine Pitch Errors:** STD is firstly stated in [84] as a measure of the accuracy of the F0 detector during voiced intervals, then slightly modified in [83]. For better analysis, STD can be calculated as

$$STD = \sqrt{\frac{1}{N_{FE}} \sum_{n=1}^{N_{FE}} (F0_{n,refined} - F0_{n,referenced})^2 - MFPE^2} \tag{41}$$

Table 3 displays the results of the evaluation of three methods based contF0, for female and male speakers, in comparison to the YANGsaf algorithm. When refining the contF0 by time-warping (contF0_TWRP) technique, the GPE score shows an improvement of 4.46% for BDL speaker whereas 1.07% for JMK speaker. Nevertheless, we did not see any enhancement for SLT speaker in case of ContF0_TWRP. However, 2.32% improvement was found in the refinement of contF0 based on StoneMask method (ContF0_STMSK). Additionally, Table 3 has shown that there are significant differences between ContF0_STMSK and the state-of-the-art YANGsaf approaches based on MFPE and STD measures in all speakers.

**Table 3:** Average performance per each speaker in clean speech.

| Method | GPE % | | | MFPE | | | STD | | |
|---|---|---|---|---|---|---|---|---|---|
| | BDL | JMK | SLT | BDL | JMK | SLT | BDL | JMK | SLT |
| baseline | 12.754 | 9.850 | 7.677 | 3.558 | 3.428 | 4.421 | 4.756 | 4.513 | 6.764 |
| contF0_AKF | 11.268 | 12.611 | **6.732** | 2.764 | 2.754 | 3.692 | 3.964 | 3.719 | 6.113 |
| contF0_TWRP | **8.294** | 8.777 | 7.827 | 2.764 | 3.024 | 3.656 | 3.873 | 4.188 | 5.788 |
| contF0_STMSK | 10.557 | **7.530** | 6.998 | **1.661** | **1.389** | **2.105** | **2.526** | **1.872** | **4.181** |
| YANGsaf | 4.231 | 2.049 | 4.592 | 1.658 | 1.452 | 2.142 | 2.239 | 1.575 | 4.160 |

In the same way, Table 4 and Table 5 tabulates the GPE, MFPE, and STD measures averaged over all utterances for BDL, JMK, and SLT speakers in the presence of additive white noise and pink noise, respectively, at 0 dB of SNR to test the robustness of the contF0 tracker. adContF0 based on Kalman filter (contF0_AKF) is more accurate for the female speaker (as measured by GPE) than the other two candidates. However, this is not the case with pink noise.

Moreover, adContF0 based on time-warping has shown better performance in terms of GPE measurement in the presence of pink noise with all speakers. In contrast, contF0_STMSK still has the lowest MFPE and STD under SNR conditions for all speakers.

It is interesting to emphasize that the baseline does not meet the performance of the other refinements trackers for BDL, JMK, and SLT speakers. The results reported in Table 3 are comparable with state-of-the-art algorithms [18], while they strongly support the use of the proposed StoneMask based method that is the most accurate contF0 estimation algorithm in Table 4 and Table 5. In other words, the findings in Table 4 and Table 5 might have demonstrated the robustness of the proposed approaches to additive Gaussian white and pink noise.

It is worth to note that the main advantage of using the adaptive Kalman filter is that we can determine our confidence in the estimates of contF0 algorithm based TTS by adjusting SQIs to update both the measurement noise covariance and the state noise covariance. For example, it can be used to replace the one studied by Li et al. [71] in the heart rate assessment application. Whereas, the time warping scheme has the ability to track the time-varying contF0 period, and reduce the amount of contF0 trajectory deviation from their harmonic locations. By considering the system processing speed, adContF0 based StoneMask is computationally inexpensive and can be useful in a practical speech processing application.

**Table 4:** Average performance per each speaker in the presence of additive white noise (SNR = 0 dB).

| Method | GPE % | | | MFPE | | | STD | | |
|---|---|---|---|---|---|---|---|---|---|
| | BDL | JMK | SLT | BDL | JMK | SLT | BDL | JMK | SLT |
| baseline | 33.170 | 40.057 | 27.502 | 4.050 | 3.901 | 3.512 | 4.393 | 4.293 | 3.912 |
| contF0_AKF | 31.728 | 40.865 | **26.122** | 3.211 | 3.241 | 2.898 | 3.465 | 3.627 | 3.448 |
| contF0_TWRP | **29.464** | 37.839 | 26.932 | 3.199 | 3.165 | 2.890 | 3.449 | 3.511 | 3.186 |
| contF0_STMSK | 31.418 | **37.052** | 26.352 | **2.128** | **1.896** | **2.067** | **2.103** | **1.658** | **2.058** |
| YANGsaf | 27.530 | 35.200 | 25.852 | 2.233 | 2.181 | 2.175 | 2.206 | 2.219 | 2.265 |

**Table 5:** Average performance per each speaker in the presence of pink noise (SNR = 0 dB).

| Method | GPE % | | | MFPE | | | STD | | |
|---|---|---|---|---|---|---|---|---|---|
| | BDL | JMK | SLT | BDL | JMK | SLT | BDL | JMK | SLT |
| baseline | 25.041 | 26.870 | 33.124 | 2.919 | 2.799 | 2.845 | 3.061 | 2.936 | 3.180 |
| contF0_AKF | 24.548 | 28.034 | 31.103 | 2.285 | 2.293 | 2.284 | 2.338 | 2.327 | 2.468 |
| contF0_TWRP | **21.512** | **22.329** | **29.893** | 2.256 | 2.482 | 2.472 | 2.253 | 2.702 | 2.787 |
| contF0_STMSK | 24.371 | 26.131 | 32.775 | **1.429** | **1.179** | **1.387** | **1.686** | **1.981** | **1.140** |
| YANGsaf | 15.401 | 12.509 | 22.186 | 1.419 | 1.307 | 1.393 | 2.282 | 2.732 | 2.022 |

### 4.3.2 Noise Robustness

I used white Gaussian noise and pink noise as the background noise to test the quality of the adContF0 and also to clarify the effects of refinement. The amount of noise is specified by the signal-to-noise ratio (SNR) ranged from 0 to 40 dB. I calculated the root mean square error (RMSE) over selected sentences for each speaker.

Figure 19a and 19b show the overall RMSE values obtained from various methods as a function of the SNR between speech signals and noise. The average RMSE over all three speakers is presented. The smaller the value of RMSE, the better the F0 estimation's performance. The results of white and pink noise suggest that the RMSE for all proposed methods is smaller than the baseline, and the stone mask method becomes the best. This means that our proposed one is: a) robust against the white and pink noise; b) superior the one based on YANGsaf. Consequently, this positive result is beneficial in TTS synthesis.

Furthermore, Figure 20 shows the power spectral density (PSD) calculated with the periodogram method for all F0 estimators compared with ground truth. In this figure, the adContF0 based StoneMask method gives similar performance to that of ground truth (F0_egg) and better than baseline [9]. It can be concluded that all refined approaches were robust against the noise and outperformed the conventional one as expected.

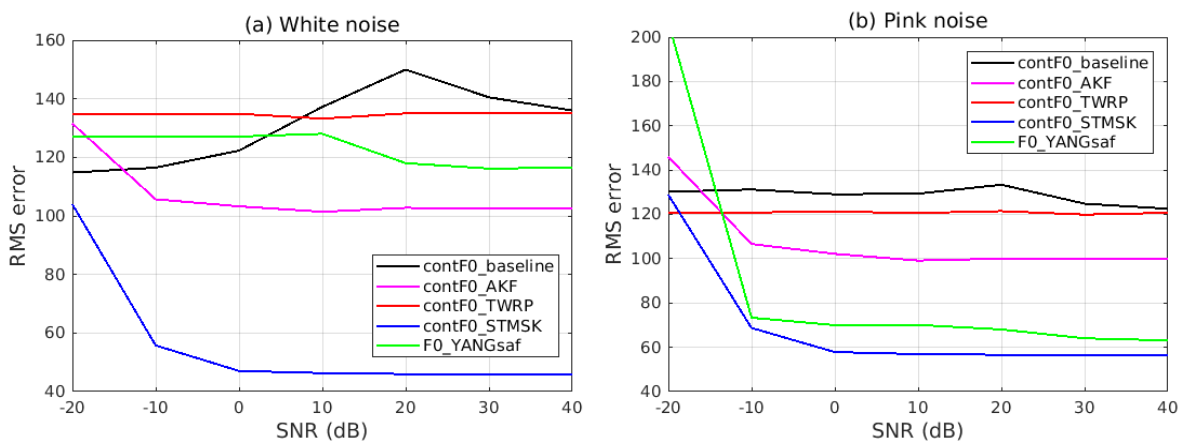

**Figure 19:** Influence of the SNR on the average RMSE with proposed refined contF0_AKF (adaptive Kalman filter), contF0_TWRP (time-warping), and contF0_STMSK (StoneMask) methods.

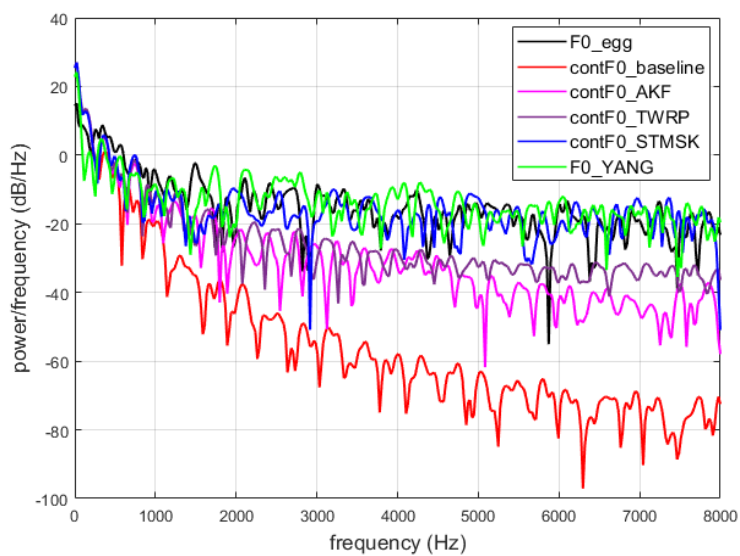

**Figure 20:** The periodogram estimate of the PSD for the extracted F0 trajectories.

## 4.4  Summary

In this chapter, a modified version of the simple continuous pitch estimation algorithm in terms of adaptive Kalman filter, time-warping, and instantaneous-frequency methods was proposed. A relatively large database containing simultaneous recordings of speech sounds and EGG was used for the performance evaluation. According to our observations of the experiments, it was found that refined contF0 methods could provide the expected results for both clean speech and speech contaminated with additive white noise and pink noise.

This thesis provides a reference for selecting appropriate techniques to optimize and improve the performance of current fundamental frequency estimation methods-based text-to-speech.

# Chapter 5

# Parametric HNR Estimation Approach

> *"An experiment is a question which science poses to Nature, and a measurement is the recording of Nature's answer."*
> Max Planck (1858 – 1947)

## 5.1 Motivation

A valid and reliable method for calculating levels of noise in human speech would be required to give appropriate information for SPSS. Existing methods of measuring noise in the human speech divide the acoustic signal into two parts: a harmonic and a noise component. Based on this assumption, estimates of the harmonic-to-noise ratio (HNR) have been calculated. I expect that adding a HNR to the voiced and unvoiced components that involve the presence of noise in voiced frames, the quality of synthesized speech in the noisy time regions will be more accurate and it is comparable to the state-of-the-art results. This method has a twofold advantage: it allows to eliminate most of the noise residuals, and it attempts to reproduce the voiced and unvoiced (V/UV) regions more precisely, that is, resembles natural sound signal based TTS synthesis.

The goal of this chapter is to further improve our proposed vocoder [85], discussed in Chapter 2, for high-quality speech synthesis. Specifically: a) it studies adding HNR as a new excitation parameter to the voiced and unvoiced segments of speech (this is shown by the dashed box (HNR) in the top left corner upper half of Figure 2); and b) it explores a different methodology for the estimation of MVF.

## 5.2 Harmonic-to-Noise Ratio

The main goal of vocoders is to achieve high speech intelligibility and naturalness. It was shown before that the mixed excitation source model yields sufficiently good quality in the synthesized speech by reducing the buzziness and breathiness [86]. Such an analysis/synthesis system may also suffer from some degradations: 1) loss of the high-frequency harmonic components, 2) high-frequency noise components, or 3) noise components in the main formants. As the degree of these losses increases, more noise appears and consequently degrade the speech quality highly [87].

This thesis proposes to add a continuous HNR as a new excitation parameter to the continuous vocoder in order to alleviate previous problems. Consequently, the excitation model in the proposed vocoder is represented by three continuous parameters: F0, MVF, and HNR. There are various methods of time and frequency domain algorithms available in the literature to estimate HNR in speech signals (for a comparison, see [88]). As we are dealing here with time domain processing, we want to follow the algorithm by [89] to estimate the level of noise in human voice signals for the following reasons: 1) the algorithm is very straightforward, flexible and robust, 2) it works equally well for low, middle, and high pitches, and 3) it is correctly tested for periodic signals and for signals with additive noise and jitter.

For a time signal $x(t)$, the autocorrelation function $r_x(\tau)$ as a function of the $lag\ \tau = t_2 - t_1$ (that are $\tau$ time periods apart) can be defined as

$$r_x(\tau) \cong \int x(t)x(t+\tau)dt \tag{42}$$

This function has a global maximum for $\tau = 0$, and a local maximum for $\tau_{max}$ (highest value among the local maxima). The fundamental period $T_0 = 1/F_0$ is defined as the value of $\tau$ corresponding to the highest maximum of the $r_x(\tau)$, and the normalized autocorrelation is

$$\acute{r}_x(\tau) = \frac{r_x(\tau)}{r_x(0)} \tag{43}$$

We could make such a signal $x(t)$ by taking a harmonic signal $H(t)$ with a period $T_0$ and adding a noise $N(t)$ to it. We can now write Equation (42) as

$$r_x(\tau) = r_H(\tau) + r_N(\tau) \tag{44}$$

Because the autocorrelation of a signal at 0 equals the power in the signal, Equation (43) at $\tau_{max}$ represents the relative power of the harmonic component of the signal, and its complement represents the relative power of the noise component:

$$\acute{r}_x(\tau_{max}) = \frac{r_H(0)}{r_x(0)} \tag{45}$$

$$1 - \acute{r}_x(\tau_{max}) = \frac{r_N(0)}{r_x(0)} \tag{46}$$

Thus, the HNR is defined at $\tau_{max} > 0$

$$HNR \triangleq \frac{\acute{r}_x(\tau_{max})}{1 - \acute{r}_x(\tau_{max})} \tag{47}$$

Accordingly, the HNR is positive infinite for purely harmonic sounds while it is very low for the noise (see Figure 21). In a continuous vocoder, our approach here is to use the HNR to weight the excitation signal in both voiced and unvoiced frames. If we define the generation of the voiced excitation frame $v[k]$ as

$$v[k] = p[k] * w_v \tag{48}$$

then, the weighted voice $w_v$ value can be determined by

$$w_v = \sqrt{\frac{hnr[i]}{hnr[i]+1}} \quad , \quad i = \frac{K}{F_{shift} * f_s} \tag{49}$$

where $p[k]$ is the residual PCA voiced signal, $F_{shift}$ is $5\ ms$ frame shift, $f_s$ is the sampling frequency, and $K$ is the location of impulse in original impulse excitation. Similarly, the unvoiced excitation frame $u[k]$ and the unvoiced weight $w_u$ value can also be computed by

$$u[k] = n[k] * w_u \tag{50}$$

$$w_u = \sqrt{\frac{1}{hnr[i] + 1}}\ ,\ \ i = k \tag{51}$$

where $n[k]$ is the additive Gaussian noise. As a result, the voiced and the unvoiced speech signal components are added in the ratio suggested by the HNR, and then used to excite the MGLSA filter as illustrated in the bottom part of Figure 2.

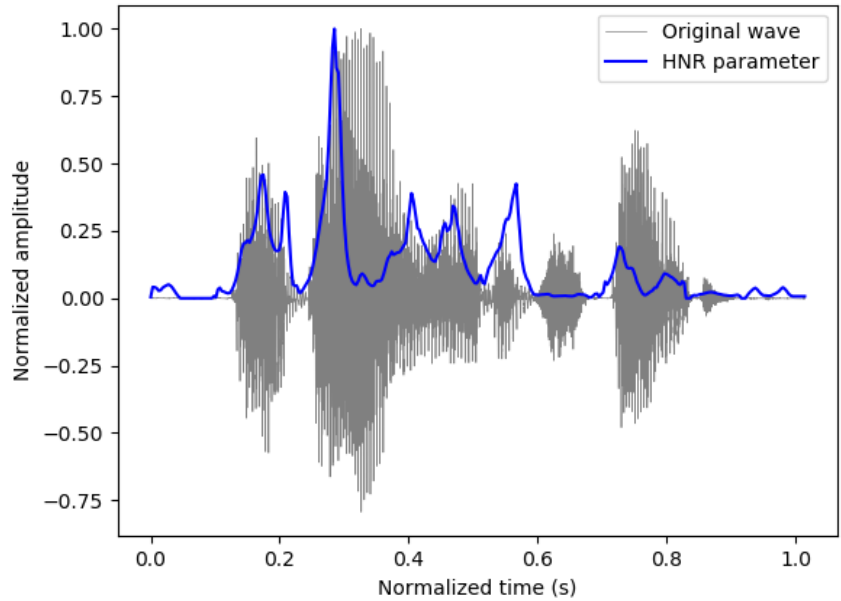

**Figure 21:** Example of a HNR parameter for the clean speech signal. Sentence: "They die out of spite." from a male speaker.

## 5.3   Maximum Voiced Frequency Estimation

In voiced sounds, MVF is used as the spectral boundary separating a low-frequency periodic and high-frequency aperiodic components. It has been used in numerous speech models, such as [24] [26] [50], that yield sufficiently good quality in the synthesized speech.

The preliminary version of our vocoder followed Drugman and Stylianou [10] approach which exploits both amplitude and phase spectra. Although this approach tends to relatively reduce the acoustic buzziness of the reconstructed signals, it cannot distinguish between a production noise and a background noise. This means that MVF might be underestimated if the speech is recorded under pseudo noisy environment. Moreover, we found that the MVF estimation based on [10] lacks to capture some components of the sound that lies in the region of the higher frequencies (especially for the females). For this reason, higher MVF is required in this work to yield even more natural synthetic speech.

Over the last few years, several attempts, with varying results, have been already made to analyze the MVF parameter. In this thesis, we used a sinusoidal likeness measure (SLM) [90] based approach to extract the MVF. A representative block diagram is shown in Figure 22 using five main functional steps:

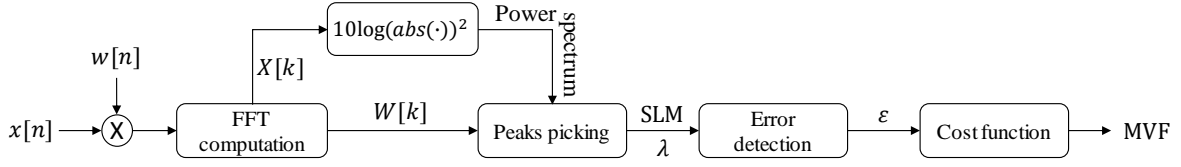

**Figure 22:** Workflow of the MVF estimation algorithm based on SLM method.

1) Consecutive frames of the input signal $x[n]$ are obtained by using a 3-period-long Hanning window $w[n]$.
2) $N$-point fast Fourier transform (FFT) of every analysis frame $m$ is computed $X^m[k]$. $N$ is equal or greater than 4 times of the frame length $L$.

$$X^m[k] = \log\left(\frac{|FFT_N\{x[n].w[n-n_m]\}|}{\sqrt{Lf_s}}\right) \quad (52)$$

3) The magnitude spectral peak detection for each frame is calculated, and their SLM score $\lambda_i$ is given through cross-correlation [90]

$$\lambda_i = \frac{|\sum S[k].W_i^*[k]|}{\sqrt{\sum |S[k]|^2 . \sum |W_i[k]|^2}} \quad (53)$$

where $W$ is the Fourier transform of $w[n]$ multiplied by $e^{-j2\pi fn}$, operator * denotes a complex conjugation, and $i$ is the index of the peak. The $\lambda$ always lies in the range [0,1]. Consequently,

$$\lambda = \begin{cases} 1, & pure\ sinusoid \\ otherwise, & presence\ of\ noise \end{cases} \quad (54)$$

4) The error of the MVF position at each peak $i$ is figured as

$$\varepsilon_i^m = \frac{1}{P}\left[\sum_{j=1}^{i-1}(1-\lambda_j^m)^2 + \sum_{j=i}^{P}(\lambda_j^m)^2\right] \quad (55)$$

where $P$ is the total number of spectral peaks.

5) To give a final sequence of MVF estimates, a dynamic programming approach is used to eliminate the spurious values and to minimize the following cost function

$$C_i^m = \sum_{k=1}^{K}\varepsilon_i^m + \gamma\sum_{k=2}^{K}\left(\frac{f_i^m - f_{i-1}^{m-1}}{\frac{f_s}{2}}\right)^2 \quad (56)$$

where $f_i^m$ is the $i_m$ candidate at frame $k$ and $\gamma = 1$ at $5\ ms$.

Figure 23 shows the spectrograms of an example of voiced speech with MVF estimation algorithm obtained by the baseline [10] (red line) and SLM (blue line). It can be seen that the MVF based SLM approach capture wide frequency segments of data (e.g. between 0.75s - 1.3s, and between 1.9s - 2.5s). This observation suggests that the baseline often underestimate some of the voicing frequency in the higher frequency regions of the spectrogram.

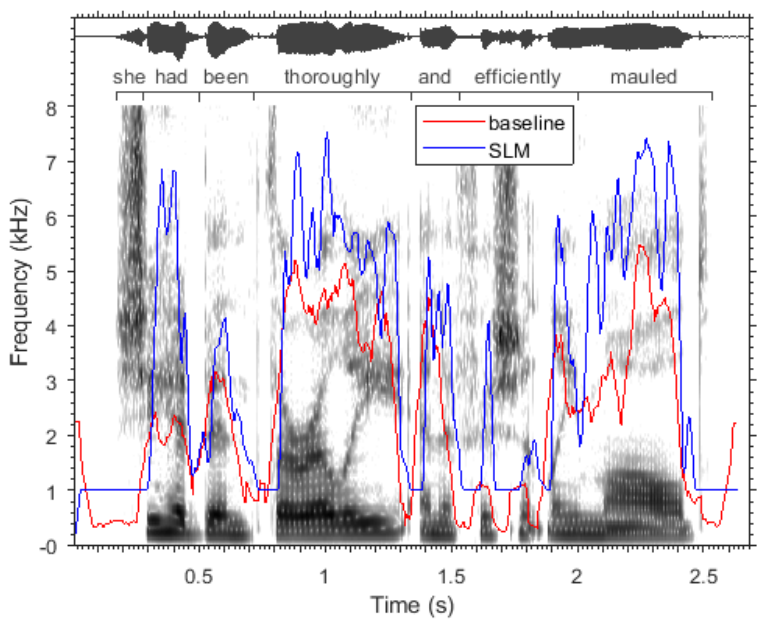

**Figure 23:** Importance of the SLM in MVF estimation. Top is the speech waveform from a female speaker, bottom is the spectrogram and MVF contours.

## 5.4 Evaluation

### 5.4.1 Objective Measurement

A range of objective speech quality and intelligibility measures are considered to evaluate the quality of synthesized speech based on the modified version of the continuous vocoder. fwSNRseg, NCM, WSS, and ESTOI have already been defined in Chapter 2. As the speech production process can be modelled efficiently with Linear Predictive Coefficients (LPC), another objective measure is called the Log-Likelihood Ratio (LLR) [91] can be introduced. It is generally a distance measure that can be directly calculated from the LPC vector of the clean and enhanced speech. The segmental LLR is

$$LLR = \frac{1}{N} \sum_{i=1}^{N} \log\left(\frac{a_{y,i}^T R_{x,i} a_{y,i}}{a_{x,i}^T R_{x,i} a_{x,i}}\right) \quad (57)$$

where $a_x$, $a_y$, and $R_x$ are the LPC vector of the natural signal frame, synthesized signal frame, and the autocorrelation matrix of the natural speech signal, respectively. The segmental LLR values were limited in the range of [0, 1].

Before I proceed to further details on examining the results, I first describe our experiments. A number of experiments based on the HNR parameter are implemented to find out the best continuous pitch tracking algorithm that works well with our continuous vocoder as well as understanding the behavior of adding a new HNR excitation parameter in both voiced/unvoiced speech frames. The three experiments can be summarized in Table 6.

**Table 6:** An overview of the three proposed methods based on HNR parameter.

| Method | Pitch algorithm |
|---|---|
| Proposed #1 | adContF0 based adaptive Kalman filter |
| Proposed #2 | adContF0 based adaptive Time-warping |
| Proposed #3 | adContF0 based adaptive StoneMask |

The performance evaluations are summarized in Table 7. For all empirical measures, a calculation is done frame-by-frame and higher value indicates better performance except for the WSS and LLR measures (a lower value is better). From this table, a number of observations can be made. First, focusing on the WSS, it is clear that all methods for refining contF0 appear to work quite well with the HNR parameter. The fact is that proposed #3 can outperform the STRAIGHT vocoder for JMK speaker. In terms of fwSNRseg, it can be also seen that all refined methods can perform well with a continuous vocoder (highest results were obtained); nevertheless, proposed #3 is shown the best. Similarly, the NCM measure shows similar performance between proposed #3 and STRAIGHT. In terms of LLR, the lowest correlation values were obtained with all proposed methods for all speakers. On the other hand, a good improvement was noted for the proposed #1, #2, and #3 in the ESTOI measure. Hence, these experiments showing that adContF0 with HNR was beneficial.

**Table 7:** Average performance based on synthesized speech signal per each speaker.

| Metric | Speaker | Baseline | Proposed#1 | Proposed#2 | Proposed#3 | STRAIGHT |
|---|---|---|---|---|---|---|
| fwSNRseg | BDL | 8.083 | 11.812 | 11.807 | **13.033** | 15.062 |
| | JMK | 6.816 | 9.505 | 9.784 | **10.621** | 13.094 |
| | SLT | 7.605 | 9.906 | 9.736 | **11.079** | 15.295 |
| NCM | BDL | 0.650 | 0.850 | 0.854 | **0.913** | 0.992 |
| | JMK | 0.620 | 0.847 | 0.860 | **0.906** | 0.963 |
| | SLT | 0.673 | 0.850 | 0.854 | **0.910** | 0.991 |
| ESTOI | BDL | 0.642 | 0.856 | 0.861 | **0.892** | 0.923 |
| | JMK | 0.620 | 0.831 | 0.847 | **0.873** | 0.895 |
| | SLT | 0.679 | 0.848 | 0.846 | **0.894** | 0.945 |
| LLR | BDL | 0.820 | 0.457 | 0.456 | **0.453** | 0.219 |
| | JMK | 0.814 | 0.635 | 0.631 | **0.628** | 0.391 |
| | SLT | 0.744 | 0.639 | 0.640 | **0.636** | 0.194 |
| WSS | BDL | 48.569 | 32.875 | 32.559 | **24.013** | 22.144 |
| | JMK | 51.788 | 36.236 | 32.175 | **26.238** | 29.748 |
| | SLT | 58.043 | 42.789 | 45.254 | **26.906** | 23.614 |

Additionally, I compared the vocoded sentences to the natural and baseline by measuring mean phase distortion deviation (M-PDD) defined in Chapter 2. Figure 24 shows the means of the PDD values of the three speakers grouped by the 6 variants. As can be seen, the M-PDD values of the baseline system are significantly lower in BDL and SLT speakers and higher in JMK speaker compared to natural speech. It can be also noted from the JMK speaker that proposed #3 appears to match the M-PDD value of the natural speech, followed by proposed #2. Similarly, the closed M-PDD value to natural speech is shown in proposed #3 and #2 for the female speaker. For the BDL speaker, only proposed #3 is not different from the natural samples, while others seem to give lower M-PDD value. In summary, the various experiments result in different M-PDD values, but in general they are almost closer to the natural speech than the STRAIGHT (not significant) and baseline vocoders.

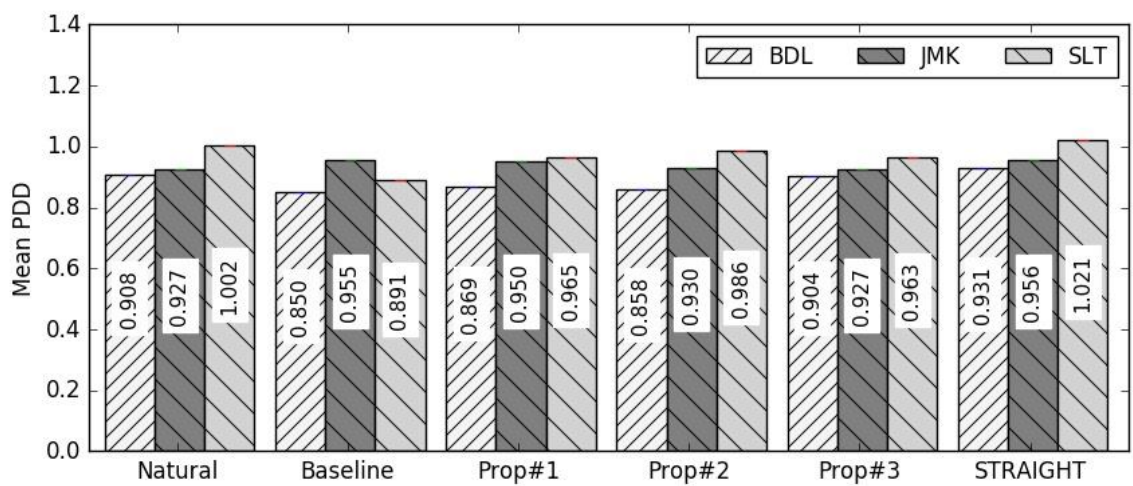

**Figure 24:** Mean PDD values by sentence type.

### 5.4.2 Subjective Listening Test

As a subjective evaluation, the idea was to select the closeness between the re-synthesized and original speech signal that fits this thesis goal. In order to evaluate which proposed system is closer to the natural speech, we conducted a web-based MUSHRA listening test. The listening test samples can be found online[7]. Twenty-one participants (12 males, 9 females) with a mean age of 29 years, were asked to conduct the online listening test. On average, the test took 10 minutes to fill. The MUSHRA scores for all the systems are shown in Figure 25, showing both speaker by speaker and overall results.

According to the results, the proposed vocoders clearly outperformed the baseline system (Mann-Whitney-Wilcoxon ranksum test, $p<0.05$). Particularly, one can see that in the case of the female speaker (SLT) all proposed vocoders are significantly better than the STRAIGHT and baseline vocoders (Figure 25c). For male speaker (JMK), we found that the proposed #3 reached the highest naturalness scores in the listening test (Figure 25b). Whereas for the BDL male speaker in Figure 25a, proposed #3 followed by proposed #2 are ranked as the second and third best choices, respectively. When taking these overall results, the difference between STRAIGHT and the proposed system is not statistically significant (Mann-Whitney-Wilcoxon ranksum test, $p<0.05$), meaning that our methods reached a point of high naturalness of synthesized speech. This positive result was confirmed by metric measures in the statistical aspects of the objective's experimental test.

---

[7] http://smartlab.tmit.bme.hu/adContF0_2019

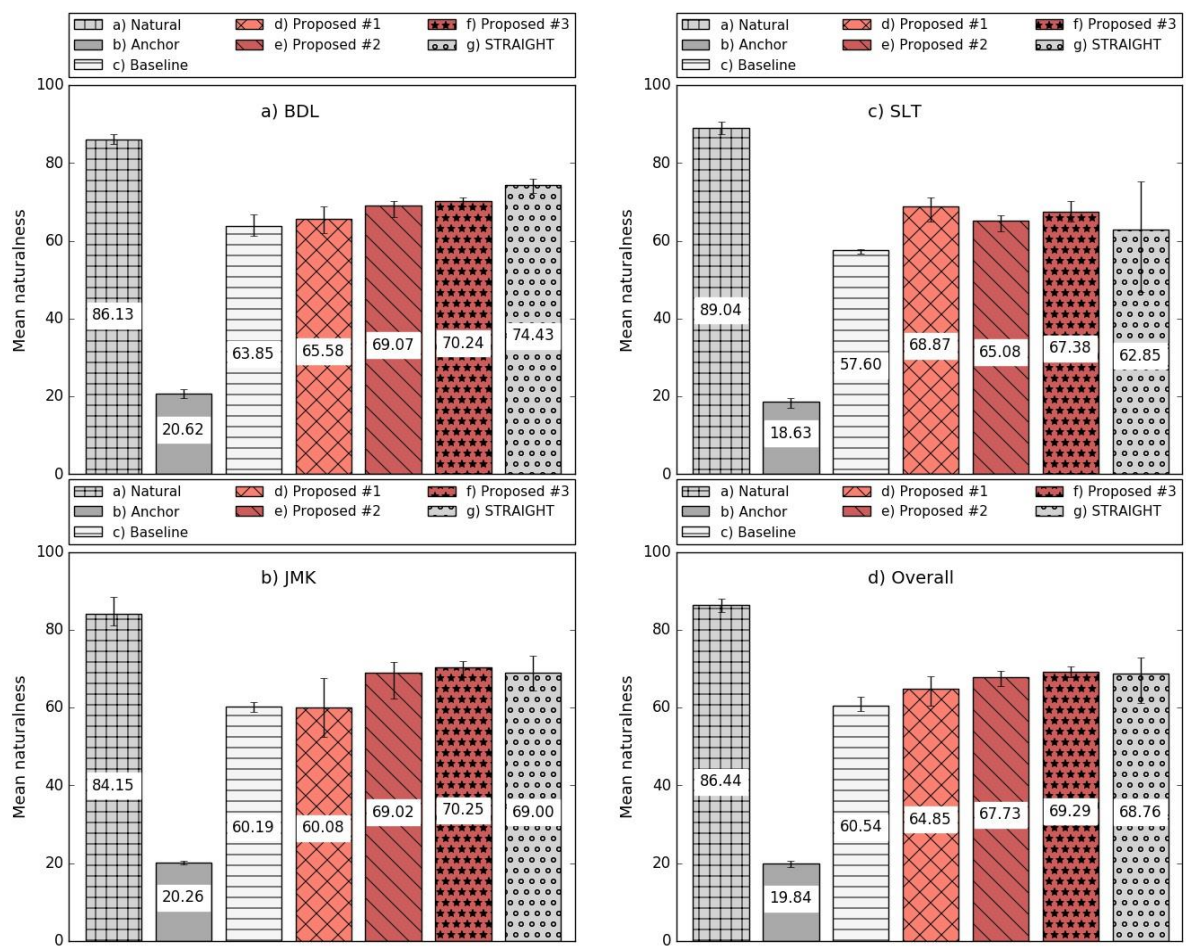

**Figure 25:** Results of the subjective evaluation for the naturalness question. A higher value means larger naturalness. Error bars show the bootstrapped 95% confidence intervals.

## 5.5 Summary

This chapter proposed a new excitation HNR parameter to the voiced and unvoiced components in order to reduce the influence of buzziness caused by the parametric vocoder. We have also presented a new method to estimate the MVF based on a sinusoidal likeness measure (SLM), and we have shown its advantages to capture some components of the sound that lie in the region of the higher frequencies and yield even more natural synthetic speech. Algorithms and examples are given for each approach. Using a variety of error measurements, the performance strengths and weaknesses of the proposed method for different speakers were highlighted. In a subjective listening test, experimental results demonstrated that our proposed methods can improve the naturalness of the synthesized speech over our earlier baseline and well-known STRAIGHT vocoders. This means that proposed #3 was rated better and more closely reached the state-of-the-art performance than the others under most objective and subjective measures. Hence, the design of the continuous vocoder leads to a very simple synthesizer, which is straightforward to understand and implement.

# Part III
# Acoustic Modelling

# Chapter 6

# Feed-Forward Deep Neural Network

*"I never did anything by accident, nor did any of my inventions come by accident; they came by work."*

Thomas Edison (1847 – 1931)

## 6.1 Introduction

The popularity of hidden Markov models (HMMs) has been growing over the past two decades, motivated by its accepted advantages of convenient statistical modelling and flexibility. Even though the quality of synthesized speech generated by HMM-TTS system has been improved recently, its naturalness is still far from that of actual human speech [92]. Moreover, these models have their limitations in representing complex, nonlinear relationships between the speech generation inputs and the acoustic features [93]. To alleviate these shortcomings, a variety of alternative models have been proposed, such as improving the model of HMM training criterion using minimum generation error (MGE) [94], reducing over-smoothing problems in both time and frequency domain [95], or using a trajectory HMM by imposing explicit relationships between static and dynamic feature vector sequences [96]. Although the above models can enhance accuracy and synthesis performance, they usually increase the amount of computational complexity with higher number of model parameters.

In the last few years, deep learning algorithms have shown in many domains their ability to extract high-level, complex abstractions and data representations from large volumes of supervised and unsupervised data [97]. More specifically, deep neural networks (DNNs) can represent functions more efficiently and achieve great improvements in various machine learning areas. Since the first DNN-TTS system in [4], a number of studies have dealt with deep learning in speech synthesis. As detailed in [98], DNNs can be viewed as a replacement for the decision tree used in HMM-based systems. Hierarchical structured deep neural networks is presented in [99]. In [100], Multi-task deep neural network with stacked bottleneck features is introduced.

The newest results in DNN-TTS have shown that it is possible to synthesize the samples of speech directly, without using the vocoders as an intermediate step [6] [101]. However, there are several drawbacks that we should take into consideration: a) it requires for each speaker a large quantity of voice data and computation power for training the neural networks which make it difficult to use in real-time applications; and b) neural models (e.g. WaveNet) are naturally serial (it needs to be repeated sequentially, one sample at a time) which cannot fully employ parallel processors (e.g. GPUs). Therefore, I believe that vocoder-based SPSS still offers a flexible and tractable solutions to TTS and voice conversion applications that could be improved in terms of quality (e.g. to be included in low resource devices like smartphones). This chapter aims to model the improved version of the continuous vocoder parameters (F0, MVF, and MGC) with FF-DNN based speech synthesis in comparison to the previous HMM-TTS system.

## 6.2 FF-DNN Based Speech Synthesis

The DNN used here is a feed-forward (FF) multilayer perceptron architecture. The input is used to predict the output with multiple layers of hidden units, each of which performs a non-linear function of the previous layer's representation, and a linear activation function was used at the output layer, as follows:

$$h_t = f(W_{xh}x_t + b_h) \tag{58}$$

$$y_t = W_{hy}h_t + b_y \tag{59}$$

where $W$ is the connection weight matrix between two layers (e.g. $W_{xh}$ is the weight matrix between input and hidden vectors), $b$ is the bias vectors, and $f(\cdot)$ denotes an activation function which is defined as:

$$f(x) = \begin{cases} \dfrac{e^{2x} - 1}{e^{2x} + 1}, & \text{in the hidden layer} \\ \\ x, & \text{in the output layer} \end{cases} \tag{60}$$

This hyperbolic tangent activation function, whose outputs lie in the range (-1 to 1), which can yield lower error rates and faster convergence than a logistic sigmoid function (0 to 1) [102]. FF-DNN aims to minimize the mean squared error function between the true output $y$ and the predicted one $\hat{y}$

$$E = \frac{1}{n}\sum_{i=1}^{n}(y_i - \hat{y}_i)^2 \tag{61}$$

This network topology consists of 6 FF hidden layers, each consisting of 1024 units. One of the important aspects through DNN training is to normalize input and output features [103]. Therefore, the input linguistic features have min-max normalization, while output acoustic features have mean-variance normalization. For the first 15 epochs, a fixed learning rate of 0.002 was chosen with a momentum of 0.3. More specifically, after 10 epochs, the momentum was increased to 0.9 and then the learning rate was halved regularly. The training procedures were conducted on a high performance NVidia Titan X GPU.

Figure 26 conceptually illustrates the main components of the continuous vocoder when applied in DNN-based training. Textual and phonetic parameters are first converted to a sequence of linguistic features as input, and neural networks are employed to predict acoustic features as output for synthesizing speech.

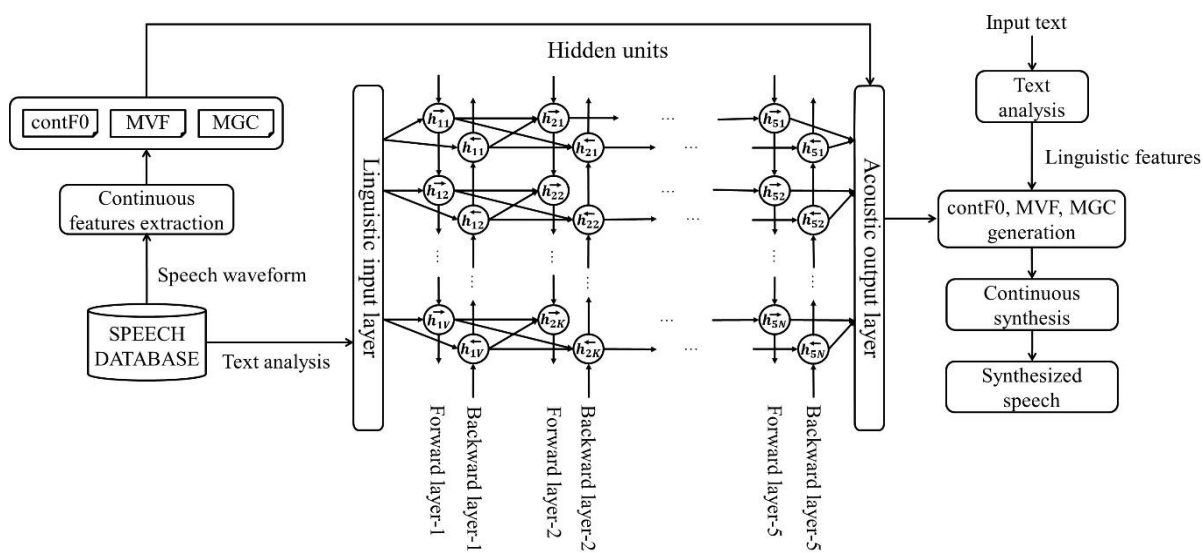

**Figure 26:** A general schematic diagram of the proposed system based text-to-speech.

## 6.3 Mel-Generalized Cepstral Algorithm

In the previous chapters, a simple spectral model represented by 24–order MGC was used [11]. Although several vocoders based on this simple algorithm have been developed, they are not able to synthesize natural sound. The main problem is that it is affected by time-varying components and it is difficult to remove them. Therefore, more advanced spectral estimation methods might increase the quality of synthesized speech.

In [104], an accurate and temporally stable spectral envelope estimation called CheapTrick was proposed. CheapTrick consists of three steps: F0-adaptive Hanning window, smoothing of the power spectrum, and spectral recovery in the quefrency domain. In a modified version of the continuous vocoder, Cheaptrick algorithm using the 60-order MGC representation with $\alpha = 0.58$ (Fs=16 kHz) will be used to achieve high-quality speech spectral estimation. A comparison of the spectral envelope between standard MGC and the CheapTrick is shown in Figure 27. Accordingly, it is clear now to see how a continuous vocoder will behave after adaptation to a more accurate spectral envelope technique than the previous MGC system.

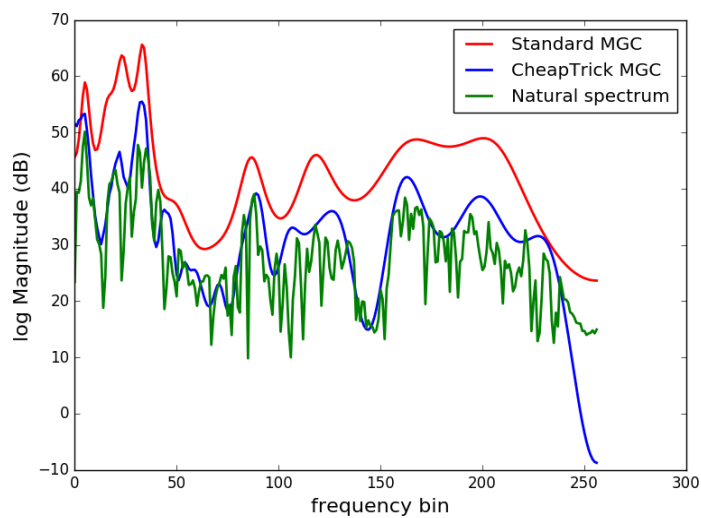

**Figure 27:** Example of the signal spectrum of a voiced segment (green) with the spectral shape (spectral envelope) estimates obtained with standard MGC (red) and CheapTrick (blue).

## 6.4 Evaluation

### 6.4.1 RMS - Log Spectral Distance

Several performance indices have been proposed for evaluating spectral algorithms. Since this Chapter dealing with speech synthesis based TTS and one major task is the refinement of the spectral envelopes in the continuous vocoder, I will focus on distance measures. Spectral distortion is among the most popular ones and plays a very important role in speech quality assessment which is designed to compute the distance between two power spectra [105].

To verify the effectiveness of the proposed vocoder using the CheapTrick algorithm in the direction of refining baseline vocoder spectral envelope [8] [85], root mean square (RMS) log spectral distance (LSD) evaluation is proposed to carry it out. $LSD_{RMS}$ is a distance measure and can be defined here by

$$LSD_{RMS} = \sqrt{\frac{1}{N}\sum_{k=1}^{N} mean[log P(f_k) - log \hat{P}(f_k)]^2} \tag{62}$$

where $P(f)$ and $\hat{P}(f_k)$ are spectral power magnitudes of the natural and synthesized speech respectively, defined at $N$ frequency points.

For a perfect synthesized speech, the ideal value of $LSD_{RMS}$ is zero, which indicates a matching frequency content. The values expressed in Table 8 refer to the average $LSD_{RMS}$ that was calculated for 20 sentences selected randomly from two categories of SLT and AWB speakers. The analysis of these results confirms that the $LSD_{RMS}$ is getting lower by using CheapTrick spectral algorithm than the simple spectral algorithm used in the baseline vocoder. This point is well illustrated in Figure 28 by three spectrograms of frequency versus time. In the middle spectrogram, the $LSD_{RMS}$ of the signal is equal to 1.6, while the bottom spectrogram has a lower $LSD_{RMS}$ equal to 0.89 that is closer to the top speech spectrogram (natural speech). Thus, we can say that our suggested scheme introduces a smaller distortion to the sound quality and approaches a correct spectral criterion.

**Table 8:** Average log spectral distance for the spectral estimation.

| Spectral algorithm | $LSD_{RMS}$ (dB) | |
|---|---|---|
| | SLT | AWB |
| Standard MGC | 1.47 | 0.94 |
| CheapTrick MGC | 0.91 | 0.89 |

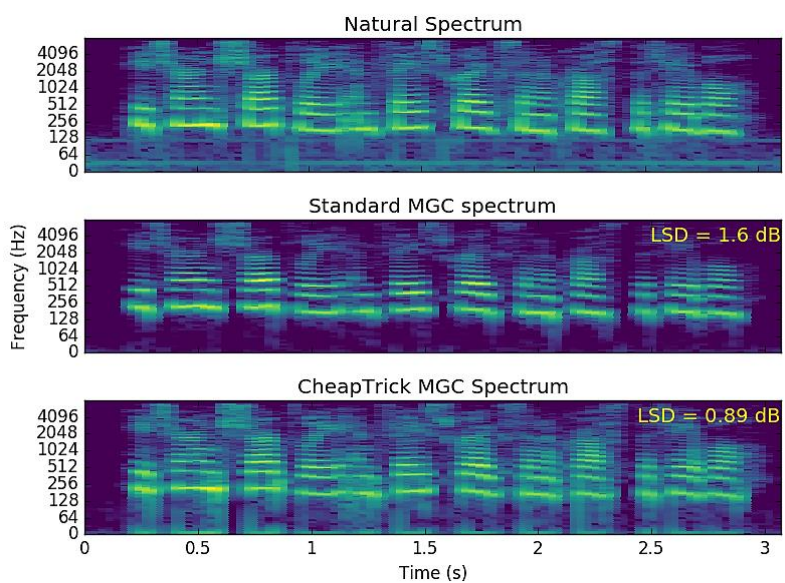

**Figure 28:** Comparison of speech spectrograms: Natural speech signal (top), synthesized speech based on a simple MGC algorithm (middle), and synthesized speech based on CheapTrick algorithm (bottom). The sentence is "He turned sharply, and faced Gregson across the table.", from speaker SLT.

### 6.4.2 Subjective Listening Test

In order to evaluate the differences in DNN-TTS synthesized samples using the above vocoders, a web-based MUSHRA listening test is performed. I compared natural sentences with the synthesized sentences from the baseline, proposed and a benchmark system. The benchmark was a DNN-TTS applied with a simple pulse-noise excitation vocoder. Also, I added samples from an earlier HMM-TTS system which was using the continuous vocoder [8]. 15 sentences were selected which were not included in the training. Altogether, 90 utterances were included in the test (6 types x 15 sentences). The utterances were presented in a randomized order (different for each participant). The listening test samples can be found online[8]. Nine participants (7 males, 2 females) with a mean age of 35 years were asked to conduct the online listening test. On average, the test took 20 minutes to fill. The results of the listening test are presented in Figure 29. DNN-TTS refers in this chapter to the DNN+Cont system (the one that based on an earlier version of a continuous vocoder [8]) and DNN+Cont+Env system (the one that discussed in Chapter 2).

Based on the overall results, the DNN-TTS with the continuous vocoder significantly outperformed baseline method based on HMM-TTS, and its naturalness is almost reached the quality of the WORLD vocoder based TTS. I can conclude that this Thesis showed the potential of the DNN-based approach for SPSS over the HMM-TTS.

---

[8] http://smartlab.tmit.bme.hu/dogs2017_vocoder_dnn

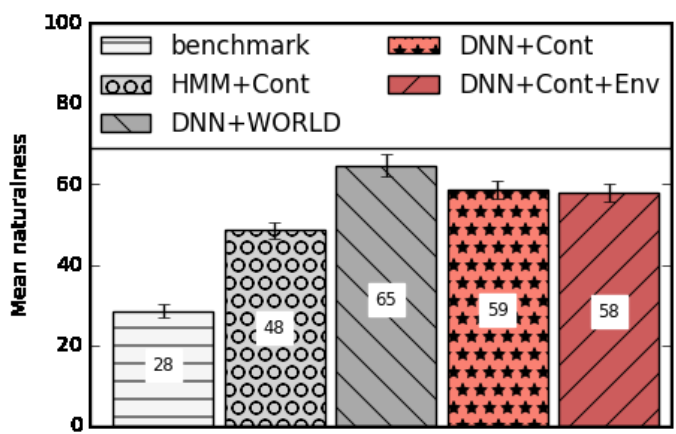

**Figure 29:** Results of the MUSHRA listening test for the naturalness question. Error bars show the bootstrapped 95% confidence intervals. The score for the reference (natural speech) is not included.

### 6.4.3 Comparison of the WORLD and Continuous Vocoders

In this study, the WORLD vocoder [14] was chosen for comparison with the optimized vocoder for the reason that it also used a CheapTrick spectral algorithm. Similarly to the continuous vocoder, the WORLD vocoder is based on source-filter separation, i.e. models separately the spectral envelope and excitation (with F0 and aperiodicity). At the beginning, WORLD estimates the F0 contour using the DIO (Distributed Inline-filter Operation) algorithm [82], the spectral envelope is estimated with the CheapTrick algorithm [104], and at the end, the excitation signal is estimated with the D4C (Definitive Decomposition Derived Dirt-Cheap) algorithm [106] and used as a band aperiodicity of speech signals.

Table 9 compares the parameters of the vocoders under study. It can be seen that the continuous vocoder uses only two one-dimensional parameters for modeling the excitation, whereas the WORLD vocoder is applying a five-dimensional band aperiodicity. Accordingly, the synthesis part is computationally feasible, therefore speech generation can be performed in real-time. For the DNN-TTS training with the WORLD vocoder, it is necessary to interpolate F0 and add a new voiced/unvoiced (V/UV) binary feature.

**Table 9:** Parameters of applied vocoders.

| Vocoder | Parameter per frame | Excitation |
| --- | --- | --- |
| Continuous | F0: 1 + MVF: 1 + MGC: 60 | Mixed |
| WORLD | F0: 1 + Band aperiodicity: 5 + MGC: 60 | Mixed |

To evaluate the performance of the proposed vocoder, the F0 modeling capability and the V/UV transitions were tested the following way. Although the WORLD vocoder can achieve a good quality when applied in speech synthesis, it is worth noting here that the WORLD vocoder (which is using the DIO pitch tracking algorithm and results in a discontinuous F0 track) can make V/UV decision errors (i.e. setting voiced that should be unvoiced, or vice versa) and also sometimes contains errors at boundaries (at the V/UV or UV/V transitions). This is not the case with the continuous vocoder, which is using a continuous pitch detection algorithm. In the latter, the voicing feature is modeled by the continuous MVF parameter; therefore, V/UV errors do not occur, but errors in MVF estimation might cause some audible issues. It can be seen in Figure 30 (showing the F0 contour of a synthesized speech sample) that the continuous vocoder interpolates the F0 contour even in unvoiced regions of speech. For that reason, the V/UV error was 5.35% for the WORLD vocoder in case of the SLT speaker. In informal listening tests, I also observed that the WORLD vocoder often synthesizes speech with clicks which are the result of false V/UV decisions.

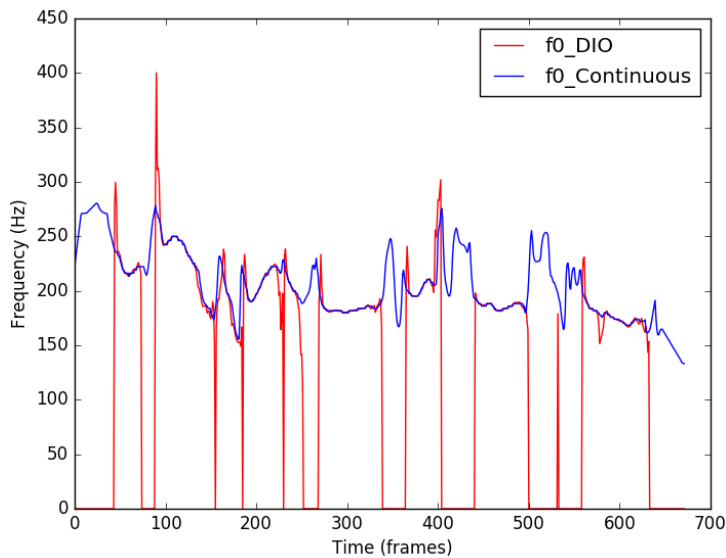

**Figure 30:** F0 trajectories of a synthesized speech signal using the DIO algorithm (red), and continuous algorithm (blue) for continuous and WORLD vocoders respectively. (sentence: "Author of the danger trail, Philip Steels, etc.", from speaker AWB).

## 6.5 Summary

This chapter presented a novel approach by employing continuous vocoder in deep neural network based speech synthesis. The experiments were successful and we were able to add the continuous features to the training of the DNNs. The motivation for using a continuous vocoder arises from our observation that the state-of-the-art WORLD vocoder has often V/UV errors and boundary errors due to the DIO F0 estimation algorithm. In a subjective MUSHRA test, it was found that the DNN-TTS using the continuous vocoder was rated better than an earlier HMM-TTS system.

Consequently, the benefit of this continuous vocoder is that it has only two 1-dimensional parameters for modeling excitation (F0 and MVF), and the synthesis part is a computationally feasible solution. In the next chapter, we will discuss how to apply the proposed vocoder into sequence-to-sequence recurrent neural networks for further improving the quality of the TTS synthesis.

# Chapter 7

# Sequence-to-Sequence Recurrent Neural Network

*"We often think that when we have completed our study of one, we know all about two; because 'two' is 'one and one.' We forget that we still have to make a study of 'and'."*

Arthur Eddington (1882 – 1944)

## 7.1 Introduction

Deep neural networks have had a tremendous influence on speech synthesis in the last few years. In the previous chapter, I proposed a vocoder which was successfully used with a feed-forward deep neural network and outperformed the baseline based HMM-TTS. However, Zen and Senior [107] comprehensively listed several limitations of the conventional DNN-based acoustic modeling for speech synthesis, e.g. its lack of ability to predict variances, unimodal nature of its objective function, and the sequential nature of speech is ignored. Therefore, the use of sequence-to-sequence modeling with the recurrent neural networks (RNNs) is investigated in this Thesis chapter to overcome the limitations of the FF-DNN.

RNN is a more popular and effective acoustic model architecture which can process sequences of inputs and produces sequences of outputs. In particular, the RNN model is different from the DNN the following way: RNN operates not only on inputs (like the DNN) but also on network internal states that are updated as a function of the entire input history. In this case, the recurrent connections are able to map and remember information in the acoustic sequence, which is important for speech signal processing to enhance prediction outputs.

RNNs vary from main FF-DNNs in their hidden layers. Every RNN hidden layer gets inputs not only from its previous layer but also from activations of itself for previous inputs. A basic version of this architecture is displayed in Figure 31, in which every node in the hidden layer is connected to the previous activation of every node in that layer.

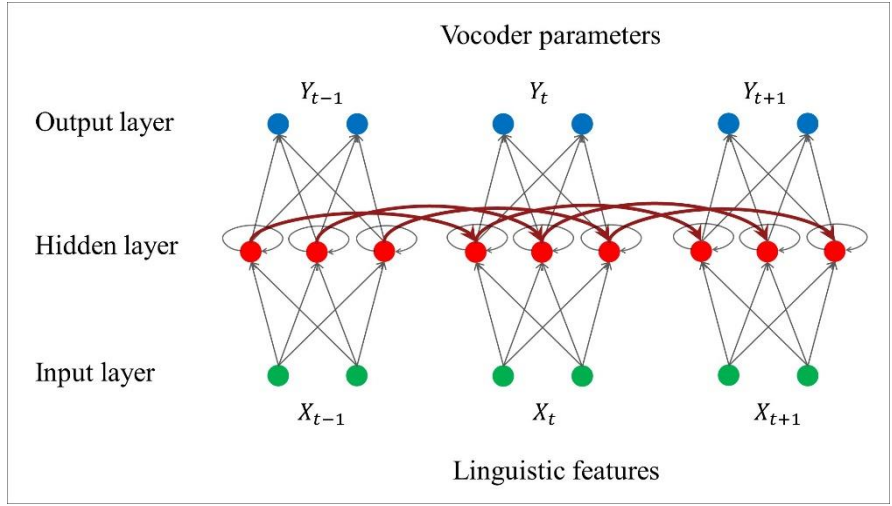

**Figure 31:** A basic version of RNN.

## 7.2 RNN Acoustic Models

In this section, the performance of recently proposed recurrent units is evaluated on sequence modelling using continuous vocoder.

### 7.2.1 Long Short-Term Memory

As originally proposed in [108] and recently used for speech synthesis [20], long short-term memory network (LSTM) is a class of recurrent networks composed of units with a particular structure to cope better with the vanishing gradient problems during training and maintain potential long-distance dependencies. This makes LSTM applicable to learn from history in order to classify, process and predict time series. Unlike the conventional recurrent unit which overwrites its content at each time step, LSTM have a special memory cell with self-connections in the recurrent hidden layer to maintain its states over time, and three gating units (input, forget, and output gates) which are used to control the information flows in and out of the layer as well as when to forget and recollect previous states. LSTM is formulated as follows:

$$i_t = \delta(W_i x_t + R_i h_{t-1} + p_i \odot c_{t-1} + b_i) \tag{63}$$

$$f_t = \delta(W_f x_t + R_f h_{t-1} + p_f \odot c_{t-1} + b_f) \tag{64}$$

$$c_t = f_t \odot c_{t-1} + i_t \odot tanh(W_c x_t + R_c h_{t-1} + b_c) \tag{65}$$

$$o_t = \delta(W_o x_t + R_o h_{t-1} + p_o \odot c_t + b_o) \tag{66}$$

$$h_t = o_t \odot tanh(c_t) \tag{67}$$

where $i_t$, $f_t$, and $o_t$ are the input, forget, and output gates, respectively; $c_t$ is the so-called memory cell; $h_t$ is the hidden activation at time $t$; $x_t$ is the input signal; $W$, and $R$ are the weight matrices applied on input and recurrent hidden units, respectively; $p$ and $b$ are the peep-hole connections and biases, respectively; $\delta(\cdot)$ and $tanh$ are the sigmoid and hyperbolic tangent activation functions, respectively; $\odot$ means element-wise product.

### 7.2.2 Bidirectional LSTM

The main concept of the Bi-LSTM was proposed in [109], and is a frequently used architecture for speech synthesis [110]. For a given input vector sequence $x = (x_1, ..., x_T)$, a regular RNN based Bi-LSTM calculates hidden state vector sequence $h = (h_1, ..., h_T)$ and outputs vector sequence $y = (y_1, ..., y_T)$. More specifically, Bi-LSTM separates the state neurons in a forward state sequence $\vec{h}$ (positive time direction), and backward state sequence $\overleftarrow{h}$ (negative time direction); which means that both forward and backward outputs are not connected. This can be observed in Figure 26. The iterative process of the Bi-LSTM can be defined here as

$$\vec{h}_t = tanh(W_{x\vec{h}}x_t + W_{\vec{h}\vec{h}}\vec{h}_{t-1} + b_{\vec{h}}) \quad (68)$$

$$\overleftarrow{h}_t = tanh(W_{x\overleftarrow{h}}x_t + W_{\overleftarrow{h}\overleftarrow{h}}\overleftarrow{h}_{t-1} + b_{\overleftarrow{h}}) \quad (69)$$

$$y_t = W_{\vec{h}y}\vec{h}_t + W_{\overleftarrow{h}y}\overleftarrow{h}_t + b_y \quad (70)$$

where $W$ is the connection weight matrix between two layers (e.g. $W_{xh}$ is the weight matrix between input and hidden vectors), $b$ is the bias vectors, and $tanh$ denotes a tangent activation function which is defined in Chapter 6.

### 7.2.3 Gated Recurrent Unit

A slightly more simplified variation of the LSTM, the gated recurrent unit (GRU) architecture was recently defined and found to achieve a better performance than LSTM in some cases [111]. GRU has two gating units (update and reset gates) to modulate the flow of data inside the unit but without having separate memory cells. The update gate supports the GRU to capture long term dependencies like that of the forget gate in LSTM. Moreover, because an output gate is not used in GRU, the total size of GRU parameters is less than that of LSTM, which allow that GRU networks converge faster and avoid overfitting. GRU is formulated as follows:

$$h_t = (1 - z_t)h_{t-1} + z_t\, h_t \quad (71)$$

$$z_t = \delta(W_z x_t + U_z h_{t-1}) \quad (72)$$

$$h_t = \tanh(W x_t + U(r_t \odot h_{t-1})) \quad (73)$$

$$r_t = \delta(W_r x_t + U_r h_{t-1}) \quad (74)$$

where $h_t$ and $z_t$ are the output and update gates, respectively; $U$ projects the input into a hidden space, $\delta$ is a logistic sigmoid function, $r_t$ is a set of reset gates and $\odot$ is an element-wise multiplication.

### 7.2.4 Hybrid Model

The advantage of RNNs is that they are able to make use of previous context. In particular, RNN based Bi-LSTM acoustic model has been shown to give state-of-the-art performance on speech synthesis tasks [110]. Obviously, there are two important drawbacks to use fully Bi-LSTM hidden layers. Firstly, the speed of training becomes very slow due to iterative multiplications over time that leads to network paralysis problems. A second problem is that the training process can be tricky and sometimes expensive undertaking due to gradient vanishing and exploding [112].

In an attempt to overcome these limitations, I propose a modification to the fully Bi-LSTM layers by using Bi-LSTM for lower layers and unidirectional RNN for upper layers to reduce complexity and to make the training easier while all the contextual information from past and future have been already saved in the memory. Consequently, reducing memory requirements and the potential of being suitable for real-time applications are the main advantages of using this topology.

## 7.3 Evaluation

In order to achieve our goals and to verify the effectiveness of the proposed methods, objective and subjective evaluations were carried out.

### 7.3.1 Network Topology

I trained a feed-forward DNN and four different recurrent neural network architectures, each having either LSTM, Bi-LSTM, GRU, or Hybrid. The objective of these experiments is to find out the best network type to model the continuous vocoder parameters. The topologies implemented in this experiment are as follows:

- **DNN:** 6 feed-forward hidden layers; each one has 1024 hyperbolic tangent units.
- **LSTM:** 3 feed-forward hidden lower layers of 1024 hyperbolic tangent units each, followed by a single LSTM hidden top layer with 1024 units. This recurrent output layer makes smooth transitions between sequential frames while the 3 bottom feed-forward layers intended to act as feature extraction layers.
- **Bi-LSTM:** Similar to the LSTM architecture, but replacing the top hidden layer with a Bi-LSTM layer of 1024 units.
- **GRU:** Similar to the Bi-LSTM architecture, but replacing the top hidden layer with a GRU layer of 1024 units.
- **Hybrid:** 2 Bi-LSTM hidden lower layers followed by another 2 standard RNN top layers each of which has 1024 units.

### 7.3.2 Empirical Measures

To get an objective picture of how these four recurrent network models perform against the DNN using the continuous vocoder, performance of these systems is assessed by five metrics:

1) **MCD (dB):** Mel-Cepstral Distortion to measure 60-dimensional mel-cepstral coefficients, as follow

$$MCD = \frac{1}{N}\sum_{j=1}^{N}\sqrt{\sum_{i=1}^{K}(x_{i,j} - y_{i,j})^2} \qquad (75)$$

where $x$ and $y$ are the $i^{th}$ cepstral coefficients of the natural and synthesized speech signals, respectively.

2) **RMSE$_{MVF}$ (Hz):** Root mean squared error to measure maximum voiced frequency parameter performance.
3) **RMSE$_{F0}$ (Hz):** Root mean squared error to measure fundamental frequency prediction performance.
4) **Overall validation error:** A validation loss between valid and train sets from last epoch (iteration).
5) **CORR:** The correlation measures the degree to which reference and generated data are close to each other (linearly related).

$$CORR = \frac{\sum_{i=1}^{n}(x_i - \overline{x})(y_i - \overline{y})}{\sqrt{\sum_{i=1}^{n}(x_i - \overline{x})^2}\sqrt{\sum_{i=1}^{n}(y_i - \overline{y})^2}} \qquad (76)$$

Where $\overline{x}$ and $\overline{y}$ are the mean of the natural $x_i$ and synthesized $y_i$ speech frames; respectively.

For all empirical metrics, a calculation is done frame-by-frame and a lower value indicates better performance except for the CORR measure where +1 is better. Overall validation error throughout the training decreases with epochs which indicates a convergence.

The test results for the baseline (DNN) and the proposed recurrent models are listed in Table 10. Compared to the DNN, the Bi-LSTM reduces all four experimental measures, and obtain similar performance for the male and female speakers. Although the Hybrid system is not better than Bi-LSTM, it slightly drops the validation error in case of AWB speaker from 1.632 in Bi-LSTM to 1.627. Interestingly, the Hybrid system does not outperform the baseline model. This indicates that increasing the number of recurrent units in the hidden layers is not helpful. We also see that using GRU system has no positive effect on the objective metrics. In summary, these empirical outcomes demonstrate that using Bi-LSTM systems to train continuous vocoder parameters improves the synthesis performance and outperforms DNN and other recurrent topologies.

**Table 10:** Objective measures for all training systems based on synthesized speech signal using proposed Continuous vocoder for SLT and AWB speakers.

| Systems | MCD (dB) | | MVF (Hz) | | F0 (Hz) | | CORR | | Validation error | |
|---|---|---|---|---|---|---|---|---|---|---|
| | SLT | AWB | SLT | AWB | SLT | AWB | SLT | AWB | SLT | AWB |
| DNN | 4.923 | 4.592 | 0.044 | 0.046 | 17.569 | 22.792 | 0.727 | 0.803 | 1.543 | 1.652 |
| LSTM | 4.825 | 4.589 | 0.046 | 0.047 | 17.377 | 23.226 | 0.732 | 0.793 | 1.526 | 1.638 |
| GRU | 4.879 | 4.649 | 0.046 | 0.047 | 17.458 | 23.337 | 0.731 | 0.791 | 1.529 | 1.643 |
| Bi-LSTM | **4.717** | **4.503** | **0.042** | **0.044** | **17.109** | **22.191** | **0.746** | **0.809** | **1.517** | 1.632 |
| Hybrid | 5.064 | 4.516 | 0.046 | 0.044 | 18.232 | 22.522 | 0.704 | 0.805 | 1.547 | **1.627** |

### 7.3.3 Subjective Listening Test

This test compared natural sentences with the synthesized sentences from the baseline (DNN), proposed (Bi-LSTM, Hybrid), and an anchor system. The anchor was an HMM-TTS using a simple pulse-noise excitation vocoder. From the four proposed recurrent systems, I only included Hybrid and Bi-LSTM, because in informal listening we perceived only minor differences between the four variants of the sentences. We evaluated ten sentences from speaker AWB, and ten sentences from speaker SLT. The listening test samples can be found online[9].

Another 13 participants (6 males, 7 females) with a mean age of 29 years were asked to conduct the online listening test. On average, the test took 23 minutes to fill. The MUSHRA scores for all the systems are shown in Figure 32. For speaker AWB, both recurrent networks outperformed the DNN system, and the Bi-LSTM and Hybrid networks are not significantly different from each other (Mann-Whitney-Wilcoxon ranksum test, $p<0.05$). For speaker SLT, we found that the Bi-LSTM system reached the highest naturalness scores in the listening test, consistent with objective errors reported above. In case of the female speaker, this difference between the Bi-LSTM and Hybrid systems is statistically significant.

From both objective and subjective evaluation metrics, experimental results demonstrated that the proposed RNN models can improve the naturalness of the speech synthesized significantly over our DNN baseline. These experimental results showed the potential of the recurrent networks based approaches for SPSS. In particular, the Bi-LSTM network achieves better performance than others.

---

[9] http://smartlab.tmit.bme.hu/vocoder2019

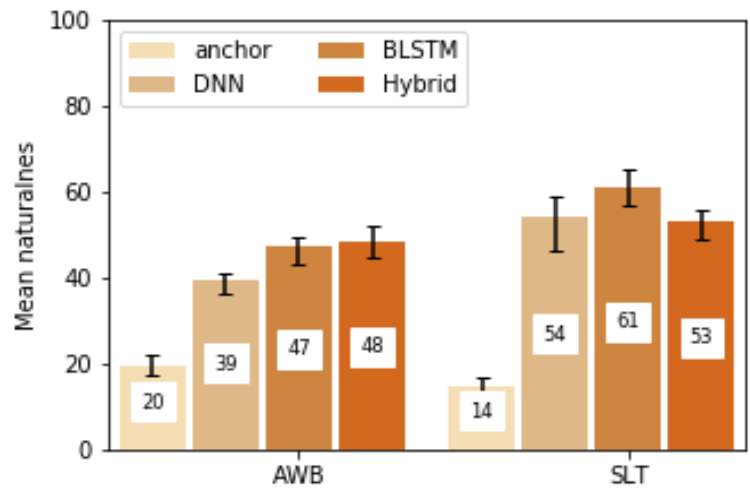

**Figure 32:** Results of the MUSHRA listening test for the naturalness question. Error bars show the bootstrapped 95% confidence intervals. The score for the reference (natural speech) is not included.

## 7.4 Summary

This chapter focused on the task of sequence modeling based on continuous vocoder, which was ignored in the conventional feed-forward neural network. Four different deep recurrent architectures (LSTM, BLSTM, GRU, and Hybrid models) have been implemented to train our acoustic features. From objective evaluation metrics, experimental results demonstrated that the proposed RNN-TTS model can improve the naturalness of the speech synthesized significantly over our DNN-TTS baseline. Preference tests show the proposed method gives further improved performance.

These experimental results showed the potential of the recurrent networks based approaches for SPSS. In particular, the Bi-LSTM network achieves better performance than others.

# Part IV
# Sinusoidal Modelling

# Chapter 8

# Continuous Sinusoidal Model

*"Nothing in life is to be feared, it is only to be understood. Now is the time to understand more, so that we may fear less."*

Marie Curie (1867 – 1934)

## 8.1 Introduction

Parametric speech synthesis based on TTS systems have steadily advanced in terms of naturalness during the last two decades. Even though the quality of synthetic speech is still unsatisfying, the benefits of flexibility, robustness, and control denote that SPSS stays as an attractive proposition. Besides, vocoder performance is the most important factor limiting the impact of overall voice quality in SPSS [2]. Vocoders attempt to produce a decoded signal that sounds like the original speech. Therefore, several approaches based on mathematical and physical models have been suggested to model the overall speech signal.

In recent years, a number of sophisticated source-filter based vocoders have been proposed and extensively used in speech synthesis. Specifically, for example, STRAIGHT (Speech Transformation and Representation using Adaptive Interpolation of weiGHT spectrum) vocoder [5] is probably the most used vocoder for SPSS which decomposes signals into spectral envelope, excitation, and aperiodicity parameters. For real-time processing, the computational issue is expensive in STRAIGHT. Furthermore, the Deterministic plus Stochastic Model (DSM) proposed by Drugman et al. [113] is based on a two-band mixed excitation in which the upper band was treated as noise and the lower band was modeled through a set of deterministic waveforms. More recently, a high-quality vocoder named WORLD was developed in [14] to meet the requirements of real-time processing.

Sinusoidal vocoder is an alternative category for the source-filter model of speech and has been successfully applied to a broad range of speech processing problems such as speech modification and conversion. Sinusoidal modeling can be characterized by the amplitudes, frequencies, and phases of the component sine waves; and synthesized as the sum of a number of sinusoids that can generate high quality speech. For each frame, a set of those parameters is estimated corresponding to peaks in the short-term Fourier transform. Concisely, voiced speech can be modeled as a sum of harmonics (quasi periodic) spaced at F0 with instantaneous phases, whereas unvoiced speech can be represented as a sum of sinusoids with random phases [114].

Various sinusoidal model formulations have been discussed in the literature. In particular, Harmonic plus Noise Model (HNM) was developed in [115] and has shown the capability of providing high-quality copy synthesis and prosodic modifications. Based on time-varying frequency, HNM decomposes speech into deterministic lower band where the signal is modeled as a sum of harmonically related sinusoids and stochastic upper band where the signal is modeled by colored noise. Another sinusoidal based speech vocoder is being developed by Degottex and Stylianou [116] in which an adaptive Quasi-Harmonic vocoder (aQHM) and Adaptive Iterative Refinement (AIR) method combined as an intermediate model to iteratively minimize the mismatch of harmonic frequencies. Hence, the full system is called aHM-AIR. Similarly, Perception based Dynamic sinusoidal Model (PDM) and Harmonic Dynamic Model (HDM) have been proposed in [117] and have both been applied during analysis and synthesis to be modelled in hidden Markov models (HMM) based speech synthesis.

Thus, from a point of view of either objective or subjective measures, sinusoidal vocoders were preferred in terms of quality. However, these models have usually more parameters (each frame has to be represented by a set of frequencies, amplitude, and phase) than in the source-filter models. Consequently, more memory would be required to code and store the speech segments. Although some experiments have been made to use either an intermediate model [116] or intermediate parameters (regularized cepstral coefficients) [117] to overcome these issues, the computational complexity of SPSS can be quite high once additional algorithms are including [2].

By keeping the number of our vocoder parameters unchanged [85], which are simpler to model than traditional vocoders with discontinuous F0, the goal of the work reported in this Thesis was to develop a new sinusoidal model as an alternative synthesis technique in a continuous vocoder, which can provide a high quality sinusoidal model with a fixed and low number of parameters.

## 8.2 Proposed Method

The sinusoidal model assumes the excitation-filter is modeled by a sum of sine waves. Continuous vocoder based Sinusoidal Model (CSM) was designed to overcome shortcomings of discontinuity in the speech parameters and the computational complexity of modern vocoders. The novelty behind this vocoder is to use harmonic features to facilitate and improve the synthesizing step before speech reconstruction.

By keeping the number of our previous source-filter vocoder parameters unchanged [85] and similarly to [115] [50], the synthesis algorithm implemented in this Thesis decomposes the speech frames into a lower-band voiced component $s_v(t)$ and an upper-band noise component $s_n(t)$ based on MVF values. We define these components here as

$$s(t) = s_v(t) + s_n(t) \tag{77}$$

In order to avoid discontinuities at the frames boundaries, Overlap-add (OLA) technique is used to reconstruct the speech signal from their corresponding parameters estimated from our analysis model in [85]. If the current frame is voiced, the harmonic part can be expressed as:

$$s_v^i(t) = \sum_{k=1}^{K^i} A_k^i \cos(k w_0^i t + \varphi_k^i) \tag{78}$$

$$w_0^i = 2\pi \frac{contF0^i}{F_s} \tag{79}$$

where $A_i$ and $\varphi_i$ are the synthetic harmonic amplitudes and phases at frame $i$, respectively, $F_s = 16\ kHz$ is the sampling frequency, $t = 0, 1, \ldots, L$ and $L$ is the frame length. $K$ is the time-varying frequency components or harmonics that depends on the contF0 and MVF as:

$$K^i = \begin{cases} round\left(\dfrac{MVF^i}{contF0^i}\right) - 1, & voiced\ frames \\ 0, & unvoiced\ frames \end{cases} \tag{80}$$

$$A_k^i = 2\sqrt{contF0^i} \cdot H_h^i(k\,contF0^i) \cdot exp(Re\{C_k^i\}) \tag{81}$$

where $H_h$ is complementary low-pass filter for the harmonic part, $C_i$ is complex harmonic log-amplitude obtained by resampling the MGC [104] envelope

$$C_k^i = c_0^i + 2\sum_{n=1}^{N} c_n^i \cos(n\beta_\alpha^i) \tag{82}$$

$$\beta_\alpha(w_0^i) = tan^{-1} \frac{(1-\alpha^2)\sin w_0^i}{(1+\alpha^2)\cos w_0^i - 2\alpha} \tag{83}$$

where $\alpha$ is the all-pass factor takes 0.42 for $F_s = 16\ kHz$. The phases are obtained recursively in a minimum phase response between harmonics in adjacent frames

$$\varphi_k^i = Im\{C_k^i\} + k\gamma^i \tag{84}$$

$$\gamma^i = \gamma^{i-1} + \frac{T}{2}(w_0^i + w_0^{i-1}) \tag{85}$$

where $k\gamma^i$ is a linear-in-frequency term which can be attributed to the underlying excitation, and $T$ is the frame shift measured in samples (typically it corresponds to a $5ms$ interval).

The synthetic noise signal $n(t)$ is filtered by a high-pass filter $f_h(t)$ with a cutoff frequency equal to the local MVF, and then modulated by its time-domain envelope $e(t)$ as we described it in Chapter 2 [85]

$$s_n^i(t) = e^i(t)\left[f_h^i(t) * n^i(t)\right] \tag{86}$$

If the current frame is unvoiced, the harmonic part is zero and the synthetic frame is usually equal to the produced noise. Hence, the synthesized speech signal is obtained by adding the harmonic and noise components. A block diagram of the proposed architecture is depicted in Figure 33.

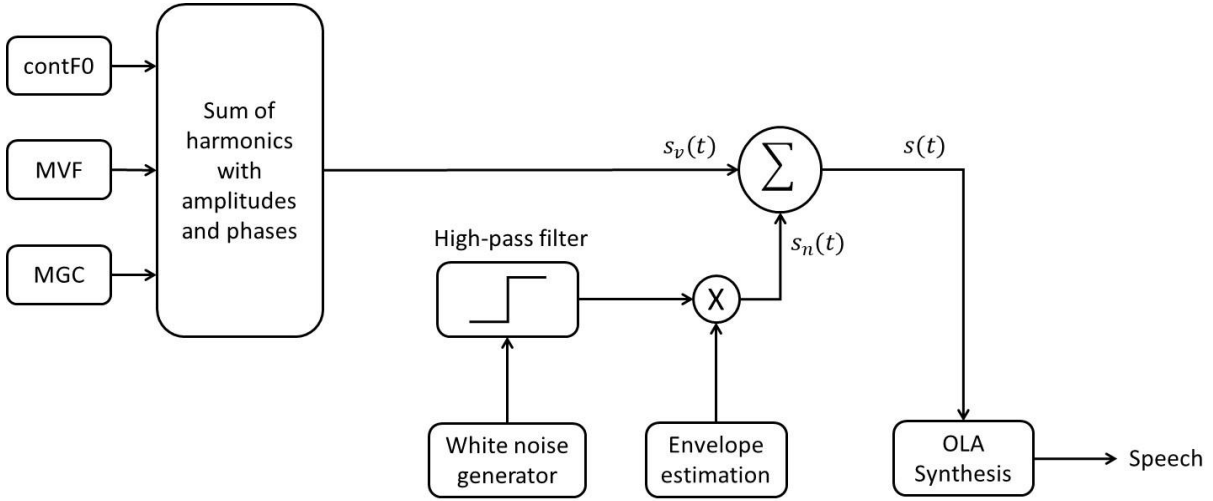

**Figure 33:** Block diagram of the sinusoidal-synthesis part in a continuous vocoder.

## 8.3 Evaluation

### 8.3.1 Objective Evaluation

A range of acoustic objective measures are considered to evaluate the quality of synthesized speech based on the proposed sinusoidal vocoder. We adopt the fwSNRseg for the error criterion since it is said to be much more correlated to subjective speech quality than classical SNR [38]. Moreover, Extended Short-Time Objective Intelligibility (ESTOI) measure is used to calculate the correlation between the natural and processed speech. We also measure the Itakura-Saito (IS) distance that has played a key role in speech analysis and synthesis [118]. To a large extent, most studies (such as [119]) confirmed that when the IS distance is below 0.1, the two spectra would be perceptually nearly identical. IS can be defined based on the linear predication coefficients (LPC) as

$$IS = \frac{1}{N} \sum_{i=1}^{N} \frac{\sigma_{x,i}^2}{\sigma_{y,i}^2} \left( \frac{a_{y,i}^T R_{x,i} a_{y,i}}{a_{x,i}^T R_{x,i} a_{x,i}} \right) + \log\left(\frac{\sigma_{y,i}^2}{\sigma_{x,i}^2}\right) - 1 \qquad (87)$$

where $a_x$, $a_y$, and $R_x$ are the LPC vector of the natural speech frame, synthesized speech frame, and the autocorrelation matrix, respectively; $\sigma_x^2$ and $\sigma_y^2$ are the LPC all-pole gains. For all objective measures, a calculation is done frame-by-frame and a higher value indicates better performance except for the IS measure (lower value is better). The results were averaged over the selected utterances (50 sentences) for each speaker.

As Table 11 shows, the proposed vocoder tends to significantly outperform the baseline approach among all metrics. In particular, it can be seen from IS measure that the proposed vocoder is slightly better than STRAIGHT in the AWB speaker whereas this is not the case with the SLT speaker. It can be concluded that the CSM presented in this Thesis has similar, or only slightly worse, performance to the reference vocoders.

**Table 11:** Average scores performance based on synthesized speech for male and female speakers. The bold font shows the best performance.

| Vocoder | IS | | fwSNRseg | | ESTOI | |
|---|---|---|---|---|---|---|
| | AWB | SLT | AWB | SLT | AWB | SLT |
| Baseline | 0.148 | 0.447 | 6.987 | 7.940 | 0.517 | 0.676 |
| Proposed | **0.058** | **0.082** | **9.560** | **11.034** | **0.749** | **0.867** |
| WORLD | **0.016** | **0.014** | **13.312** | 13.336 | **0.808** | **0.951** |
| STRAIGHT | 0.065 | 0.042 | 11.840 | **14.641** | 0.772 | 0.933 |

In addition, Table 12 compares the parameters of the vocoders under study. It can be seen that the continuous sinusoidal model uses only two one-dimensional parameters for modeling the excitation, the WORLD vocoder is applying a five-dimensional band aperiodicity, whereas STRAIGHT use high-dimensional parameters which makes the statistical modelling approach progressively complex and computationally intensive. The findings also point out that the CSM has few parameters compared to the WORLD and STRAIGHT vocoders, and it is computationally feasible; therefore, it is suitable for real-time operation.

**Table 12:** Parameters and excitation type of applied vocoders

| Vocoder | Parameter per frame | Excitation |
|---|---|---|
| CSM | F0: 1 + MVF: 1 + MGC: 24 | Mixed |
| WORLD | F0: 1 + Band aperiodicity: 5 + MGC: 60 | Mixed |
| STRAIGHT | F0: 1 + Aperiodicity : 1024 + Spectrum: 1024 | Mixed |

### 8.3.2 Subjective Evaluation

In order to evaluate the perceptual quality of the proposed systems, we conducted a web-based MUSHRA listening test. I compared natural sentences with the synthesized sentences from the baseline, proposed, STRAIGHT, WORLD, and an anchor system. The anchor type was the re-synthesis of the sentences with a standard pulse-noise excitation vocoder. The utterances were presented in a randomized order. The listening test samples can be found online[10].

13 participants (7 males, 6 females) with an age range of 20-42 years (mean: 31 years), were asked to conduct the online listening test. We evaluated ten sentences (five from each speaker). Altogether, 60 utterances were included in the test (2 speaker x 6 types x 5 sentences). On average, the test took 15 minutes to fill. The MUSHRA scores for all the systems are showed in Figure 34. According to the results, the proposed vocoder outperformed the baseline system for both speakers, and preferred over STRAIGHT (not significant), showing that the sinusoidal extension of the CSM reached an adequate level to the naturalness of speech.

---

[10] http://smartlab.tmit.bme.hu/specom2018

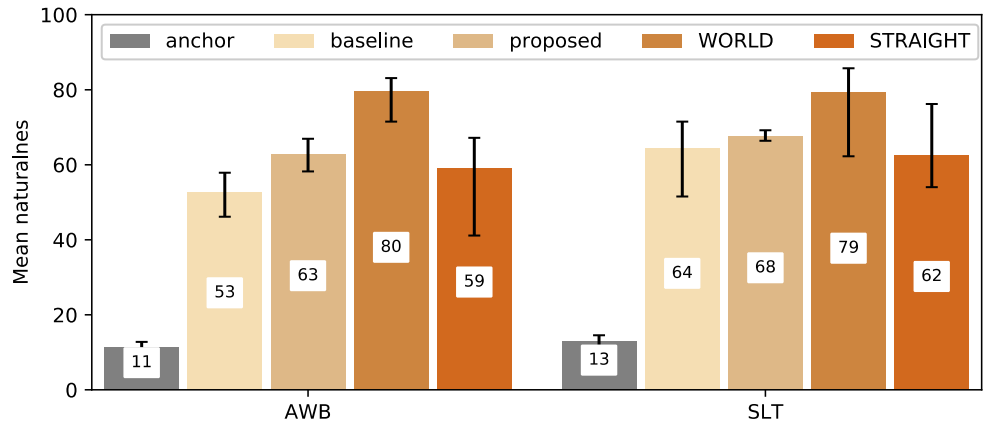

**Figure 34:** Results of the MUSHRA listening test for the naturalness question. Error bars show the bootstrapped 95% confidence intervals. The score for the reference (natural speech) is not included.

## 8.4 Summary

This chapter has proposed a new approach with the aim of designing a high quality continuous vocoder using a sinusoidal model. The performance of the systems has been evaluated through objective and subjective listening tests. Experiments demonstrate that our proposed model generates higher output speech quality than the baseline, that is a source-filter based vocoder. It was also found that the results obtained with the proposed vocoder were preferred over STRAIGHT and somewhat worse than with WORLD vocoders. Moreover, the findings point out that the continuous vocoder has few parameters and is computationally feasible; therefore, it is suitable for real-time applications.

The applicability of the proposed CSM will be studied for two major fields of speech processing: text-to-speech (Chapter 9) and voice conversion (Chapter 11).

# Chapter 9

# CSM with Deep Learning

*"Success in creating AI would be the biggest event in human history. Unfortunately, it might also be the last, unless we learn how to avoid the risks."*

Stephen Hawking (1942 – 2018)

## 9.1 Related Work

In an earlier work, a computationally feasible residual-based vocoder was proposed in [8], using a continuous F0 model [9], and MVF [10]. In this method, the voiced excitation consisting of pitch synchronous PCA residual frames is low-pass filtered and the unvoiced part is high-pass filtered according to the MVF contour as a cutoff frequency. The approach was especially successful for modelling speech sounds with mixed excitation. However, we noted that the unvoiced sounds are sometimes poor due to the combination of continuous F0 and MVF. In [85], the time structure of the high-frequency noise component was further controlled by estimating a suitable temporal envelope.

In [120], I successfully modelled all vocoder parameters (continuous F0, MVF, and MGC) with FF-DNNs and shown that the FF-DNN have higher naturalness than HMM based text-to-speech. Furthermore, modeling the parameters of continuous vocoder using RNN, LSTM, BLSTM, and GRU variants is extended in [121]. Experimental results demonstrate that using Bi-LSTM systems to train continuous vocoder parameters improves the synthesis performance and outperforms FF-DNN and other recurrent topologies. The advantage of a continuous vocoder in this scenario is that vocoder parameters are simpler to model than conventional vocoders with discontinuous F0.

Previous studies have shown that human voice can be modelled effectively as a sum of sinusoids and has shown the capability of providing high-quality copy synthesis and prosodic modifications [115] [116] [117]. Therefore in [122], I proposed a continuous sinusoidal model (CSM) that is applicable in statistical frameworks by keeping the number of our vocoder parameters unchanged [85]. Experimental results from objective and subjective evaluations have shown that the CSM gives state-of-the-art vocoders performance in analysis-synthesis while outperforming the previous work based on source-filter vocoder. Therefore, in this Thesis, I study the interaction between CSM and Bi-LSTM based RNN by feeding linguistic features to the Bi-LSTM based neural network to predict acoustic features, which are then passed to a CSM to generate the synthesized speech. I expect that the new model gives better performance using RNN and enhances the quality of synthesized speech.

## 9.2  Proposed Method

The mathematical background for both Bi-LSTM and CSM approaches have already been given and explained in Chapter 7 and 8, respectively. Based on them, the overall architecture is depicted in the block diagram as shown in Figure 35. Consequently, 4 feed-forward hidden layers each consisting of 1024 units and performs a non-linear function of the previous layer's representation, followed by a single Bi-LSTM layer with 385 units, will be used in this work to train the CSM parameters. In the RNN-TTS experimental tests, 132 sentences from each speaker were analyzed and synthesized with the WORLD, baseline (that is our source-filter model [121]) and proposed vocoders. For WORLD and baseline vocoders, I used the same RNN architecture as for the proposed sinusoidal model.

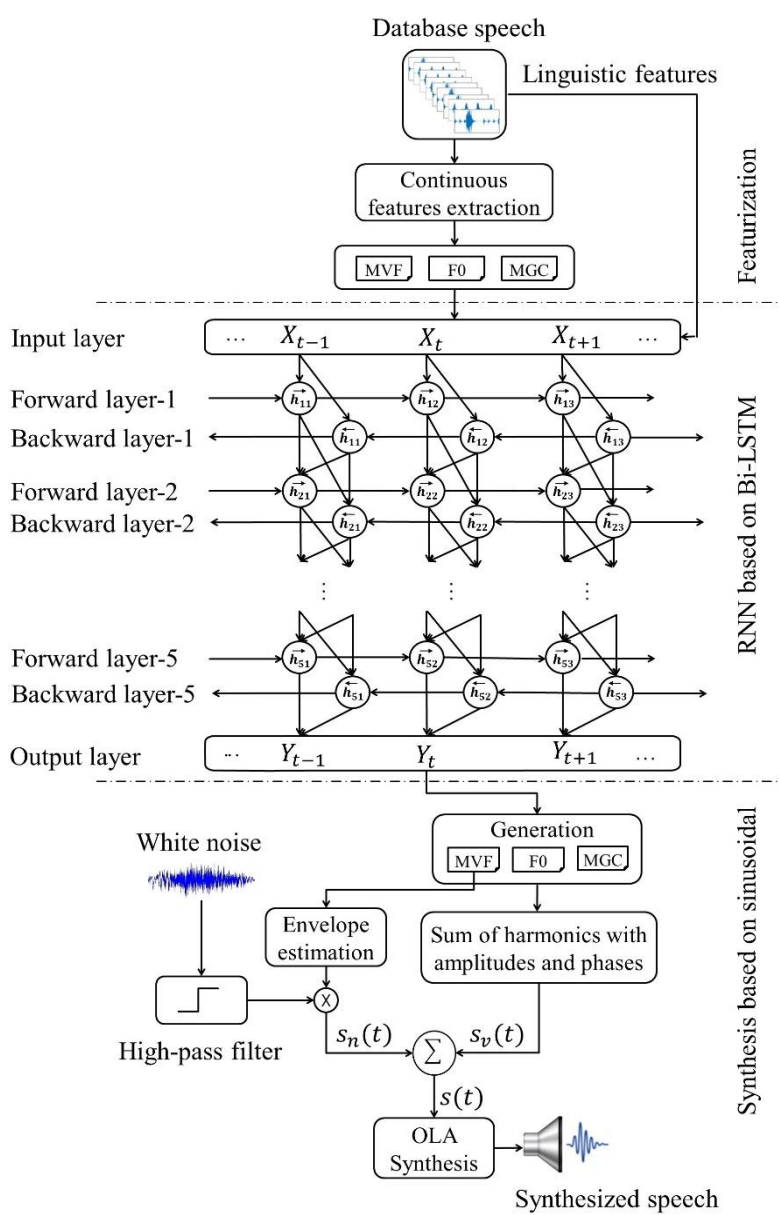

**Figure 35:** Block diagram of the CSM based Bi-LSTM.

## 9.3 Evaluation

### 9.3.1 Error Metrics

A range of objective speech quality and intelligibility measures are considered to evaluate the quality of the proposed model. The results were averaged over the test utterances for each speaker. Log-Likelihood Ratio (LLR), frequency-weighted segmental SNR (fwSNRseg), and Log Spectral Distortion (LSD) were used in this experiment. LSD can be defined as the square difference carried over the logarithm of the spectral envelopes of natural $X(f)$ and synthesized $Y(f)$ speech signals at $N$ frequency points

$$LSD = \sqrt{\frac{1}{N}\sum_{i=1}^{N} mean\ (\log X(f_i) - log Y(f_i))^2} \tag{88}$$

The results were averaged over 132 test sentences, and the best value in each column of Table 13 is bold faced. For WORLD [14] and baseline [121] (that is a source-filter) vocoders, we used the same RNN architecture as for the proposed one.

It is good to note that the findings in Table 13 showed that the proposed vocoder based sinusoidal model succeeded in the Bi-LSTM training. Moreover, the CSM framework provides satisfactory results in terms of naturalness and intelligibility comparable to the high-quality WORLD vocoder and baseline. In particular, LLR between natural and synthesized speech frame is smaller than those using the baseline and WORLD methods. Focusing on the fwSNRseg, it indicates that the WORLD model outperformed the proposed one only in the male speaker. While in terms of LSD, lowest correlation values were obtained with baseline method. However, a slightly improvement was noted for the CSM over the WORLD model.

As a result, these experiments showing that the proposed model with continuous sinusoidal model was beneficial in the statistical deep recurrent neural networks.

**Table 13:** Average scores performance based on synthesized speech signal using proposed CSM for Male and Female speakers.

| Metrics | Model | AWB | SLT |
|---|---|---|---|
| **LLR** | Baseline | 1.4309 | 1.6966 |
| | Proposed | **1.4178** | **1.6791** |
| | WORLD | 1.5008 | 1.7516 |
| **fwSNR$_{seg}$** | Baseline | 2.514 | 1.1882 |
| | Proposed | 2.4972 | **1.2278** |
| | WORLD | **2.5802** | 0.81389 |
| **LSD** | Baseline | **2.0739** | **2.2254** |
| | Proposed | 2.0995 | 2.2391 |
| | WORLD | 2.108 | 2.3373 |

### 9.3.2 Subjective Test

To demonstrate the efficiency of our proposed model, we performed a web-based MUSHRA listening test. We compared natural sentences with the synthesized sentences from the baseline, proposed, WORLD, and an anchor system. The anchor type was the re-synthesis of the sentences with a standard pulse-noise excitation vocoder. 100 utterances were included in the test (2 speakers x 5 types x 10 sentences). 13 participants between the age of 24-38 (mean age: 31 years) were asked to conduct the online listening test. Five of them were males and eight were females. On average, each test was completed within 17 minutes. The listening test samples can be found online[11].

The results of the listening test are presented in Figure 36 for the two speakers separately. For speaker AWB, it can be observed that the proposed framework significantly outperforms the baseline vocoder (Mann-Whitney-Wilcoxon ranksum test, with a 95% confidence level), while for speaker SLT, this difference is not significant. In both cases, the WORLD vocoder was rated slightly better than the CSM, but this difference is also not significant. This means that CSM based RNN-TTS is closer to the level of the state-of-the-art high quality vocoder than the baseline system.

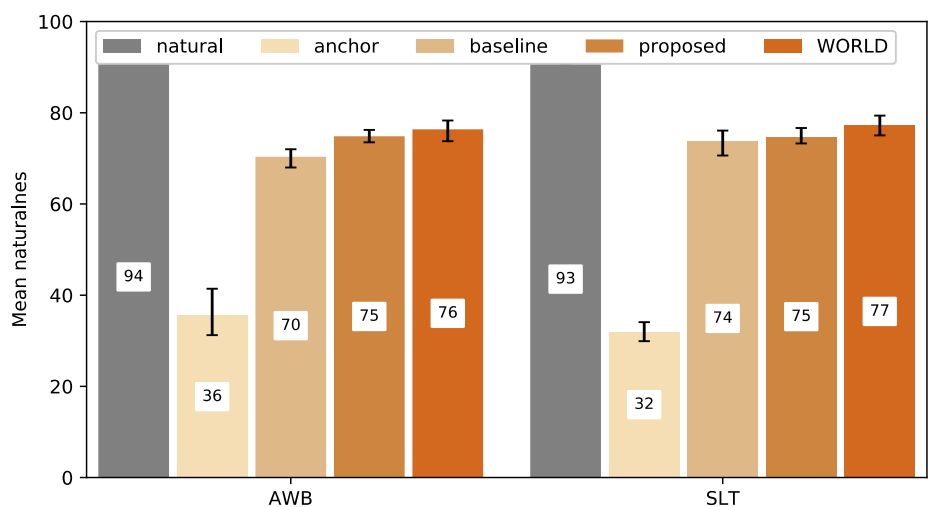

**Figure 36:** MUSHRA scores for the naturalness question. Higher value means better naturalness. Errorbars show the bootstrapped 95% confidence intervals.

---

[11] http://smartlab.tmit.bme.hu/ijcnn2019_vocoder

## 9.4 Summary

In this Chapter, we have introduced a novel simple approach to the statistical parametric speech synthesis using continuous sinusoidal model. The main idea was to integrate the CSM into sequence-to-sequence Bi-LSTM deep neural network. The experiment was successful and we were able to add the continuous features (F0, Maximum Voiced Frequency, and Mel-Generalized Cepstrum) to the training framework based RNN. That means that the acoustic model has fewer acoustic features to predict from the input text features than commonly used by conventional vocoders. Using a variety of measurements, the performance strengths and weaknesses of the proposed method for two different speakers were highlighted. From both objective and subjective evaluation metrics, the performance of the proposed system clearly tends to perform better than the baseline. The naturalness achieved by the proposed waveform generator was also found to be closed to the state-of-the-art model that uses the WORLD vocoder.

# Part V
# Voice Conversion

# Chapter 10

# Statistical VC with Source-Filter Model

*"Measure what can be measured, and make measureable what cannot be measured."*

Galileo Galilei (1564 – 1642)

## 10.1 Introduction

Statistical voice conversion (SVC) is an effective technique for flexibly synthesizing several kinds of speech. While keeping the linguistic content and environmental conditions unchanged, the goal of SVC is to change and modify speaker individuality; i.e., the source speaker's voice is transformed to sound like that of the target speaker [7]. There are several applications within the concept of voice conversion, such as converting speech from impaired to normal voice [123], from normal to singing sound [124], electro-laryngeal to normal speech [125], etc.

Over the years, voice conversion frameworks have mostly focused on spectral conversion between source and target speakers [126] [127]. In the sense of the statistical parametric approaches, such as Gaussian mixture model (GMM) [128] and exemplar based on non-negative matrix factorization [129] [130], SVC marked a success in spectrum linear conversion. Nonlinear transformation approaches, such as hidden Markov models (HMMs) [131], deep belief networks (DBNs) [132] and restricted Boltzmann machines (RBMs) [133], have been shown to be effective in modeling the relationship between source-target features more accurately. The DBN and RBM were used to replace GMM to model the distribution of spectral envelopes [134]. However, the resulting speech parameters from these models tend to be over-smoothed and affect the similarity and quality of generated speech. To cope with these problems, some approaches attempt to reduce the difference between natural and the converted speech parameters by using Global variance [128], modulation spectrum [135], dynamic kernel partial least squares regression [136], or generative adversarial networks [137]. Even though these techniques achieve some improvements, the accuracy of the converted voice still deteriorates compared to the source speaker. Therefore, improving the performance of converted voice is still a challenging research question.

There seem to be four factors that degrade the quality of SVC: 1) speech parameters (i.e. vocoder features), 2) mapping function between the source and target speakers, 3) learning model, and 4) vocoder synthesis quality. To capture the quality of these factors, feed-forward deep neural networks (FF-DNNs) were proposed as an acoustic modeling solution of different research areas [4] [98] [138]. FF-DNNs have shown their ability to extract high-level, complex abstractions and data representations from large volumes of supervised and unsupervised data [97], and achieve significant improvements in various machine learning areas including the ability to model high-dimensional acoustic parameters [139], and the availability of multi-task learning [100]. In this chapter, I predict acoustic features using a FF-DNN, which are then passed to a vocoder to generate the converted speech waveform. Thus, FF-DNN is used to improve both the converted acoustic parameters and the vocoder performance.

Recent studies are still considering some of these vocoders [23] in voice conversion, such as STRAIGHT [140] [128] [141], mixed excitation [142], Harmonic plus Noise Model [143], glottal source modeling [144], or even with more complex vocoders like adaptive WAVENET [145], or Tacotron [146]. Consequently, simple and uniform vocoders, which would handle all speech sounds and voice qualities (e.g. creaky voice) in a unified way, are still missing in SVC. Therefore, it is still worth to develop advanced vocoders for achieving high-quality converted speech. In our recent work in statistical parametric speech synthesis, a novel continuous vocoder using continuous parameters is proposed, which was shown to improve the performance under a FF-DNN [120]. However, in SVC, the effectiveness of the continuous vocoder has not been confirmed yet. Thus, we are developing a solution in this Thesis to achieve higher sound quality and conversion accuracy, while the SVC remains computationally efficient.

Unlike the methods referenced above, the proposed structure implicates two major technical developments. First, I build a voice conversion framework that consists of a FF-DNN and a continuous vocoder to automatically estimate the mapping relationship between the parameters of the source and target speakers. Second, I apply a geometric approach to spectral subtraction (GA-SS) to improve the signal-to-noise ratio of the converted speech. I expect that the new voice conversion model gives high-quality synthesized speech compared to the source voice.

## 10.2 Voice Conversion Model

The framework of the proposed SVC system is shown in Figure 37. It consists of feature processing, training and conversion-synthesis steps. MVF, contF0, and MGC parameters are extracted from the source and target voices using the analysis function of the continuous vocoder. A training process based on a FF-DNN is applied to construct the conversion phase.

The purpose of the conversion function is to map the training features of the source speaker $X = \{x_i\}_{i=1}^{I}$ to the corresponding training features of the target speaker $Y = \{y_j\}_{j=1}^{J}$. Here, $X$ ad $Y$ vector sequences are time-aligned frame by frame by the Dynamic Time Warping (DTW) algorithm [147] [148] since both vectors differ in the durations and have different-length recordings. DTW is a technique for deriving a nonlinear mapping between two vectors to minimize the overall distance $D(X, Y)$ between the source and target speakers. So that the time events (sequence of phonemes) in $x_i$ can be aligned to corresponding events in $y_j$ using the warp path $W_k$ to form such a new joint vector sequence $z_t$ of equal length. Specifically, the $k^{th}$ element in $W$ can be constructed as

$$W_k = (i_k, j_k) \tag{89}$$

$$\max(I, J) \leq K < I + J \tag{90}$$

where $K$ is the length of the $W$. The minimum-distance warp path between two feature vectors is

$$d(W) = d(i,j) = \|x_i - y_j\| \qquad (91)$$

Thus,

$$D(X,Y) = min \sum_{k=1}^{K} d(W_k) \qquad (92)$$

Then, the time-aligned acoustic feature sequences of both speakers are trained and used for the conversion function in order to predict the target features from the features of the source speaker. Finally, the converted $con\grave{t}F0$, $M\grave{V}F$, and $M\grave{G}C$ are synthesized to get the converted speech waveform by the synthesis function of the Continuous vocoder.

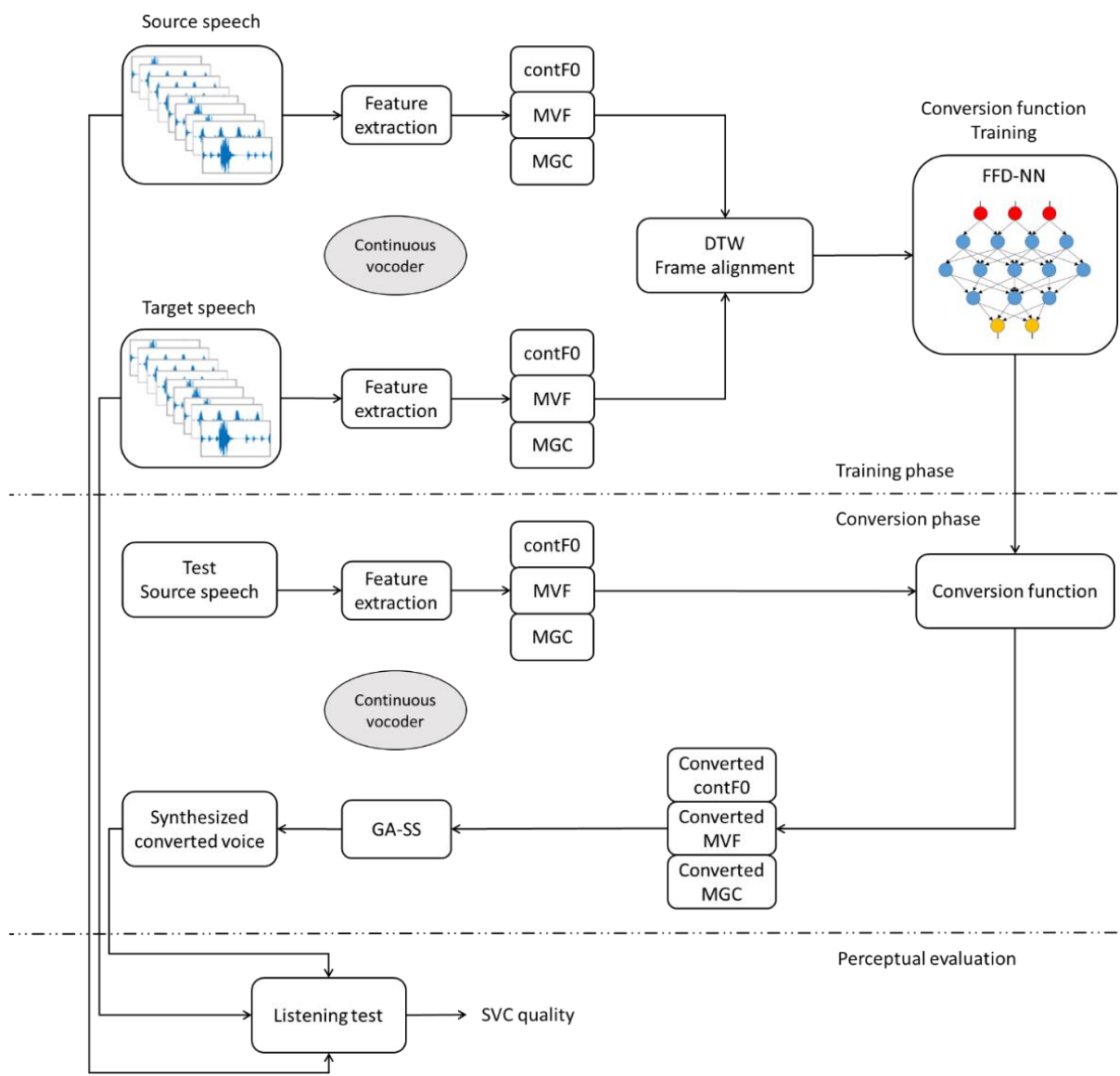

**Figure 37:** Flowchart of the proposed SVC algorithm.

## 10.3 Reducing Unwanted Frequencies

The goal of this section is to remove or reduce the level of unwanted high-frequency components from the converted features, that may be generated during training or conversion phase. Therefore, we apply the GA-SS approach proposed by [149] in order to improve the performance of the converted speech signal. This approach consistently outperforms other conventional spectral subtractions particularly at low SNRs. Besides, GA-SS is more suitable for our work because of its simplicity and low computational cost. Here, GA-SS can be applied in each frame signal $f(n)$ by letting $y(n) = f(n) + e(n)$ be the sampled speech signal with the estimation error $e(n)$, assuming that the first 3 frames are noise/silence. Taking the short-time Fourier transform of $y(n)$

$$Y(w_k) = F(w_k) + E(w_k) \tag{93}$$

where $w_k = 2\pi k/N$, $k = 0,1,2,\ldots,N-1$, and $N$ is the frame length in samples. Then, we can rewrite Equation (93) in polar form as

$$A_Y e^{j\theta_Y} = A_F e^{j\theta_F} + A_E e^{j\theta_E} \tag{94}$$

where $A$ and $\theta$ are the magnitude and phase of the frame spectra respectively. Taking into account the trigonometric principles in Equation (94), the gain function $H_G$ can be derived as always real and positive

$$H_G = \frac{A_F}{A_Y} = \sqrt{\frac{1 - \cos^2(\theta_Y - \theta_E)}{1 - \cos^2(\theta_F - \theta_E)}} \quad , (H_G \geq 0) \tag{95}$$

Obtain the enhanced magnitude spectrum of the signal by

$$\hat{A}_F = H_G * A_Y \tag{96}$$

Using the inverse discrete Fourier transform of $\hat{A}_F \cdot e^{j\theta_Y}$, the enhanced frame signal $\hat{f}(n)$ can be obtained.

To clarify the effects of this approach, white Gaussian noise is added to the natural and synthetic speech waveforms. The amount of noise is specified by signal-to-noise ratio (SNR) in the range of -20 to 10 dB. The root mean square (RMS) error was calculated over 20 sentences selected randomly from each speaker. The smaller the value of RMS, the better performance. The overall RMS error values obtained as a function of the SNR between clean speech (natural or synthesized) sample and the noisy one (the same speech sample, with noise added) is shown in Figure 38. The results suggest that the RMS for the synthesized signal with GA-SS approach is smallest and close to the natural signal than without GA-SS. Nevertheless, the differences were very small. But adding this approach as an extra step to our proposed model does help to some extent in improving the overall sound quality, especially in noisy conditions.

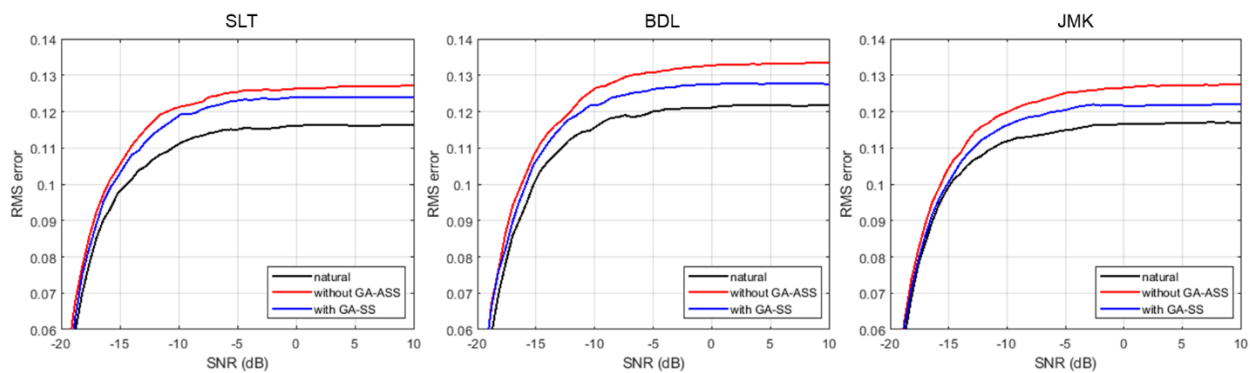

**Figure 38:** Influence of the GA-SS approach on the average RMS error. We present the average RMS error over 20 synthesized sentences per each speaker. "SLT" is an American English female speaker, whereas "BDL" and "JMK" are American and Canadian English male speakers, respectively.

## 10.4 Evaluation

I used three main speakers, namely SLT, BDL, and JMK for source and target speakers. With the aim of seeing the statistical behavior of the proposed model, four cross-gender ("male-to-female" and "female-to-male") conversions experiments are carried out for evaluation:
- SLT to BDL
- BDL to SLT
- SLT to JMK
- JMK to SLT

### 10.4.1 Error Measurement Metrics

It is well-known that the efficient method for evaluating speech quality is typically done through subjective listening tests. However, there are various issues related with the use of subjective testing. It can be sometimes very expensive, time consuming, and hard to find a sufficient number of suitable volunteers [91] [150]. For that reason, it can often be useful in this work to run objective tests in addition to listening tests. Seven objective measures are considered to evaluate the quality of the proposed model.

A reference (baseline) system with high quality performance is required to demonstrate the effectiveness and performance of the proposed methodology. Since the WORLD vocoder has a high-quality speech synthesis system and better than several high-quality vocoders (such as STRAIGHT), we use it as our state-of-the-art baseline within SVC and used the same architecture as for the proposed vocoder.

It is interesting to emphasize that the findings in Table 14 showed that the baseline does not meet the performance of our proposed model. That, in other words, the results reported in Table 14 strongly support the use of the proposed vocoder for SVC. In particular, the fwSNRseg between converted and target speech frames using the proposed method with continuous vocoder are higher than those using the baseline method. Nevertheless, the WORLD vocoder is shown to be better only for the SLT-to-JMK speaker conversion.

The comparison of the spectral envelope of one speech frame converted by the proposed method is given in Figure 39. The converted spectral envelope is plotted along with the source and the preferred target. It may be observed that the converted spectral envelope is more similar in general to the target one than the source one. Even though, these two trajectories seem similar, they are moderately smoothed compared with the target one; that can affect the quality of the converted speech. It can also be seen in Figure 40 that the converted contF0 trajectories generated from proposed method follow the same shape of the target confirming the similarity between them and can provide better F0 predictions. Similarly, when looking at Figure 41, it makes apparent that the proposed framework produces converted speech with MVF more similar to the target trajectories rather than to the source ones.

As a result, these experiments show that the proposed model with continuous vocoder is competitive for the SVC task, and superior to the reference WORLD model.

**Table 14:** Average scores on converted speech signal per each of the speaker pairs conversion.

| Error metrics | Model | SLT-to-BDL | BDL-to-SLT | SLT-to-JMK | JMK-to-SLT |
|---|---|---|---|---|---|
| **MCD** | Reference | 5.624 | 5.355 | 5.856 | 5.765 |
| | Proposed | **5.609** | **5.341** | **5.846** | **5.754** |
| **fwSNR$_{seg}$** | Reference | 1.660 | 1.119 | **2.162** | 0.558 |
| | Proposed | **3.072** | **1.873** | 1.970 | **1.312** |
| **LSD** | Reference | 2.423 | 2.208 | 2.506 | 2.557 |
| | Proposed | **2.214** | **2.107** | **2.368** | **2.401** |
| **IS** | Reference | 33.005 | 24.887 | 33.060 | 39.418 |
| | Proposed | **15.183** | **21.212** | **13.973** | **29.137** |
| **WSS** | Reference | 8.842 | 16.299 | 8.068 | 17.310 |
| | Proposed | **7.723** | **13.683** | **7.783** | **14.046** |
| **LLR** | Reference | 1.718 | 1.724 | 1.610 | 1.744 |
| | Proposed | **1.451** | **1.581** | **1.442** | **1.640** |
| **NCM** | Reference | 0.103 | 0.102 | 0.024 | 0.030 |
| | Proposed | **0.115** | **0.124** | **0.028** | **0.035** |

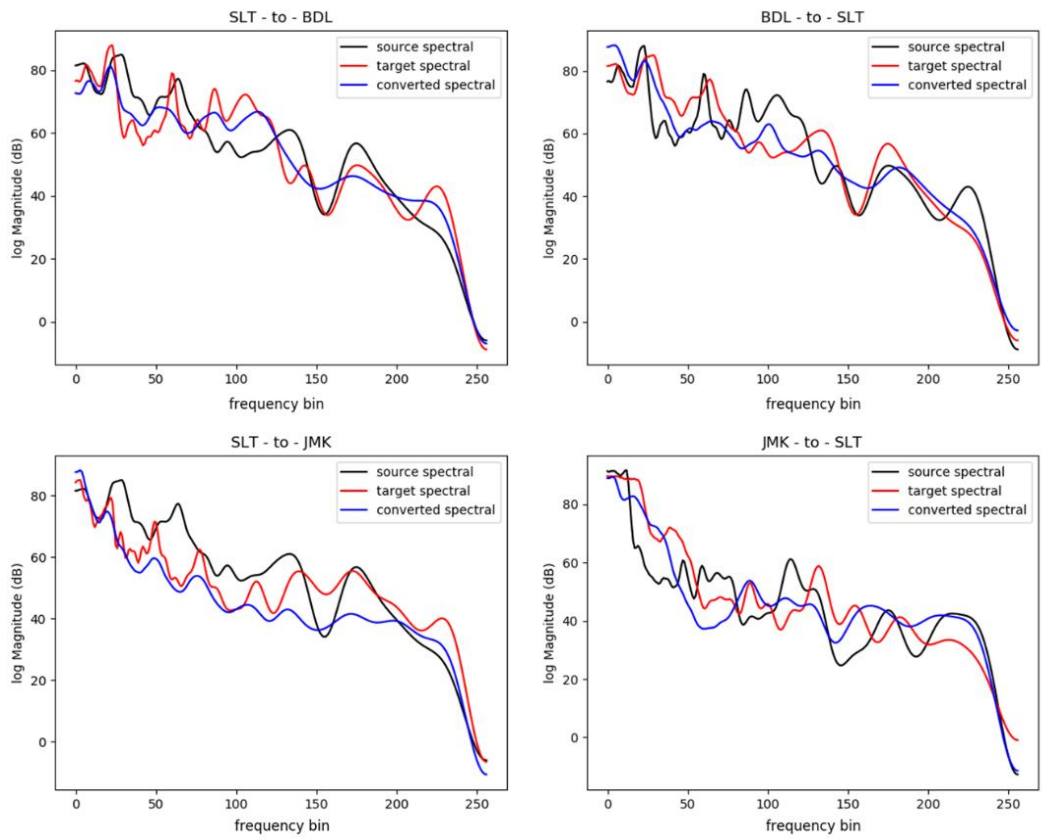

**Figure 39:** Example of one shorter segment /e/ from the natural source, target, and converted spectral envelopes using proposed method. Sentence: "Gad, your letter came just in time."

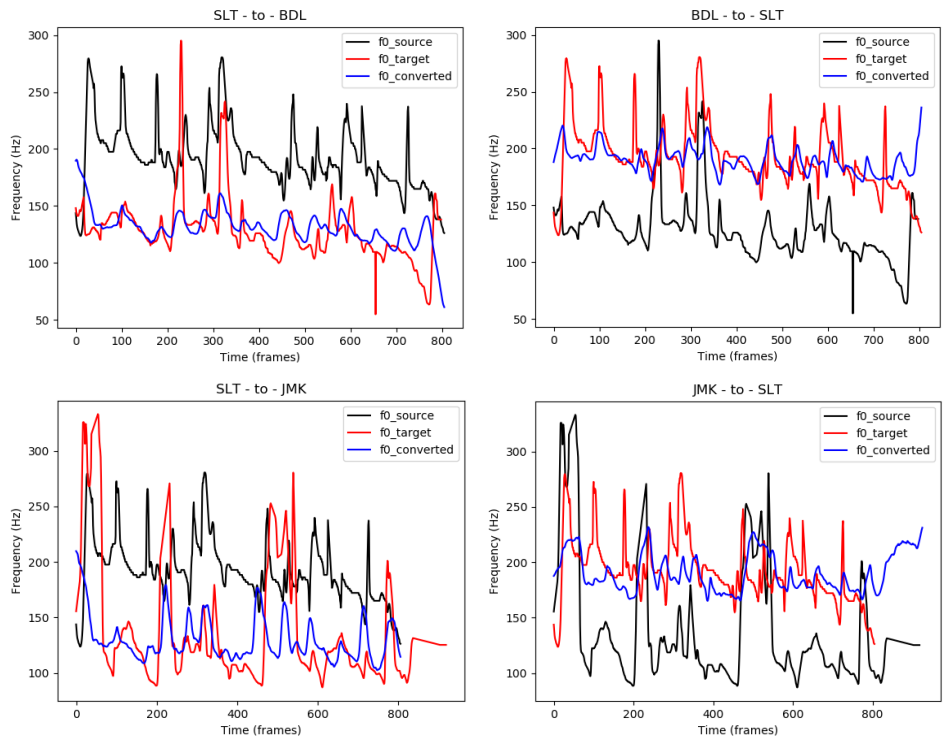

**Figure 40:** Example of the natural source, target, and converted contF0 trajectories using proposed method. Sentence: "From that moment his friendship for Belize turns to hatred and jealousy."

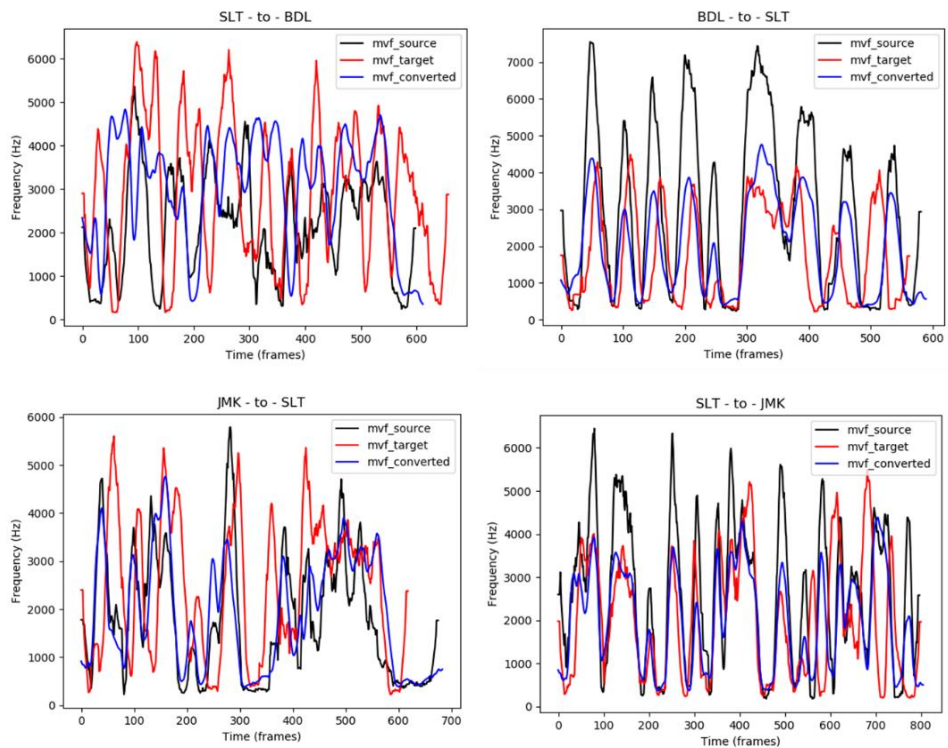

**Figure 41:** Example of the natural source, target, and converted MVF contours using the proposed method. Sentence: "Gregson shoved back his chair and rose to his feet."

### 10.4.2 Perceptual Test

To demonstrate the efficiency of our proposed model, we conducted two different perceptual listening tests. First, in order to evaluate the similarity of the converted speech to a reference target voice (which was the natural voice), we performed a web-based MUSHRA-like listening test. The advantage of MUSHRA is that it enables evaluation of multiple samples in a single trial without breaking the task into many pairwise comparisons, and it is a standard method for speech synthesis evaluations. Within the MUSHRA test, I compared four variants of the sentences: 1) Source, 2) Target, 3) Converted speech using the high-quality baseline (WORLD) vocoder, 4) Converted speech using the proposed (Continuous) vocoder. 48 utterances were included in the MUSHRA test (4 types x 12 sentences).

Second, in order to evaluate the overall quality and identity of the synthesized speech from both proposed and baseline systems, a Mean Opinion Score (MOS) test was carried out. In the MOS test we compared three variants of the sentences: 1) Target, 2) Converted speech using the baseline (WORLD) vocoder, and 3) Converted speech using the proposed vocoder. 36 utterances were included in the MOS test (3 types x 12 sentences).

19 participants between the age of 23-40 (mean age: 30 years) were asked to conduct the online listening test. 12 of them were males and 7 were females. On average, the MUSHRA test took 13 minutes, while the MOS test was 12 minutes long. The listening tests samples can be found online[12].

---

[12] http://smartlab.tmit.bme.hu/vc2019

The MUSHRA similarity scores of the listening test are presented in Figure 42. It can be seen that both systems achieve almost similar performance to the target voice across all gender combinations. This means that our proposed model has successfully converted the source voice to the target voice on cross-gender cases. In case of SLT-to-JMK conversion, the difference between the baseline and the proposed systems is statistically significant (Mann-Whitney-Wilcoxon ranksum test, with a 95% confidence level), while the other differences between the baseline and proposed are not significant.

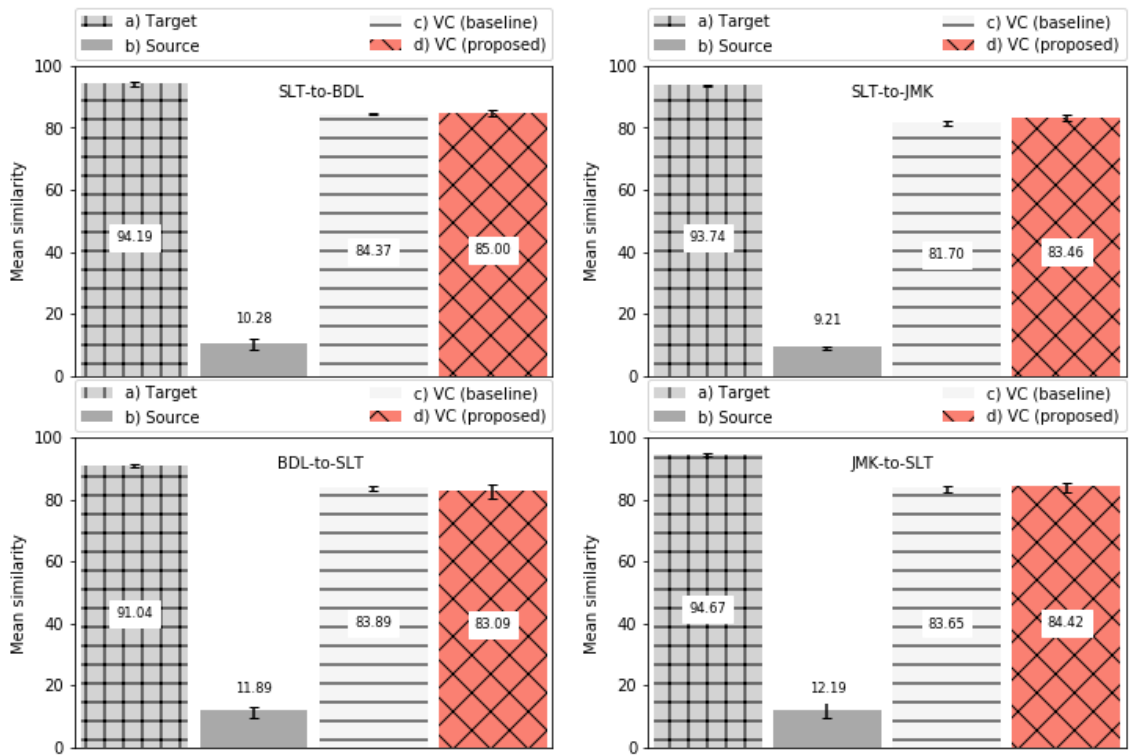

**Figure 42:** MUSHRA scores for the similarity question. Higher value means larger similarity to the target speaker. Errorbars show the bootstrapped 95% confidence intervals.

Additionally, Figure 43 shows the results of the MOS test. We can see that both the baseline and proposed systems achieved low naturalness scores compared to the target speaker, showing that the listeners clearly differentiated the utterances resulting from the voice conversion. It can be also found that the listeners preferred the baseline system compared to the proposed one.

As the final result of the listening tests investigating similarity to the target speaker and overall quality, we can conclude that the proposed continuous vocoder within the SVC framework performed well when compared to the voice conversion using the WORLD vocoder.

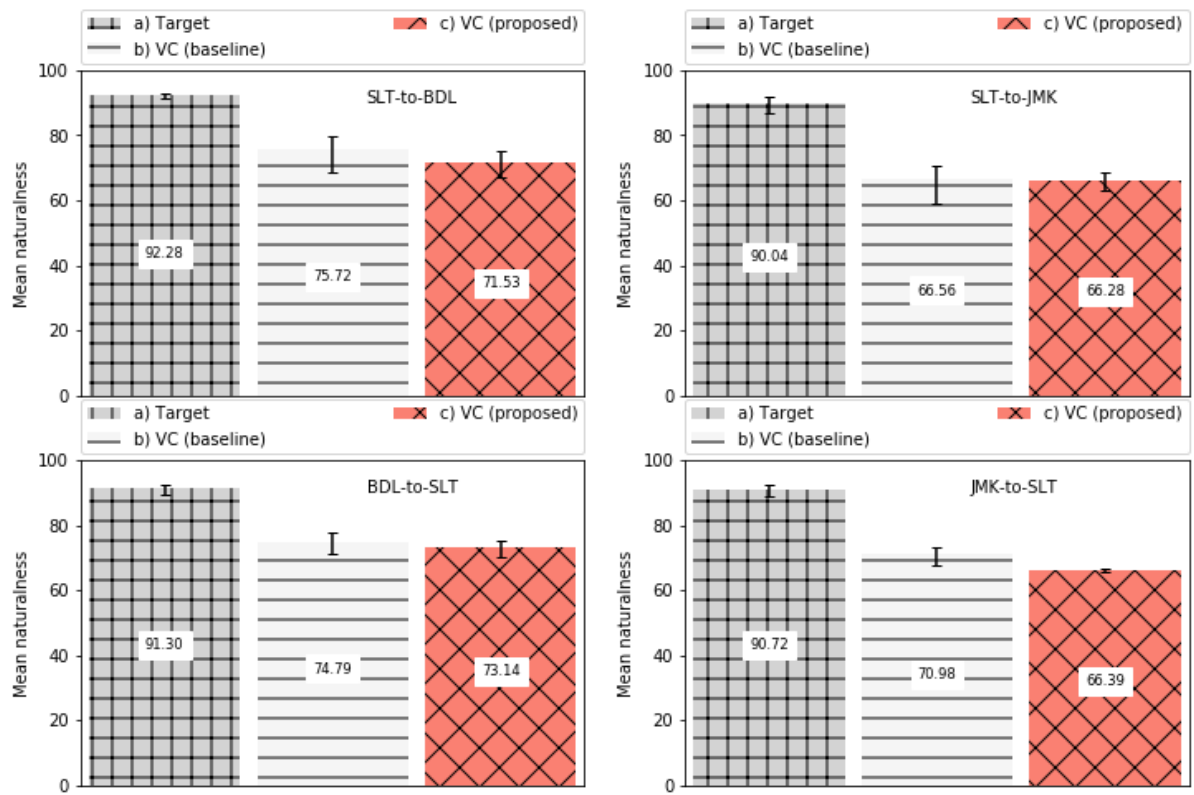

**Figure 43:** MOS scores for the naturalness question. Higher value means better overall quality. Errorbars show the bootstrapped 95% confidence intervals.

## 10.5 Summary

This Chapter has proposed a novel approach to statistical voice conversion using continuous acoustic features. The main goal was to integrate the continuous vocoder into the SVC framework, which provides an advanced model of the excitation signal, by converting its contF0, MVF, and spectral features within a statistical conversion function. The advantage of this vocoder is that it does not require to have a voiced/unvoiced decision, which means that the alignment error will be avoided in SVC between voiced and unvoiced segments. Therefore, its simplicity and flexibility allows us to easily construct a voice conversion framework using a FF-DNN.

Using a variety of measurements, the performance strengths and weaknesses of the proposed method for different speakers were highlighted. From the objective experiments, the performance of the proposed system was superior in most cases to that of the reference system (using the WORLD vocoder). Moreover, two listening tests have been performed to evaluate the effectiveness of the proposed method. The similarity test showed that the reference and proposed systems are both similar to the target speaker. This also confirms our findings, that are reported in the objective evaluations. Significant differences were not found compared to the reference system during the quality (MOS) test. This means that the proposed approach is capable of converting speech with higher naturalness and perceptual speech intelligibility.

# Chapter 11

# Parallel VC with Sinusoidal Model

*"We view things not only from different sides, but with different eyes; we have no wish to find them alike."*

Blaise Pascal (1623 - 1662)

## 11.1 Introduction

Most of the voice conversion systems found in the literature can be built either using a parallel framework in which source and target speakers read out the same set of utterances, or using a non-parallel framework in which the target speaker's utterances are different from those of the source speaker. However, in practice, the subjective experiment results in [151] [152] yield that the average performance of the non-parallel VC system is not outperformed by the parallel VC system. The main reason behind this challenging issue is that it is usually hard to achieve an accurate non-parallel frame alignment between speaker utterances and, therefore, a parallel data-driven approach will be used in this Thesis.

In essence, a well-designed VC system often consists of analysis, conversion, and synthesis modules. The process of parametrizing the input waveform into acoustic features and then synthesizing the converted waveform based on the converted features is one of the major factors that may degrade the performance of VCs. For this, the characteristics of the speech vocoder (analysis/synthesis system) given to the VC are of paramount importance. I can group the state-of-the-art vocoders based VC into three categories. a) Source-filter models: STRAIGHT [153] and mixed excitation [142]; b) Sinusoidal models: Harmonic plus Noise Model [143] is the only model has been found in the literature based VC; c) end-to-end complex models: WaveNet-based waveform generation [145] and Tacotron [146]. In the face of their clear differences, each model has advantages to work reasonably well, for a particular speaker or gender conversion task, which make them attractive to researchers. Nonetheless, such mismatch between the trained, converted, and tested features still exist, which often causes significant quality and similarity degradation.

There seem to be three important factors that should be taken into consideration in the design and development of a VC system. Firstly, the most common feature in most of the above-mentioned VC techniques is the fact that they are based on the spectral envelope (SE). Although SE contains enough information to convert the original speech signal onto that of the target speaker, SE is not enough alone to achieve the desired converted results, for particular applications, even with a better SE estimation method. Secondly, traditional conversion systems focus on the prosodic feature represented by the discontinuous fundamental frequency (F0) assumption that depends on a binary voicing decision. Therefore, modelling of F0 in VC applications is problematic because of the differing nature of F0 observations between voiced and unvoiced speech regions. An alternative solution of increasing the accuracy of the acoustic VC model is using a continuous F0 (contF0) to avoid alignment errors that may happen in voiced and unvoiced segments and can degrade the converted speech. It should be pointed out to the third issue that leads to the degradation of the performance of VC is that most of the existing VC techniques discard or does not typically preserve phase spectrum information. However, the effectiveness of phase information in detecting synthetic speech has recently been proved by [154]. Hence, one possible way of enhancing the accuracy of VC models is to incorporate phase information in order to achieve superior synthesized speech. Therefore, it is still worth to develop advanced vocoder based VC for achieving high-quality converted speech. To tradeoff between the complexity of the model and conversion accuracy in statistical VC, a sinusoidal type synthesis model based on contF0 is proposed. A number of recently developed VC methods have been applied and compared with the proposed model.

## 11.2 Voice Conversion Based on CSM

In [155] [156], the neural network based VC reaches higher performance on conversion than the GMM-based solution. In this work, a feed-forward deep neural network (FF-DNN) is used to model the transformation between source and target speech features as described in Chapter 6. It consists of 6 feed-forward hidden layers, each consisting of 1024 units. The framework of the proposed VC system is shown in Figure 44. Similarly to Chapter 10 (Section 10.2), conversion model is performed; whereas feature processing and synthesis steps are achieved with the CSM system (see Chapter 8).

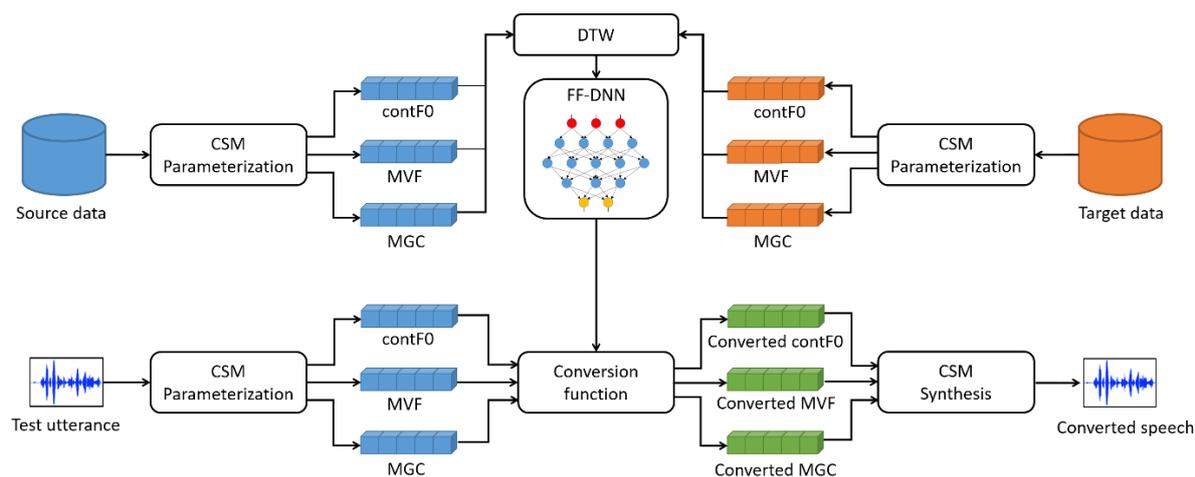

**Figure 44:** Voice conversion process with CSM based waveform generation.

## 11.3 Evaluation

Intra-gender and cross-gender pairs have been conducted in this experiment. The number of combinations of the source and target speaker was 12 pairs. Note that I trained the conversion models for every speaker pair independently. The proposed CSM based VC system was evaluated by comparing it with three systems, namely WORLD, MagPhase, and Sprocket.

To fairly compare all systems mentioned above, I used the same nonlinear conversion function architecture (FF-DNN) as for the proposed system, except Sprocket that is a linear function based on GMM. Thus, I ran 48 experiments in order to measure the performance of these VC-systems.

### 11.3.1 Objective Evaluation

Two objective speech quality measures are considered to evaluate the quality of the proposed model. A more detailed case-by-case analysis by fwSNRseg and LLR are shown in Table 15.

First, it could be observed that our proposed method gives significantly better LLR scores than other systems in female-to-male voice conversion. In other words, the CSM can convert voice characteristics more accurately than other methods when a female is a source speaker. Similar observations can be found in male-to-female voice conversions (in particular, BDL-to-SLT, BDL-to-CLB, and JMK-to-CLB), where the fwSNRseg measure tended to have the highest scores in our proposed model. In a sense, there is a tendency to an increased fwSNRseg when considering continuous F0 in the proposed method. Second, for the same-gender speaker pairs, the LLR values in Table 15 indicate that the proposed system obviously outperforms the baseline systems in female-to-female conversions. On the other hand, in terms of male-to-male voice conversions, our proposed system achieves the second highest sound quality. Overall, these findings demonstrate that the CSM can yield a good performance comparable to other systems.

As a result, these experiments show that the proposed model with continuous sinusoidal vocoder is competitive for the VC task and superior to the reference WORLD model.

### 11.3.2 Subjective Evaluation

A perceptual listening test was designed to test and evaluate the quality of our proposed model. First, we performed a web-based MUSHRA-like listening test to evaluate the speaker identity/similarity of the converted speech to a natural-reference target voice. Twelve utterances were randomly chosen and presented in a randomized order. Altogether, 72 utterances were included in the MUSHRA test (6 types x 12 sentences). Twenty listeners (11 males and 9 females) participated in the experiment. On average, the MUSHRA test took 10 minutes to fill. The listening tests samples can be found online[13].

---

[13] http://smartlab.tmit.bme.hu/sped2019_vc

The MUSHRA similarity scores of the listening test are presented in Figure 45. An interesting note is that the listeners overall preferred our system compared to others developed earlier. According to Mann-Whitney-Wilcoxon ranksum tests (with a 95% confidence level), all differences are statistically significant. This means that our proposed model has successfully converted the source voice to the target voice on the same-gender and cross-gender cases. Moreover, Figure 45 shows that the WORLD and Sprocket systems get higher scores in the MUSHRA test for only the JMK-to-SLT, JMK-to-BDL, and CLB-to-SLT speaker conversions, respectively.

Overall, these results suggest that the best conversion technique is the CSM, while the WORLD is also a good option, having the second highest similarity scores.

**Table 15:** Average scores on converted speech signal per each of the speaker pairs conversion

| Model | WORLD | | MagPhase | | Sprocket | | Proposed (CSM) | |
|---|---|---|---|---|---|---|---|---|
| | fwSNRseg | LLR | fwSNRseg | LLR | fwSNRseg | LLR | fwSNRseg | LLR |
| BDL → JMK | 2.19 | 1.57 | **3.21** | **1.37** | 2.20 | 1.48 | 2.47 | 1.50 |
| BDL → SLT | 1.12 | 1.72 | 1.25 | 1.69 | 1.04 | **1.49** | **2.33** | 1.57 |
| BDL → CLB | 0.79 | 1.83 | 1.65 | 1.72 | 0.37 | **1.69** | **1.66** | 1.74 |
| JMK → BDL | 1.31 | 1.76 | **2.49** | **1.56** | 1.73 | 1.63 | 2.15 | 1.57 |
| JMK → SLT | 0.55 | 1.74 | **1.93** | **1.56** | 0.11 | 1.64 | 1.54 | 1.65 |
| JMK → CLB | 1.45 | 1.74 | 1.75 | 1.66 | 0.69 | **1.60** | **1.81** | 1.67 |
| SLT → BDL | 1.65 | 1.71 | 1.60 | 1.70 | 1.80 | 1.51 | **2.95** | **1.49** |
| SLT → JMK | 2.16 | 1.61 | **2.71** | 1.42 | 0.713 | 1.56 | 2.59 | **1.39** |
| SLT → CLB | 1.51 | 1.75 | **2.89** | 1.59 | 2.32 | 1.56 | 2.51 | **1.50** |
| CLB → BDL | 0.97 | 1.81 | 1.60 | 1.70 | 0.95 | 1.72 | **1.92** | **1.60** |
| CLB → JMK | 2.50 | 1.49 | 2.74 | 1.40 | 0.98 | 1.46 | **3.00** | **1.30** |
| CLB → SLT | 0.98 | 1.70 | **2.17** | 1.53 | 1.96 | 1.54 | 2.12 | **1.47** |

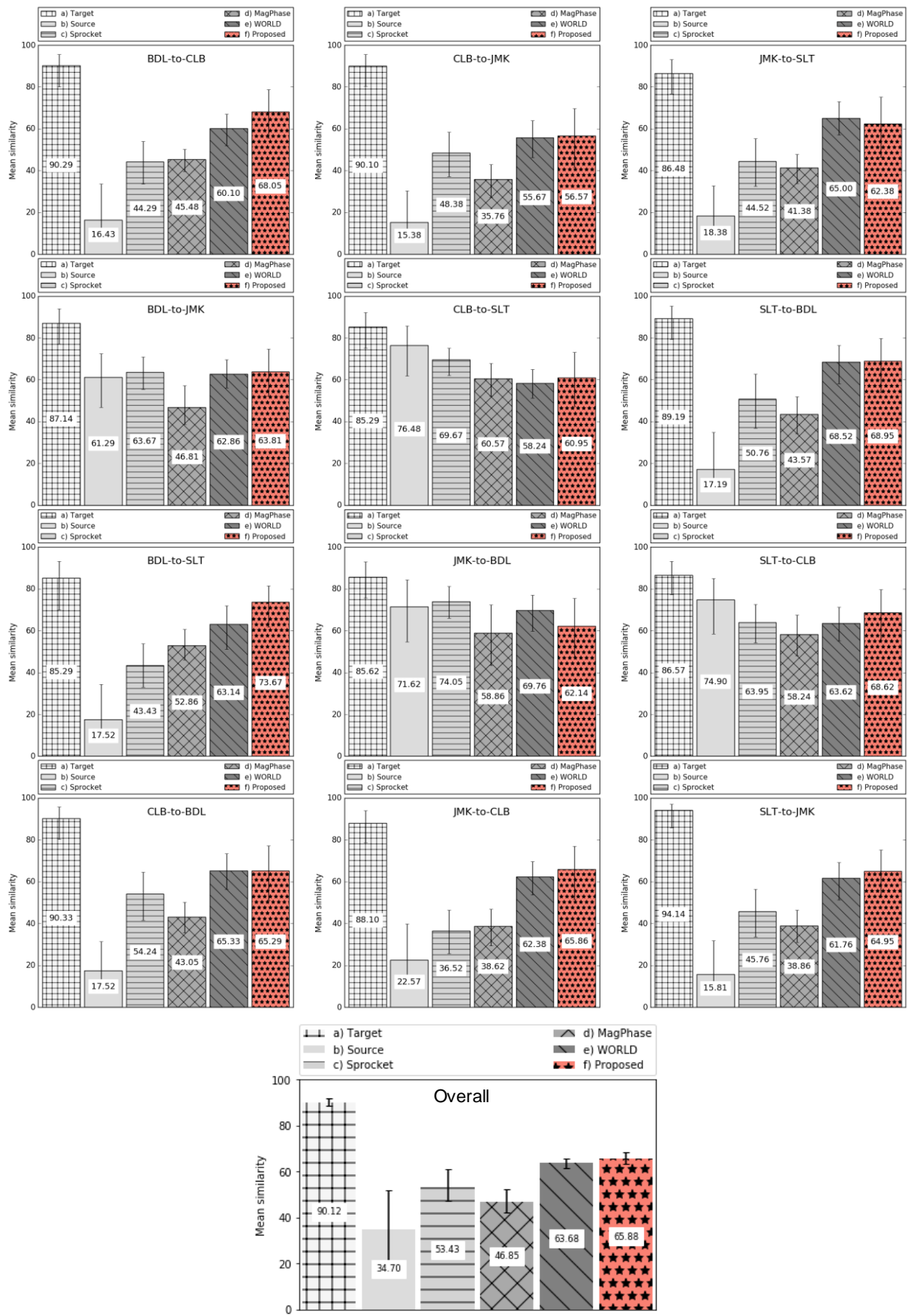

**Figure 45:** MUSHRA scores for the similarity question. Higher value means better overall quality. Errorbars show the bootstrapped 95% confidence intervals.

## 11.4 Summary

This Chapter has developed a voice conversion framework based on sinusoidal modelling. A number of recently developed VC methods have been applied and compared with the proposed model. The performance of the methods was statistically analyzed with two error metrics and subjectively evaluated by the use of expert opinion. The results discussed in previous section show the effectiveness of the proposed method in terms of speaker similarity. The advantage of the CSM is that it gives the closest results to the target speaker in both objective and similarity tests compared to other approaches.

# Chapter 12

# Conclusion and Future Work

*"Why will people go astray when they have this blessed book to guide them?"*

Michael Faraday (1791 - 1867)

## 12.1 Summary of New Scientific Results

This dissertation mainly focuses on developing a system for high-quality statistical parametric speech synthesis (SPSS) and voice conversion (VC) based on a vocoder and modeling aspects. The provided reliable solutions, which outperform other methods, can achieve higher sound quality and conversion similarity with deep learning advances, while its approach remains computationally efficient. The results achieved in the dissertation are summarized in five thesis groups. The author's publications where the actual thesis groups were published are indicated in square brackets.

**Thesis Group I: Modulating the Noise Component of the Excitation Signal**

Since the design of a vocoder-based SPSS depends on speech characteristics, the preservation of voice quality in the analysis/synthesis phase and the irregular "buzzy" synthetic speech sounds are the main problems of the vocoding approach. Therefore, this Thesis group considers the above issues by suggesting robust methods for advanced modeling of the noise excitation which can yield an accurate noise component of the excitation to remove the buzzy quality (Part I).

**Thesis I.1: Temporal envelopes [C1, J1, J5]** *I designed and implemented a new method to shape the high-frequency component of the unvoiced excitation by estimating the time envelope of the residual signal. I showed that this approach is helpful in achieving accurate approximations compared to natural speech and produce synthetic voice with significantly better quality than the baseline (Chapter 2).*

**Thesis I.2: Continuous Noise Masking [J4]** *I proposed an algorithm based on noise masking to reduce the perceptual effect of the residual noise and allowing a proper reconstruction of noise characteristics. I proved experimentally that continuous noise masking gives better quality resynthesized speech than traditional binary masking techniques (Chapter 3).*

**Thesis Group II: Excitation Harmonic Modelling**

This Thesis group concerns with estimating the fundamental frequency on clean and noisy speech signals. In particular, continuous F0 is still sensitive to additive noise in speech signals and suffers from short-term errors (when it changes rather quickly over time). Moreover, contF0 can cause some tracking errors when the speech signal amplitude is low or the voice is creaky. Therefore, I described novel approaches which can be used to enhance and optimize some other existing F0 estimator algorithms. Additionally, the Harmonic-to-Noise ratio technique is added as a new vocoded-parameter to the voiced and unvoiced excitation signal in order to reduce the buzziness caused by the vocoder (Part II).

***Thesis II.1: Adaptive Continuous Pitch Tracking Algorithm [C5, J2]*** *I developed and applied an adaptive approach based on Kalman filtering, time warping, and instantaneous frequency to optimize the performance of the continuous F0 estimation algorithm in clean and noisy speech. I showed that a clear advantage of the proposed approach is its robustness to additive noise; and the voice built with the proposed framework gives state-of-the-art speech synthesis performance while outperforming the previous baseline (Chapter 4).*

***Thesis II.2: Parametric HNR Estimation Approach [J2]*** *I proposed the addition of a new excitation HNR parameter to the voiced and unvoiced components. I proved that it can indicate the degree of voicing in the excitation and reduce the buzziness caused by the vocoder (Chapter 5).*

**Thesis Group III: Acoustic Modelling Based on Deep Learning**

Although the quality of synthesized speech generated by HMM-based speech synthesis has been improved recently, its naturalness is still far from that of actual human speech. These models have their limitations in representing complex and nonlinear relationships between the speech generation inputs and the acoustic features. Therefore, this Thesis group applies the continuous vocoder in deep neural network based TTS (Part III).

***Thesis III.1: Feed-Forward Deep Neural Network [C2, J5]*** *I built and implemented deep learning based acoustic modeling using FF-DNN with the continuous vocoder. The proposed DNN-TTS significantly outperformed the baseline method based on HMM-TTS, and its naturalness approaches the high-quality WORLD vocoder based TTS (Chapter 6).*

***Thesis III.2: Sequence-to-Sequence Recurrent Neural Network [C3, J5]*** *I investigated and examined sequence-to-sequence modelling using recurrent neural networks for the continuous vocoder. I showed that the performance of the vocoder can be significantly enhanced by the RNN framework and confirmed its superiority against the FF-DNN solution (Chapter 7).*

**Thesis Group IV: Sinusoidal Modelling**

In this Thesis group, the goal was to develop a new sinusoidal model based on continuous parameters in order to reach the level of the state-of-the-art high quality vocoders. The findings also point out that the proposed model has few parameters and is computationally feasible; therefore, it is suitable for real-time operation (Part IV).

**Thesis IV.1: Continuous Sinusoidal Model [C4]** *I designed a new vocoder based on the sinusoidal model that is applicable in statistical frameworks. I validated the efficiency and quality of the proposed model and proved that the proposed CSM vocoder gives state-of-the-art performance in resynthesized speech while outperforming the source-filter vocoder (Chapter 8).*

**Thesis IV.2: CSM with Deep Learning [C5]** *Based on the above results, I built and developed a deep learning based bidirectional LSTM version of the continuous sinusoidal model to generate high-quality synthesized speech. I showed that the proposed framework converges faster and provides satisfactory results in terms of naturalness and intelligibility comparable to the high-quality WORLD model based TTS (Chapter 9).*

**Thesis Group V: Voice Conversion**

Voice conversion aims to modify the speech signal of a source speaker into that of a target speaker. A well-designed VC system often consists of analysis, conversion, and synthesis modules. The process of parametrizing the input waveform into acoustic features and then synthesizing the converted waveform based on the converted features is one of the major factors that may degrade the performance of VCs. Therefore, this Thesis group aims to find new techniques to increase the efficiency of VC models. I performed extensive measurements to compare the proposed one to the earlier reference systems and proved its efficiency (Part V).

**Thesis V.1: Statistical VC with Source-Filter Model [J3]** *I proposed a novel VC system using the source-filter based continuous vocoder. I demonstrated that using continuous parameters provide accurate and efficient system that convert source speech signal to the target one. I experimentally proved that the new method improves similarity compared to the conventional method (Chapter 10).*

**Thesis V.2: Parallel VC with Sinusoidal Model [C7, J5]** *I proposed a new approach to develop a voice conversion system using the continuous sinusoidal model, which decomposes the source voice into harmonic components and models contF0 to improve VC performance. I have validated the new system on parallel training data and showed its superiority against state-of-the-art solutions (Chapter 11).*

## 12.2 Applicability of New Results

This Thesis work and its results are not only scientifically evaluable; they are useful for the current state of the speech technology applications and profession. I have proposed and thoroughly examined the concept of vocoding for speech synthesis and voice conversion. Certainly, the proposed frameworks and algorithms are independent of any spoken language, which can directly be used in many speech applications to provide significantly better synthetic speech performance. Here, the practical applications of the results of this dissertation are summarized.

The results of Thesis group I – excitation noise modelling – is expected to be used for making speech synthesis more natural and expressive. In particular, the temporal envelope of Thesis I.1 is a relevant acoustic feature which can contribute to get a reliable estimation of the speech features in human study, whereas it can be used to avoid artifacts near the voicing boundaries in order to improve the quality of statistical parametric speech synthesis system. The novel outcomes of Thesis I.2 play a role to improve speech intelligibility and enhance voice qualities (hoarseness, breathiness, and creaky voice) in various speech synthesis systems; for example, creaky voice segments are not properly reconstructed in both HMPD and STRAIGHT vocoders [25]. Thus, Thesis group I attempted to further assist in reducing the effects of residual noise caused by the inaccurate excitation modelling.

The results of Thesis group II – excitation harmonic modelling – provides a reference for selecting appropriate techniques to optimize and improve the performance of current fundamental frequency estimation methods based on speech synthesis. Thesis II.1 can be used to reduce the fine error that the voiced section is wrongly identified as the unvoiced section, and improving temporal resolution of the estimated F0 trajectory. The time warping scheme has the ability to track the time-varying F0 period, and reduce the amount of F0 trajectory deviation from their harmonic locations. Whereas the instantaneous frequency approach is computationally inexpensive and can be highly useful in real-time processing speech synthesis applications. Thesis II.2 can be used to indicate the degree of voicing in the excitation, to detect the pitch exactly in various speech applications, and hence subsequently reducing the influence of buzziness caused by the vocoder.

The results of Thesis group III – acoustic modelling – is the possibility of creating new DNN-TTS voices automatically (e.g., from a speech corpus of telephone conversations) for simple devices (e.g. smartphones). This Thesis demonstrates the superiority of DNN acoustic models over the decision tree used in the HMM. Thesis III.1 based on DNN and Thesis III.2 based on RNN have already been applied in TTS with a developed vocoder as a simple, flexible, and powerful alternative acoustic model for SPSS to significantly improve the synthetic speech quality.

The results of Thesis group IV – sinusoidal modelling – introduces a vocoder-based speech synthesis system to improve the sound quality of real-time applications. This new speech synthesis system can be used in various speech technology, such as voice conversion, speech manipulation, and singing synthesizers. Thesis IV.1 can handle a wide variety of speakers and speaking conditions and give natural sounding speech comparable to the state-of-the-art STRAIGHT vocoder. Besides significant quality improvements over the baseline, the resulting system in Thesis IV.2 can be used in many speech applications, including message readers (SMS, e-mail, e-book, etc.), and navigation systems.

The results of Thesis group V – voice conversion – give an advanced novel approach to improve the conversion performance. The systems from both Thesis V.1 and V.2 have already been applied in the speaker conversion application using the continuous vocoder based on source-filter and sinusoidal model, respectively. These methods were tested with English speech corpora; however, it could be easily extended to other languages as well. This Thesis can also be applied in emotion conversion, virtual-augmented reality systems (voice avatars), accent conversion in language learning [157], and other speech assistance for overcoming speech impairments [158].

In addition to the individuals mentioned above, such an application is already under development within cooperation with an Egyptian university to create Arabic text-to-speech synthesis engine, in which continuous vocoder was applied on a modern standard Arabic audio-visual corpus which is annotated both phonetically and visually to produce a high-quality Arabic emotion TTS system [J1]. The general application of this TTS engine is to make a screen reader for Arabic's blind users. Moreover, continuous vocoder has already been applied in silent speech interfaces [C6], which is a form of spoken communication where an acoustic signal is not produced [159]. Continuous parameters were predicted form ultrasound tongue image by using the automatic articulatory-to-acoustic mapping, in which deep convolution neural network was used to learn the mapping task. Such an application can be applied to help the communication of the speaking impaired (e.g. patients after laryngectomy).

## 12.3 Future Research Directions

I believe that the work presented in each chapter opens up a number of interesting research directions to increase the overall SPSS quality. Here, I highlight key directions for future research.

- **Noise Generation:** In this dissertation, I generally used white noise for generating aperiodic components in unvoiced speech and aperiodic noise in voiced speech. Since short-term white noise includes a zero-frequency component and inaudible components below 20 Hz, these components must be reduced in advance when synthesizing to prevent degradation in the synthesized speech. In [160], a new variant of the velvet noise generation algorithm is proposed and shown to be superior to white noise in the perceived smoothness and stability of short-term power. Therefore, a work to do next is to use a new noise generation algorithm based on velvet noise for vocoder-based speech synthesis.

- **Systematic Test:** As both source-filter and sinusoidal vocoders proposed here are highly suited for SPSS, reached the state-of-the art results in speech synthesis and voice conversion, and outperformed other interesting vocoders (e.g. STRAIGHT that requires considerable computational resources.), one of the future areas of research that I suggest is to test them systematically for speech recognition, speech manipulation, or speaker verification.

- **Voice Conversion:** Although the proposed model introduced in Part V achieved the best result in similarity test, I observe quality degradation in the MUSHRA test. This is caused by the feed-forward neural network. So one direction for future work is to try to use recurrent neural network (e.g. LSTM). Further research could also lead to consider the non-parallel training data using generative adversarial networks for cross-lingual applications.

- **Joint Training:** One of the most important merits of using I suggest to build an architecture for both TTS and VC tasks based on the developed vocoders. This model can accomplish these two different tasks using both source-filter and sinusoidal vocoders according to the type of input. An end-to-end TTS task is conducted when the model is given text as the input while a sequence-to-sequence VC task is conducted when it is given the speech of a source speaker as the input.

- **Pitch Tracking:** Considering the suitability of adaptive continuous F0 discussed Chapter 4, the speech signal can be reconstructed from other high-quality vocoders (such as PML, MagPhase, HNM, and WORLD) with the adContF0 transformed from continuous vocoder.

- **Sampling Rate:** A small further examination could be done regarding the sampling frequency selection. The vocoder is designed under the assumption of 16kHz. With an increase in sampling rate (e.g. 48kHz), continuous vocoder and CSM need to be further studied. Listening tests with higher sampling rates are also necessary.

# Acknowledgements

First and foremost, I would like to utmost gratitude to my supervisors, Professor Géza Németh and Dr. Tamás Gábor Csapó, for their precious guidance, advices, dynamism, and friendship. Without their help this research would never have come to fruition. I learned a lot from the regular discussion every week and their wisdom, insight, diligence and passion in research. It was an honor for me to work under their supervisions. Also, this dissertation would not have been possible unless the support from the Stipendium Hungaricum scholarship.

I would like to thank my examiners for their insightful comments and valuable suggestions to improve this dissertation. In addition, I would like to extend my appreciation to the head of doctoral schools Professor János Levendovszky and Professor József Bíró who provided me with useful tips, advices, motivations, or support during my PhD process. I also have to acknowledge all the members of staff at the Department of Telecommunications and Media Informatics in the Faculty of Electrical Engineering and Informatics. I am thankful to all my colleagues at the Speech Technology and Smart Interactions Laboratory, it was a real pleasure to work in such a familiar and productive atmosphere. I am thankful to my friends for their help and all the good moments we shared.

The biggest thanks go to my father and my deceased mother. For many years, they have offered everything possible to support me. Without their encouragements, I would not be here. This Thesis is dedicated to them. I would also like to specially thank my dear sisters and brother for their help and taking care of my father during my long absence. Special thanks to my relatives Jameel and Firdews for their support in so many ways. I am thankful to all the people who prayed for me when I encountered various obstacles.

Last but not least, the greatest "Thank you" to Professor Khalil Alkadhimi of the Portsmouth University for standing beside me with his encouragement and unconditional support. Finally, and on a more personal note, an infinity of thanks to my wonderful wife Safa, our princess daughter Ruqaiya, and our precious son Hussain; they are simply a sunshine, and through their joy of life brought an ocean of love and happiness in me.

# Figures

# Tables

# Publications

## Publications related to Ph.D. Thesis

**International journals (peer-reviewed)**

[J1]   Mohammed Salah Al-Radhi, Omnia Abdo, Tamás Gábor Csapó, Sherif Abdou, Géza Németh, Mervat Fashal, A continuous vocoder for statistical parametric speech synthesis and its evaluation using an audio-visual phonetically annotated Arabic corpus, *Computer Speech and Language*, ScienceDirect Elsevier, 60, pp. 1-15, 2020.
(WoS, IF = 2.12, Q1), [50% · 6p = 3 points]

[J2]   Mohammed Salah Al-Radhi, Tamás Gábor Csapó, Géza Németh, Adaptive refinements of pitch tracking and HNR estimation within a vocoder for statistical parametric speech synthesis. *Applied Sciences*, 9(12), 2460, pp. 1-23, 2019.
(WoS, IF = 2.47, Q1), [50% · 6p = 3 points]

[J3]   Mohammed Salah Al-Radhi, Tamás Gábor Csapó, Géza Németh, Continuous vocoder applied in deep neural network based voice conversion, *Multimedia Tools and Applications*, 78 (23), Springer, pp. 1-24, 2019.
(WoS, IF = 2.31, Q1), [50% · 6p = 3 points]

[J4]   Mohammed Salah Al-Radhi, Tamás Gábor Csapó, Géza Németh, Continuous noise masking based vocoder for statistical parametric speech synthesis, *IEICE Transactions on Information and* Systems, E103-D (5), 2020.
(WoS, IF = 0.58, Q3), [50% · 6p = 3 points]

[J5]   Mohammed Salah Al-Radhi, Tamás Gábor Csapó, Géza Németh, Noise and acoustic modeling with waveform generator in text-to-speech and neutral speech conversion, *Multimedia Tools and Applications*, 79, Springer, pp. 1-26, 2020.
(WoS, IF = 2.31, Q1), [50% · 6p = 3 points]

**International conferences (peer-reviewed)**

[C1]   Mohammed Salah Al-Radhi, Tamás Gábor Csapó, Géza Németh, Time-domain envelope modulating the noise component of excitation in a continuous residual-based vocoder for statistical parametric speech synthesis, *in Proceedings of the 18th Interspeech conference*, pp. 434-438, Stockholm, Sweeden, 2017.
(Scopus), [50% · 3p = 1.5 points]

[C2]  Mohammed Salah Al-Radhi, Tamás Gábor Csapó, Géza Németh, Continuous vocoder in feed-forward deep neural network based speech synthesis, *in Proceedings of the 11$^{th}$ Digital speech and image processing conference*, pp. 1-4, Novi Sad, Serbia, 2017.                               (SemanticScholar), [50% · 3p = 1.5 points]

[C3]  Mohammed Salah Al-Radhi, Tamás Gábor Csapó, Géza Németh, Deep recurrent neural networks in speech synthesis using a continuous vocoder. In: Karpov A., Potapova R., Mporas I. (eds) *Speech and Computer. SPECOM. Lecture Notes in Computer Science*, vol 10458. Springer, pp. 282-291, Hatfield, England, 2017.

(Scopus, chapter book), [50% · 3p = 1.5 points]

[C4]  Mohammed Salah Al-Radhi, Tamás Gábor Csapó, Géza Németh, A continuous vocoder using sinusoidal model for statistical parametric speech synthesis. In: Karpov A., Jokisch O., Potapova R. (eds) *Speech and Computer. SPECOM. Lecture Notes in Computer Science*, vol 11096. Springer, pp. 11-20, Leipzig, Germany, 2018.

(Scopus, chapter book), [50% · 3p = 1.5 points]

[C5]  Mohammed Salah Al-Radhi, Tamás Gábor Csapó, Géza Németh, RNN-based speech synthesis using a continuous sinusoidal model, *in Procedings of the 28$^{th}$ IEEE International Joint Conference on Neural Networks (IJCNN)*, pp. 1-8, Budapest, Hungary, 2019.                                          (IEEE), [50% · 3p = 1.5 points]

[C6]  Tamás Gábor Csapó, Mohammed Salah Al-Radhi, Géza Németh, Gábor Gosztolya, Tamás Grósz, László Tóth, Alexandra Markó, Ultrasound-based silent speech interface built on a continuous vocoder, *in Proceedings of the 20$^{th}$ Interspeech conference*, pp. 894-898, Graz, Austria, 2019.

(Scopus), [17% · 3p = 0.51 points]

[C7]  Mohammed Salah Al-Radhi, Tamás Gábor Csapó, and Géza Németh, Parallel voice conversion based on a continuous sinusoidal model, *in Proceedings of the 10$^{th}$ IEEE Speech Technology and Human-Computer Dialogue conference*, pp. 1-6, Timisoara, Romania, 2019.                                       (IEEE), [50% · 3p = 1.5 points]

**International abstract conferences (peer-reviewed)**

[C8]  Mohammed Salah Al-Radhi, Tamás Gábor Csapó, Géza Németh, Effects of adding a Harmonic-to-Noise Ratio parameter to a continuous vocoder, *in Proceedings of the 6$^{th}$ of the UKspeech*, Cambridge University, England, 2017.

(poster, 1 page), [50% · 1p = 0.5 points]

[C9]  Mohammed Salah Al-Radhi, Tamás Gábor Csapó, Géza Németh, High quality continuous vocoder in deep recurrent neural network based speech synthesis, *in Eastern European Machine Learning*, Bucharest, Romania, 2019.

(poster, 2 pages), [50% · 0p = 0 points]

[C10] <u>Mohammed Salah Al-Radhi</u>, Tamás Gábor Csapó, Géza Németh, Improving continuous F0 estimator with adaptive time-warping for high-quality speech synthesis, *in Beszédkutatás (conference of the speech reseacrch)*, Budapest, Hungary, 2018.                                          (oral, 2 pages), [50% · 0p = 0 points]

**International doctoral consortium conferences (peer-reviewed)**

[C11] <u>Mohammed Salah Al-Radhi</u>, High quality continuous residual-based vocoder for statistical parametric speech synthesis, *International Speech Communication Association (ISCA-SAC), Interspeech*, KTH Royal Institute of Technology, Stockholm, Sweeden, 2017.
                                 (Googlescholar, oral, 3 pages), [100% · 0p = 0 points]

[C12] <u>Mohammed Salah Al-Radhi</u>, High-quality vocoding for speech synthesis and voice conversion, *International Joint Conference on Neural Networks (IJCNN)*, Budapest, Hungary, 2019.

                                           (oral, 3 pages), [100% · 0p = 0 points]

**Additional publications (my contribution related to deep learning)**

**International journals (peer-reviewed)**

[J6]  W. I. Hameed, B. A. Sawadi, Safa J. Al-Kamil, <u>Mohammed Salah Al-Radhi</u>, Y. I. Al-Yasir, A. L. Saleh, R. A. Abd-Alhameed, Prediction of solar irradiance based on artificial neural networks. *Inventions*, 4(3), 45, pp. 1-10, 2019.
                                                  (Scopus), [15% · 6p = 0.9 points]

**Independent citations**

[C4-1] Jiang C., Chen Y., Cho C., A Novel Genetic Algorithm for Parameter Estimation of Sinusoidal Signals, *12th International Congress on Image and Signal Processing, BioMedical Engineering and Informatics (CISP-BMEI)*, Suzhou, pp. 1-5, 2019.

[C3-1] Bittner Ksenia, Building Information Extraction and Refinement from VHR Satellite Imagery using Deep Learning Techniques, *PhD thesis, Osnabrück University*, Germany, 2020.

## Total Ph.D. publications score: 25.91 points